\documentclass[useAMS,usenatbib]{aa}
\usepackage{txfonts}
\usepackage[authoryear]{natbib}
\bibpunct{(}{)}{;}{a}{}{,}
\usepackage[dvips]{graphicx}

\newcommand{\simgt}{\lower.5ex\hbox{$\; \buildrel > \over \sim \;$}}
\newcommand{\simlt}{\lower.5ex\hbox{$\; \buildrel < \over \sim \;$}}

\begin{document}

\title{HIFLUGCS: Galaxy cluster scaling relations between X-ray luminosity,
  gas mass, cluster radius, and velocity dispersion}

\author{Yu-Ying Zhang\inst{1,2}, 
Heinz Andernach\inst{1,3},
C\'esar A. Caretta\inst{3},
Thomas H. Reiprich\inst{1},
Hans B\"ohringer\inst{4},
Ewald Puchwein\inst{5},
Debora Sijacki\inst{6},
\and
Marisa Girardi\inst{7,8}}

\institute{Argelander-Institut f\"ur Astronomie, Universit\"at Bonn, Auf dem
  H\"ugel 71, 53121 Bonn, Germany 
\and National Astronomical
    Observatories, Chinese Academy of Sciences, Beijing, 100012, China
\and Departamento de Astronom\'{\i}a, Universidad de Guanajuato, AP 144, Guanajuato CP 36000, Mexico
\and Max-Planck-Institut f\"ur extraterrestrische Physik, Giessenbachstra\ss e, 85748 Garching, Germany
\and Max-Planck-Institut f\"ur Astrophysik, Karl-Schwarzschild-Stra\ss e 1,
85741 Garching, Germany
\and Kavli Institute for Cosmology, Cambridge and
Institute of Astronomy, Madingley Road, Cambridge, CB3 0HA, United Kingdom
\and Dipartimento di Fisica dell' Universit\'a degli Studi di Trieste – Sezione di Astronomia, via Tiepolo 11, 34143 Trieste, Italy
\and INAF – Osservatorio Astronomico di Trieste, via Tiepolo 11, 34143 Trieste, Italy 
}

\authorrunning{Zhang et al.}

\titlerunning{HIFLUGCS: Galaxy cluster scaling relations between $L_{\rm
    bol,500}$, $M_{\rm gas,500}$, $r_{500}$, and $\sigma$}

\date{Received 28/09/2010 / Accepted 10/11/2010}

\offprints{Y.-Y. Zhang}

\abstract{We present relations between X-ray luminosity and velocity
  dispersion ($L-\sigma$), X-ray luminosity and gas mass ($L-M_{\rm gas}$),
  and cluster radius and velocity dispersion ($r_{500}-\sigma$) for 62 galaxy
  clusters in the HIFLUGCS, an X-ray flux-limited sample minimizing bias
  toward any cluster morphology. Our analysis in total is based on $\sim
  1.3$~Ms of clean X-ray \emph{XMM-Newton} data and 13439 cluster member
  galaxies with redshifts. Cool cores are among the major contributors to the
  scatter in the $L-\sigma$ relation. When the cool-core-corrected X-ray
  luminosity is used the intrinsic scatter decreases to 0.27~dex. Even after
  the X-ray luminosity is corrected for the cool core, the scatter caused by
  the presence of cool cores dominates for the low-mass systems. The scatter
  caused by the non-cool-core clusters does not strongly depend on the mass
  range, and becomes dominant in the high-mass regime. The observed $L-\sigma$
  relation agrees with the self-similar prediction, matches that of a
  simulated sample with AGN feedback disregarding six clusters with $<45$
  cluster members with spectroscopic redshifts, and shows a common trend of
  increasing scatter toward the low-mass end, i.e., systems with $\sigma \le
  500{\rm km}\;{\rm s}^{-1}$. A comparison of observations with simulations
  indicates an AGN-feedback-driven impact in the low-mass regime. The best
  fits to the $L-M_{\rm gas}$ relations for the disturbed clusters and
  undisturbed clusters in the observational sample closely match those of the
  simulated samples with and without AGN feedback, respectively. This suggests
  that one main cause of the scatter is AGN activity providing feedback in
  different phases, e.g., during a feedback cycle. The slope and scatter in the
  observed $r_{500} - \sigma$ relation is similar to that of the simulated
  sample with AGN feedback except for a small offset but still within the
  scatter.
\begin{keywords}
  Cosmology: observations --- (Cosmology:) dark matter --- Galaxies:
  clusters: general --- Methods: data analysis --- Surveys --- X-rays:
  galaxies: clusters
\end{keywords}
}

\maketitle

\section{Introduction}

Galaxy clusters have been suggested as a potential probe of the dark energy
equation of state parameter ($w=p/\rho$, where $\rho$ is the energy density
and $p$ is the pressure), through the evolution of the mass function (e.g.,
Schuecker et al. 2003; Predehl et al. 2007; Henry et al. 2009; Vikhlinin et
al. 2009a, 2009b; Mantz et al. 2010). Observational surveys select galaxy
clusters by their observables rather than by their mass. Therefore, a
relationship between the cluster total mass and an observable such as X-ray
luminosity is required to recover the selection function of an X-ray survey in
terms of cluster masses and predict the cluster mass, hence the cluster mass
function. During the past, there have been a large number of studies of X-ray
luminosity scaling relations along with their applications to constrain
cosmological parameters in galaxy cluster surveys and the physical state of
the hot intracluster medium (ICM) in galaxy clusters (e.g., Henry \& Tucker
1979; Henry \& Arnaud 1991; Edge \& Stewart 1991; David et al. 1993; Fabian et
al. 1994; Girardi et al. 1996; Mushotzky \& Scharf 1997; Cavaliere et
al. 1997; White et al. 1997; Markevitch 1998; Wu et al. 1998, 1999; Allen \&
Fabian 1998; Arnaud \& Evrard 1999; Reiprich \& B\"ohringer 2002; Ota et
al. 2006; Chen et al. 2007; Zhang et al. 2006, 2008; Pratt et al. 2009;
Leauthaud et al. 2010; Stanek et al. 2010). Large X-ray cosmology surveys,
e.g., by \emph{eROSITA}, are expected to substantially improve cosmological
constraints using a large number of galaxy clusters. For \emph{eROSITA}, the
use of X-ray mass proxies has been proposed, specifically X-ray luminosity, to
infer the total mass and construct the selection function in the forthcoming
wide survey of the satellite (Predehl et al. 2007). The superb quality X-ray
data in the \emph{XMM-Newton} archive provide us with an excellent opportunity
to calibrate the luminosity scaling relations and more clearly understand the
X-ray selection method.

Simulations show that the formation of galaxy clusters is not a purely
gravitational process; The galaxy velocity dispersions of clusters
appear to indicate that heating is present when compared to the cold
dark matter (CDM) velocity dispersion normalized to the
\emph{Wilkinson Microwave Anisotropy Probe} (\emph{WMAP}) and
large-scale structure (LSS) distributions (Evrard et al.
2008). Cluster mergers not only change the cluster X-ray luminosity
(e.g., Ricker \& Sarazin 2001; Poole et al. 2006), but also affect the
properties of the cluster galaxies (e.g., Sun et al.  2007; Smith et
al. 2010). Although the hot gas and galaxies are not pure tracers of
the gravitational potential of galaxy clusters, they are indeed
sensitive probes of the dynamical properties of galaxy clusters, and
react on different timescales during a merger in simulations
(e.g., Roettiger et al. 1999). In particular, the optical information
about the line-of-sight velocity of cluster galaxies complements X-ray
information about the cluster morphology projected onto the sky. The
luminosity -- velocity dispersion ($L-\sigma$) relation of galaxy
clusters is thus crucial to understanding the dynamical properties of
galaxy clusters and their impact on the scaling relations and possibly
the X-ray selection bias (e.g., Wu et al. 1999; Ortiz-Gil et
al. 2004).

To carry out the $L-\sigma$ studies, one requires a representative sample with
a well-defined selection function and minimal bias toward any cluster
morphology, as well as superb quality X-ray data and large amount of cluster
galaxy redshifts. The HIghest X-ray FLUx Galaxy Cluster Sample (HIFLUGCS,
Reiprich \& B\"ohringer 2002) of 64 galaxy clusters selected from the
\emph{ROSAT} All-Sky Survey (RASS; Ebeling et al. 2000; B\"ohringer et
al. 2004) is such a sample. In the HIFLUGCS, we analyzed all available X-ray
data in the \emph{XMM-Newton} archive for 63 clusters which represents nearly
$4$~Ms of data. After cleaning and selecting the longest observation closest
to the cluster center for clusters with multiple observations, we still have
$\sim 1.3$~Ms \emph{XMM-Newton} data for 59 clusters. For 62 clusters in the
HIFLUGCS, we obtained a sum of 13439 cluster member galaxies based on
spectroscopic redshifts and performed a careful exclusion of non-members. In
the end, we were able to measure X-ray observables, combining \emph{XMM-Newton}
and \emph{ROSAT} data, and velocity dispersion, based on 13439 cluster
members, to make a cross-calibration for 62 out of 64 clusters in the
HIFLUGCS.

The outline of this paper is as follows. We describe the data analysis
in Sect.~\ref{s:ana}, present the scaling relations of the 62 clusters
in the HIFLUGCS in Sect.~\ref{s:scaling}, compare the observational
and simulated samples in Sect.~\ref{s:sim}, discuss the systematic
errors in determining the velocity dispersion in Sect.~\ref{s:bias},
and summarize our conclusions in Sect.~\ref{s:con}.  Our Appendix
provides extra information on the cross-calibration between
\emph{XMM-Newton} and \emph{ROSAT}, the iron abundance versus (vs.)
temperature correlation, results using either the 0.5--2~keV X-ray
luminosity corrected for the presence of a cool core ($\le
0.2r_{500}$), or the luminosity including or excluding the cluster
core, the \emph{XMM-Newton} 0.7--2~keV images, and the figures
illustrating systematic errors in estimates of $\sigma$. Throughout
the paper, we assume that $\Omega_{\rm m}=0.3$, $\Omega_\Lambda=0.7$,
and $H_0=70$~km\,s$^{-1}$\,Mpc$^{-1}$. Confidence intervals correspond
to the 68\% confidence level. Unless explicitly stated otherwise, we
apply the BCES regression fitting method taking into account
measurement errors in both variables (Akritas \& Bershady 1996).

\section{Data analysis}
\label{s:ana}

\subsection{Optical data analysis and velocity dispersion}
\label{s:bcgsigma}

We draw the velocity of the cluster galaxies from the literature (updated
until March 2010, including the compilation in Andernach et al. 2005). When
there is more than one velocity per galaxy, we calculate an
average\footnote{Since the individual error estimates are inhomogeneous, we
  decided not to weight the calculation of the average velocity.} of the
measurements, excluding discordant values and those with large errors when
more than one measurement is available.

Brightest cluster galaxies (BCGs) in galaxy clusters are almost invariably
giant ellipticals and are more luminous than normal galaxies. The BCGs have
line-of-sight velocities that are similar to the mean of their host clusters
and extended stellar envelopes. We identify the BCG on the basis of its
apparent magnitude and spectroscopic confirmation as a cluster member. To
define a BCG position for every HIFLUGCS cluster, we made the following
choices for clusters without a single dominant BCG. A0400 and A2065 have
dumbbell BCGs, and A3158 and A2256 have BCG pairs, for which we place the BCG
positions in the middle of the two components of indistinguishable
brightness. A3266, A3391, A0576, A2634, MKW8, and IIIZw54 have dumbbell BCGs,
and Coma and Hydra (A1060) have two brightest galaxies of similar
brightness, for which we place the BCG position on the brighter component as
the difference in the brightness is measurable. A2199 has multiple nuclei, and
we place the BCG position at the brightest nucleus. We list the BCG positions
in Table~\ref{t:sigma}.

\begin{figure*}
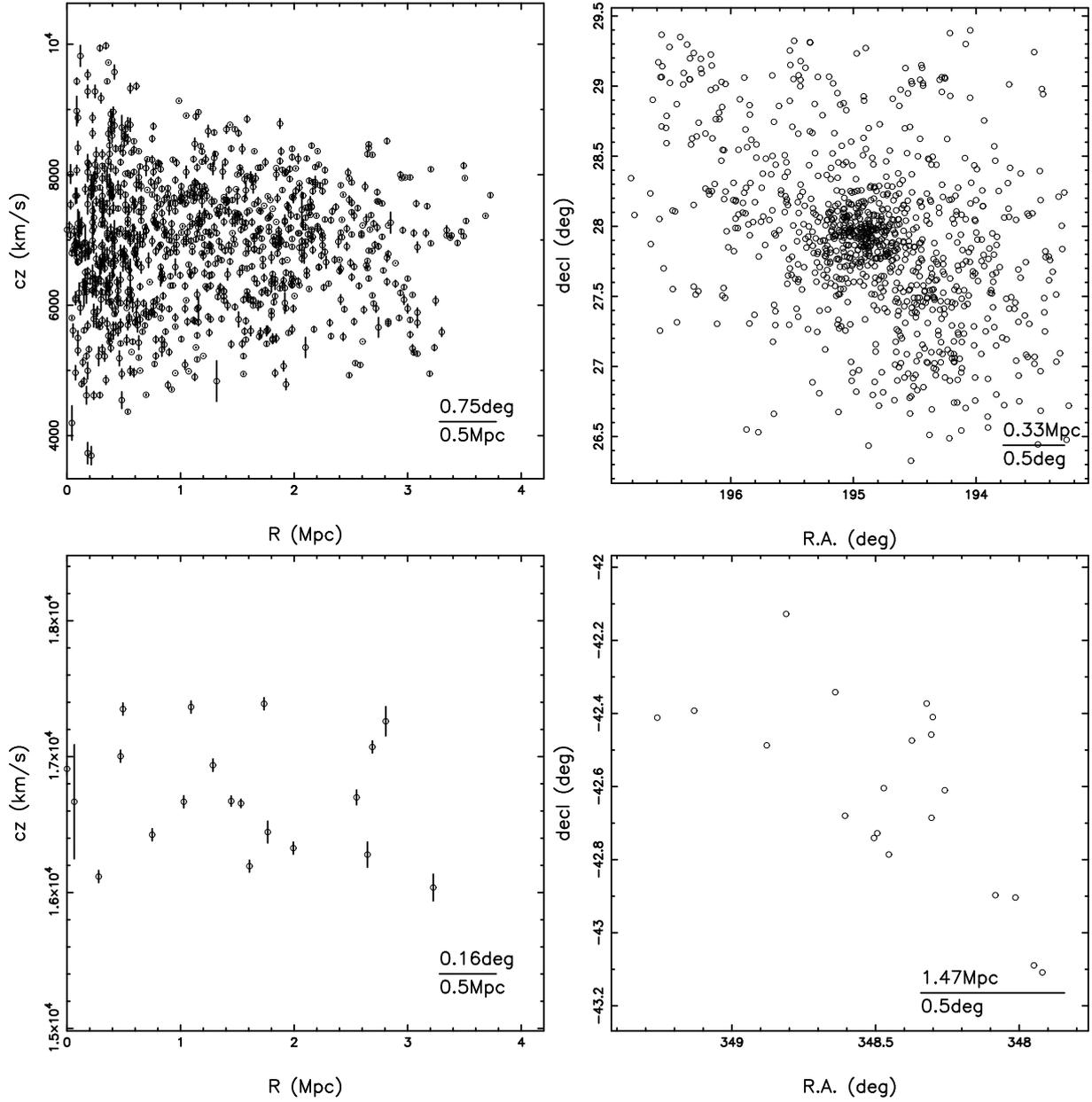

\begin{center}
\includegraphics[angle=270,width=8cm]{plots/15830f1a.ps}
\includegraphics[angle=270,width=8cm]{plots/15830f1b.ps}

\includegraphics[angle=270,width=8cm]{plots/15830f1c.ps}
\includegraphics[angle=270,width=8cm]{plots/15830f1d.ps}
\end{center}
\caption{Line-of-sight velocity vs. projected radius (left panels) and sky
  positions (right panels) of the selected galaxies in a rich cluster, i.e., 
  Coma (top panels) and
  in a poor cluster, i.e., S1101 (bottom panels).
  \label{f:caustic}}
\end{figure*}

As most BCGs are located very near the X-ray flux-weighted cluster
centers (definition see \S~\ref{s:offset}), we select preliminarily
galaxies with spectroscopic redshifts in each cluster within an
aperture of at least 1.2~Abell radius, i.e., $2.57$~Mpc, centered on
the BCG. For each cluster, we plot the line-of-sight velocity of the
selected galaxies as a function of their projected distance from the
BCG, and locate the caustic, a trumpet-shaped region, which
efficiently excludes interlopers (e.g., Diaferio 1999; Katgert et
al. 2004; Popesso et al. 2005; Rines \& Diaferio 2006). We consider
only the galaxies inside the caustic as cluster members, and exclude
the others from subsequent analysis. More than 80\% of the clusters
have a clearly evident caustic shape. In Fig.~\ref{f:caustic}, we show
as an example the caustic and sky positions of the cluster galaxies in
Coma, and as a poor example in S1101. There are eight clusters that
have fewer than 45 cluster members with spectroscopic redshifts in the
HIFLUGCS. We excluded 2A0355 and RXCJ1504 from our study since both
have at most three redshifts each. We still consider the remaining six
systems (i.e., A0478, NGC1550, EXO0422, HydraA, S1101, and A2597) with
$>12$ but $<45$ cluster members with spectroscopic redshifts, and
highlight them in our results. We gathered a total of 13439
cluster-member galaxies based on spectroscopic redshifts and a careful
exclusion of non-members, which gives a median value of 185.5 per
cluster.

For the galaxies selected as cluster members we apply the bi-weight estimator
(e.g., Beers et al. 1990) to measure the velocity dispersion. The errors are
estimated through 1000 bootstrap simulations. We list the number of cluster
members ($n_{\rm gal}$) and the velocity dispersion ($\sigma$) of the cluster
for 62 clusters in the HIFLUGCS in Table~\ref{t:sigma}. The systematic errors
in the determination of the velocity dispersion are discussed in
Sect.~\ref{s:bias}.

\subsection{X-ray data analysis} 
\label{s:screen}

There are 63 clusters in the HIFLUGCS in the \emph{XMM-Newton} archive. Only
A2244 has not yet been observed. We analyzed 150 \emph{XMM-Newton}
observations, which give 3.90~Ms for MOS1, 3.97~Ms for MOS2, and 3.68~Ms
for pn, respectively. To filter flares, we apply iterative screening similar
to Zhang et al. (2006) using both the soft (0.3--10~keV) band and the hard
(10--12~keV for MOS, 12--14~keV for pn) band but with a 3.3-$\sigma$
clipping. We found that the \emph{XMM-Newton} observations of four clusters
(i.e., A0401, A0478, A1736, A2163) are flared. For clusters with multiple
observations, we select the longest observation of which the pointing position
is the closest to the cluster center. Since 2A0355 and RXCJ1504 have at most
three redshifts each, we exclude these two clusters, and end up with nearly
1.3~Ms of \emph{XMM-Newton} data of 57 clusters for a more detailed analysis.

For the four flared clusters (i.e., A0401, A0478, A1736, A2163), as well as
for A2244, the X-ray quantities are derived from \emph{ROSAT} pointed
observations. The X-ray quantities for the remaining 57 clusters in the
  HIFLUGCS are derived from combined \emph{XMM-Newton} and \emph{ROSAT}
data. We note that the \emph{XMM-Newton} observations only cover an incomplete
sector of A2142, such that we have to use the \emph{ROSAT} data to derive its
surface brightness profile. The \emph{XMM-Newton} observations of A2142 are
only used to measure the global temperature and iron abundance. We describe in
detail the procedures we adopted to detect and subtract point-like sources and
for background treatment in Sect.~2 and Sect.~3 of Zhang et
al. (2009). Significant substructure features clearly detected in the image
are excised before we perform the spectral and surface brightness analysis. We
note that the surface brightness analysis is slightly different from that in
Sect.~4 of Zhang et al. (2009) in that we directly convert the \emph{ROSAT}
surface brightness profile to the \emph{XMM-Newton} count rate using the
best-fit spectral model obtained from the \emph{XMM-Newton} data. We then
combine the \emph{XMM-Newton} surface brightness profile within the truncation
radius, where the \emph{XMM-Newton} signal-to-noise ratio is $\sim 3$, with
the \emph{ROSAT} converted surface brightness profile beyond the truncation
radius for further analysis. We list the properties of the \emph{XMM-Newton}
observations, redshift, hydrogen column density, gas mass, X-ray morphology,
and presence of a cool core of each cluster in Table~\ref{t:xmm}.

\subsubsection{Cluster radius, i.e., $r_{500}$}
\label{s:r500}

X-ray quantities have to be derived consistently within a certain cluster
radius, e.g., $r_{500}$, the radius within which the mass density is 500 times
of the critical density\footnote{The critical density is given by $\rho_{\rm
    c}(z)=E^2(z)3H_{0}^2(8\pi G)^{-1}$, where $E^2(z)=\Omega_{\rm
    m}(1+z)^3+\Omega_{\Lambda}+(1-\Omega_{\rm
    m}-\Omega_{\Lambda})(1+z)^2$.}, at the cluster redshift. The quantity
$r_{500}$ can be measured from the X-ray measured mass distribution derived
under the assumption of hydrostatic equilibrium as we did in Zhang et
al. (2009). Observations have found evidence of deviations from hydrostatic
equilibrium (e.g., Zhang et al. 2008, 2010; Mahdavi et al. 2008). The
cross-calibration between weak lensing masses and X-ray observables instead
uncovers a tight scaling relation between gas mass and cluster total mass
(e.g., Okabe et al. 2010). We therefore use the gas mass to infer the cluster
mass and $r_{500}$. Our sample occupies a wide mass range with the gas masses
from $1.74 \times 10^{11} M_{\odot}$ to $2.12 \times 10^{14} M_{\odot}$, which
is similar to the mass range of the extended sample in Pratt et
  al. (2009) consisting of 41 groups and clusters collected from Vikhlinin et
al. (2006), Arnaud et al. (2007), B\"ohringer et al. (2007), and Sun et
al. (2008). We thus adopt their relation $E^{1.5}(z) \ln (M_{\rm
  gas,500}/M_{500})=-2.37+0.21 \ln (M_{500}/2\times 10^{14} M_{\odot})$
to derive the cluster mass and radius ($r_{500}$) from our gas mass
estimate. 

\subsubsection{X-ray luminosity}
\label{s:lx}

The X-ray luminosity is estimated by integrating the X-ray surface
brightness. At $3\sigma$ significance, the surface brightness profiles are
detected out to at least $r_{500}$ for all 62 clusters combining
\emph{XMM-Newton} and \emph{ROSAT} data (see Zhang et al. 2009). In practice,
we estimate the total count rate from the background-subtracted, flat-fielded,
point-source-subtracted, and point-spread-function (PSF) corrected surface
brightness profile in the 0.7-2~keV band, and convert this to X-ray luminosity
using the best-fit ``mekal''\footnote{For A0401, A0478, A1736, A2163, and
  A2244, which have no \emph{XMM-Newton} data, we extracted the temperature
  from Vikhlinin et al. (2009a) assuming $0.3Z_{\odot}$ iron abundance.} model
given by the spectra in XSPEC in the aperture covering all annuli defined in
Sect.~3.2 in Zhang et al. (2009).  We note that we do not study the
temperature scaling relations for this sample here because of the
inhomogeneous range of projected distances used to measure the cluster
temperature.

We show the \emph{XMM-Newton}-\emph{ROSAT} vs. \emph{ROSAT}-only measured
X-ray luminosity in the 0.1--2.4~keV band in Fig.~\ref{f:lxlr} in
Appendix~\ref{a:lxlr}. The \emph{XMM-Newton}-\emph{ROSAT} to \emph{ROSAT}-only
measured X-ray luminosity ratio is $(92\pm2)$\% with $(0.07 \pm 0.01)$~dex
scatter. The faint point sources subtracted from the \emph{XMM-Newton} data may
account for a small fraction of the difference. A systematic difference in the
flux calibration between \emph{ROSAT} and \emph{XMM-Newton} might play a major
role in the 8\% difference in the X-ray luminosity (e.g., Snowden et
al. 2002). A good fraction of the scatter may be introduced by the varying
amounts of point sources and, especially, substructures that get excluded in
the \emph{ROSAT} and \emph{XMM-Newton} analysis.

In addition, there are some low-temperature systems (i.e., NGC507, Fornax,
NGC1550, MKW4, NGC4636, NGC5044, and A3581; $kT<2$~keV) in the sample. To
examine whether the line emission becomes important and boosts the X-ray
luminosity for those systems, we show the iron abundance vs. temperature
relation for the 62 clusters in Fig.~\ref{f:TZ} in Appendix~\ref{a:TZ}. The
best fit is $Z/Z_{\odot}=10^{-(0.323\pm 0.061)} (kT/{\rm keV})^{-(0.324\pm
  0.098)}$ using the bisector method and $Z/Z_{\odot}=10^{-(0.325\pm 0.043)}
(kT/{\rm keV})^{-(0.320\pm 0.068)}$ using the orthogonal method,
respectively. This is consistent with the results found in Balestra et
al. (2007) but for clusters at higher redshifts ($z\ge 0.3$) and in a higher
temperature range (3--15~keV), though their clusters show a steeper slope than
that for our nearby clusters. The iron abundance vs. temperature correlation
indicates that a flux-limited sample tends to include low-mass systems with
high iron abundance, of which the X-ray luminosity is in part boosted by
the line emission. This may modify the scaling relations at the
low-mass end in terms of the mass dependence of the slope and the intrinsic
scatter.

Fabian et al. (1994) pointed out that some clusters are significantly above
the best fit of the luminosity scaling relation because of the presence of
cool cores. This motivates the cluster core correction in deriving the X-ray
luminosity (e.g., Markevitch 1998). We focus on the results using the X-ray
luminosity corrected for the cluster core (hereafter $L^{\rm co}$) by assuming
a constant value in the cluster core equal to the value at $0.2 r_{500}$,
$S_{\rm X}(R < 0.2r_{500})=S_{\rm X}(0.2r_{500})$ (Zhang et al. 2007). We note
that this correction is only applied in determining the X-ray luminosity, not
the gas mass. The bolometric luminosity corrected for the cluster core is
listed in Table~\ref{t:sigma}, and the bolometric luminosity within $r_{500}$
(hereafter $L^{\rm in}$) and in the $[0.2-1.0] r_{500}$ radial range
(hereafter $L^{\rm ex}$) are listed in Table~\ref{t:lx_in_ex}. To examine the
scatter in the scaling relations caused by the presence of cool cores, we also
compare the results using $L^{\rm co}$ with those using $L^{\rm in}$
(Appendix~\ref{a:lin}) and $L^{\rm ex}$ (Appendix~\ref{a:lex}), respectively.

Since the soft band X-ray luminosity is widely used in studies of the scaling
relations, we calibrate the luminosity scaling relations using both the
bolometric luminosity in the 0.01-100~keV band ($L_{\rm bol}$) and the soft
band luminosity in the 0.5-2~keV band ($L_{\rm 0.5-2keV}$, see also
Appendix~\ref{a:lx}).

\subsection{Quantification of the cluster dynamical state}
\label{s:sub}

\subsubsection{Offset between the X-ray flux-weighted center and BCG position}
\label{s:offset}

The X-ray flux-weighted center of each cluster is listed in Cols.~2--3 of
Table~\ref{t:sigma}, which is determined based on \emph{XMM-Newton} data as
described in \S~2.3 in Zhang et al. (2010). Our choice of the BCG position
(Cols.~4--5 of Table~\ref{t:sigma}) is explained in \S~\ref{s:bcgsigma}. The
angular separation between the X-ray flux-weighted center and BCG position is
converted into the physical separation at the cluster redshift, and is listed
in Table~\ref{t:sigma}, in units of both kpc ($d_{\rm offset}$) and $r_{500}$
($d_{\rm offset}/r_{500}$).

The offsets between the X-ray flux-weighted centers and BCG positions for the
62 clusters closely follow a log-normal Gaussian distribution
(Fig.~\ref{f:offset}, left panel). The best fit of $\log_{10}(d_{\rm
  offset}/r_{500})$ gives a mean value of $-(1.93 \pm 0.06)$ and
$\sigma=(0.50\pm 0.06)$. Forty-six clusters show $\le 0.037r_{500}$ offsets,
within $1\sigma$ of the mean value. The remaining 16 clusters are
  sparsely spread over the range of $[0.037 - 1]r_{500}$. The best fit of
$\log_{10}(d_{\rm offset}/{\rm kpc})$ gives a mean value of $(1.03 \pm 0.06)$
and $\sigma=(0.55\pm 0.06)$. Forty-seven clusters show $\le 38$~kpc offsets,
within $1\sigma$ of the mean value. The remaining 15 clusters are sparsely spread
over the range of $[38 - 1000]$~kpc. Thirteen of those 16 clusters with large
offsets between the X-ray flux-weighted centers (see Table~\ref{t:sigma}) and
BCG positions are disturbed clusters (see Table~\ref{t:xmm} and
Sect.~\ref{s:xraymor}).

\subsubsection{Central cooling time}

\begin{figure*}
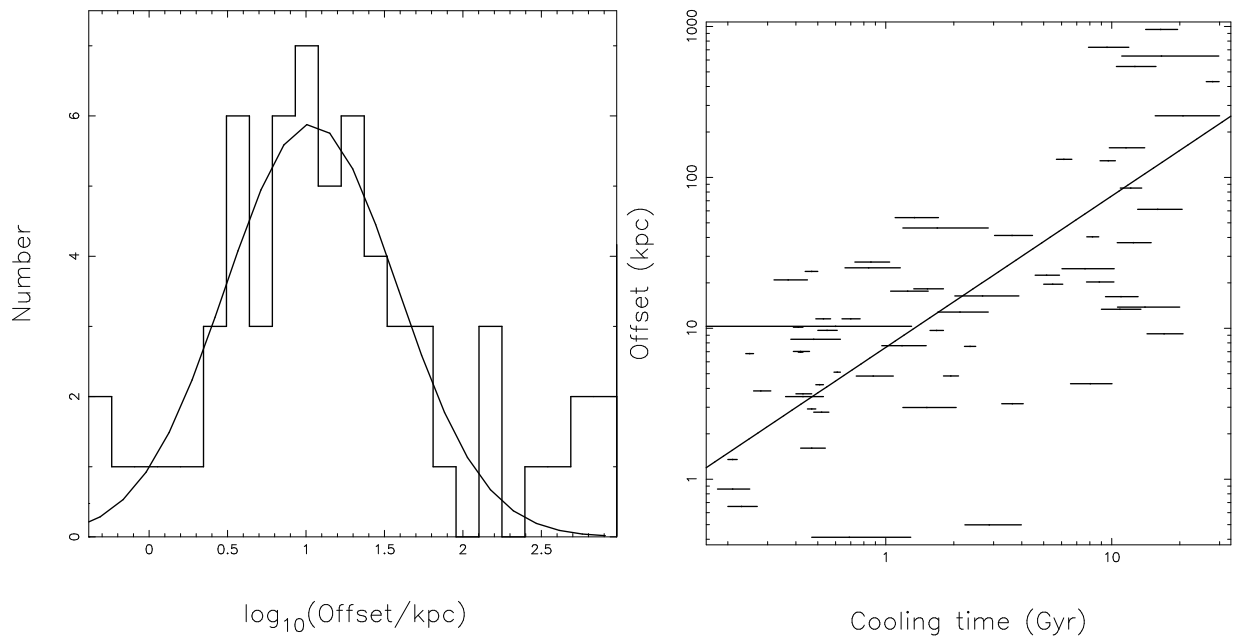

\begin{center}
\includegraphics[angle=270,width=8cm]{plots/15830f2a.ps}
\includegraphics[angle=270,width=8cm]{plots/15830f2b.ps}
\end{center}
\caption{Histogram of the offset between the X-ray flux-weighted center and
  BCG position (left panel) and central cooling time vs. offset (right panel).
  \label{f:offset}}
\end{figure*}

The central cooling time can be more accurately estimated from \emph{Chandra}
data because of its smaller PSF. We thus use the central cooling time
calculated at $0.004r_{500}$ from Eq.(15) in Sect.~2.6 in Hudson et al. (2010) to
divide the sample of the 62 clusters into 26 cool-core clusters (i.e., ``SCC''
in Hudson et al. 2010) and 36 non-cool-core clusters (i.e., ``NCC'' and
``WCC'' in Hudson et al. 2010) as listed in Table~\ref{t:xmm}. Interestingly,
we also found a correlation between the central cooling time and the offset
between the cluster center and BCG position (Fig.~\ref{f:offset}, right
panel). The best power-law fit to the relation between the offset and central
cooling time is $\log_{10}\left (\frac{\rm Offset}{r_{500}} \right)=(-2.051\pm
0.058) + (0.907 \pm 0.081)\log_{10}\left (\frac{\rm Cooling \;time}{\rm
    Gyr}\right)$ and $\log_{10}\left (\frac{\rm Offset}{\rm kpc}
\right)=(0.874\pm 0.059)+(1.003\pm 0.081)\log_{10} \left (\frac{\rm Cooling
    \;time}{\rm Gyr} \right)$.

\subsubsection{X-ray morphology}
\label{s:xraymor}

The combined MOS and pn images in the 0.7-2~keV band are shown in
Appendix~\ref{a:image}. According to their X-ray flux images, Vikhlinin et
al. (2009a) divide the 62 clusters into 41 undisturbed clusters and 21
disturbed clusters listed in Table~\ref{t:xmm}.
 
\section{Results for the observational sample}
\label{s:scaling}

We investigate the three scaling relations between the luminosity and velocity
dispersion, luminosity and gas mass, and cluster radius and velocity
dispersion, respectively, for the 62 clusters in the HIFLUGCS. To examine
possible systematic uncertainties due to the choice of the fitting method, we
apply the BCES bisector and orthogonal methods. For all 62 clusters, the best
power-law fits of all studied relations given by the bisector and orthogonal
methods are consistent (Table~\ref{t:scaling}). We therefore focus on the best
fits given by one of the two methods, i.e., the BCES bisector method, to
illustrate the results.

\subsection{$L-\sigma$ relation}
\label{s:lbv}

\begin{figure*}
\begin{center}
\includegraphics[angle=270,width=16cm]{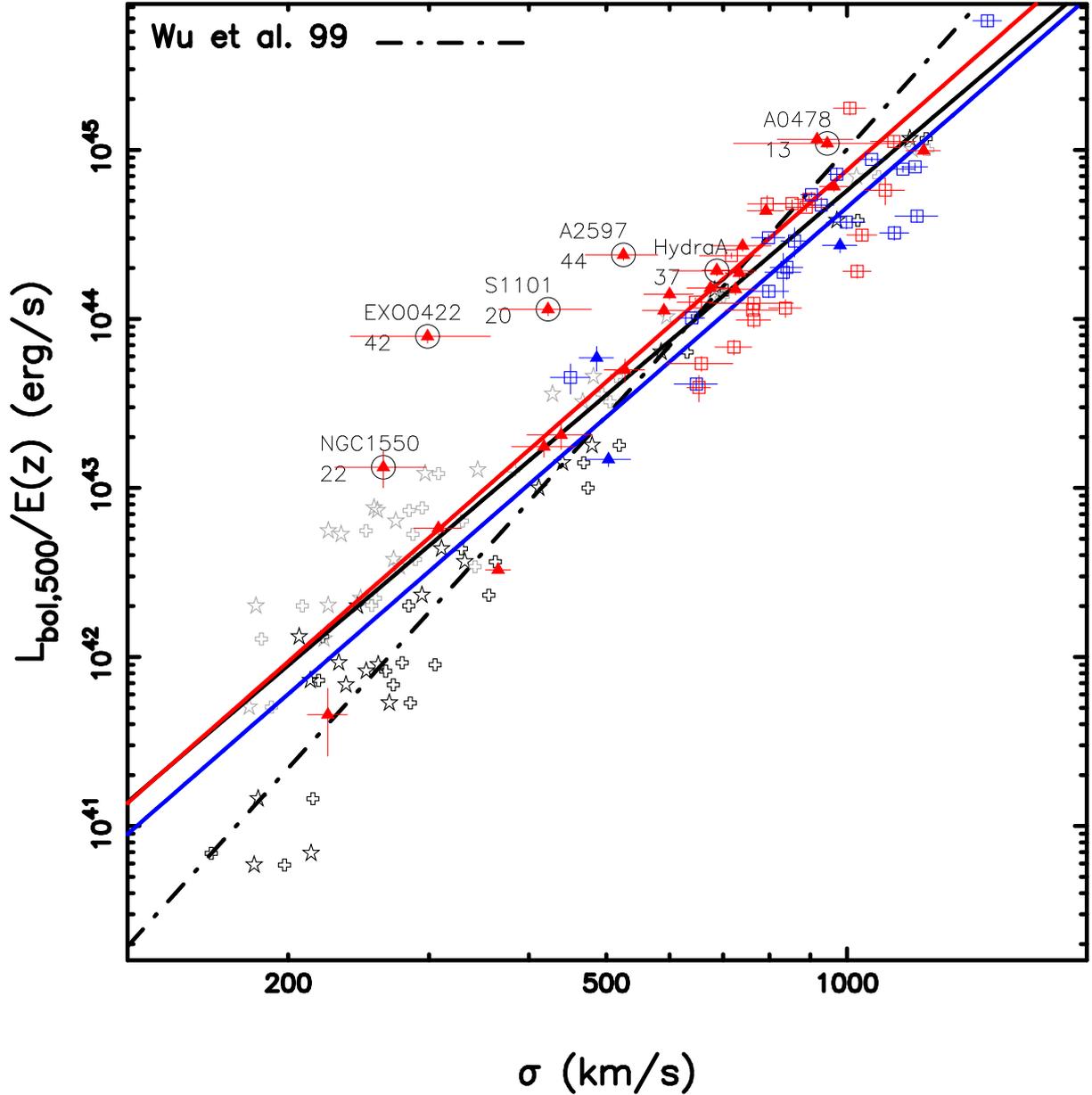}
\end{center}
\caption{X-ray bolometric luminosity vs. velocity
  dispersion with luminosity corrected for the cluster core 
($L^{\rm co}_{\rm bol}$). Our observational sample is shown in red (undisturbed)
  and blue (disturbed) colors, with filled triangles and open boxes
  denoting cool-core and non-cool-core clusters. The black circles
  highlight the six clusters with $<45$ cluster galaxy redshifts in
  the determination of the velocity dispersion. The black, red, and
  blue lines are the best fits using the BCES bisector method for the
  whole observational sample, subsample of the undisturbed clusters,
  and subsample of the disturbed clusters, respectively. The simulated
  sample is shown in black (with AGN feedback) and gray (without AGN
  feedback) stars using $\sigma_{\rm dirty}$. Crosses show the
  corresponding cases using $\sigma_{\rm clean}$, the velocity
  dispersion being based only on those galaxies within the virialized region
  of the cluster and within a projected radius of 1.2 Abell radii for
  the simulated sample. It is worth noting that no redshift correction
  and cool-core correction is applied in Wu et al. (1999).
  \label{f:lbv}}
\end{figure*}

We summarize the best power-law fits of the $L-\sigma$ relations using the
X-ray bolometric luminosity ($L^{\rm co}_{\rm bol}$) and 0.5--2~keV luminosity
($L^{\rm co}_{\rm 0.5-2keV}$), respectively, in Table~\ref{t:scaling}. In
Fig.~\ref{f:lbv}, we show the $L^{\rm co}_{\rm bol}-\sigma$ relation of the 62
clusters. 

The slope for the 62 clusters, i.e., ($4.02 \pm 0.33$), agrees with the
self-similar prediction ($L\sim \sigma^4$). The slopes for the undisturbed
clusters, disturbed clusters, cool-core clusters, and non-cool-core clusters
are statistically indistinguishable. Ignoring their measurement uncertainties,
both slopes for disturbed and undisturbed clusters are steeper than for the
combined sample. This is because most disturbed clusters are below the
best-fit relation, most undisturbed clusters are above, and hardly any
low-mass clusters are flagged as disturbed. The slope for the combined sample
is thus influenced by a number of low-mass systems, which are all undisturbed
clusters. The normalization for the undisturbed clusters is $\sim 60$\% higher
than for the disturbed ones.

The intrinsic scatter (Table~\ref{t:scaling}) of the undisturbed clusters and
cool-core clusters is only slightly larger than that of the disturbed clusters
and non-cool-core clusters. The clusters with more morphological substructure
do not show larger scatter than those with less substructure. This indicates
that the scatter driven by the presence of cool cores is comparable to that
driven by substructure using $L^{\rm co}$. The increasingly large scatter
toward the low-mass end is caused by the systems with $<45$ cluster members
with spectroscopic redshifts. In Sect.~\ref{s:bias}, we will discuss the
systematic uncertainties in the velocity dispersion measurements due to the
limited number of cluster members.

The top-left panel of Fig.~\ref{f:dlbv} shows the histogram of residuals in
logarithmic space from the $L_{\rm bol,500} - \sigma$ relation. The best
Gaussian fit gives $0.33^{+0.06}_{-0.05}$~dex scatter, dominated by the
intrinsic scatter, i.e., ($0.27\pm 0.03$)~dex. We note that the histogram does
not closely follow a symmetric Gaussian distribution, which may slightly
underestimate the scatter. The top-right, bottom-left, and bottom-right panels
of Fig.~\ref{f:dlbv} show the residuals as a function of the offset between
the X-ray flux-weighted center and BCG position, luminosity fraction within
$0.2r_{500}$, and central cooling time, respectively. There are very weak
correlations caused mainly by the systems that have fewer than 45 cluster
members with spectroscopic redshifts, for which the measurement uncertainties
in the velocity dispersion can be large and in part account for the scatter.

As shown in Fig.~\ref{f:lbv}, the normalization of the $L - \sigma$ relation
for the sample in Wu et al. (1999) is slightly higher than that of our
sample. For $L^{\rm in}$, the two samples are in better agreement (see
Fig.~\ref{f:dlbvin} in Appendix~\ref{a:lin}). Therefore, the core correction
applied when deriving the X-ray luminosity for our sample accounts for
the normalization difference between our sample and the sample of Wu et
al. (1999) in Fig.~\ref{f:lbv}. The different slopes between two samples may
be due to their different selection functions as the sample in Wu et
al. (1999) is not a flux-limited sample.

The presence of cool cores is one of the main causes of the scatter in the
$L-\sigma$ relation as the scatter using $L^{\rm in}$ for the non-cool-core
clusters is $\sim 20$\% smaller than that for the cool-core clusters
(Appendix~\ref{a:lin}). When the X-ray luminosity corrected for the central
region ($<0.2r_{500}$) is used, the intrinsic scatter is smaller by $\sim
0.05$~dex equaling $(0.27\pm 0.03)$~dex for the sample of 62 clusters. The
intrinsic scatter in the $L - \sigma$ relation is similar using $L^{\rm co}$
and $L^{\rm ex}$ (Appendix~\ref{a:lex}). The residuals of $L^{\rm in}_{\rm
  bol}-\sigma$ are more strongly correlated with the luminosity fraction
within $0.2r_{500}$ than the residuals of both $L^{\rm co}_{\rm bol}-\sigma$
and $L^{\rm ex}_{\rm bol}-\sigma$. The best-fit relation is $\Delta_{\rm
  lgL}=(0.52\pm 0.10)+ (0.51\pm 0.09)\log_{10}\left( L_{\rm bol,
    0.2r_{500}}/L_{\rm bol, 500} \right)$ with its correlation coefficient of
0.44. Correcting or excluding the central emission therefore efficiently
reduces the intrinsic scatter.

\subsection{$L - M_{\rm gas}$ relation}
\label{s:lbmg}

We summarize the best power-law fits of the $L - M_{\rm gas}$ relations using the
X-ray bolometric luminosity ($L^{\rm co}_{\rm bol}$) and 0.5--2~keV luminosity
($L^{\rm co}_{\rm 0.5-2keV}$), respectively, in Table~\ref{t:scaling}. In
Fig.~\ref{f:lbmg}, we present the $L^{\rm co}_{\rm bol}- M_{\rm gas}$ relation
of the 62 clusters. The slope of the best-fit power-law for the 62 clusters is
$(1.29\pm 0.05)$. The slopes for the undisturbed and disturbed clusters are
statistically identical. The slope for the non-cool-core clusters, i.e.,
($1.42 \pm 0.05$), is steeper than that for the cool-core clusters, i.e.,
($1.24 \pm 0.06$). The intrinsic scatter of those subsamples is comparable.

The top-left panel of Fig.~\ref{f:dlbmg} shows the histogram of
residuals in logarithmic space from the best-fit $L^{\rm co}_{\rm
bol,500} - M_{\rm gas, 500}$ relation for the 62 clusters using the
BCES bisector method. The best Gaussian fit gives ($0.07\pm 0.01$)~dex
scatter in logarithmic space, comparable to the intrinsic
scatter. We note that the histogram does not closely follow a symmetric
Gaussian distribution, which may slightly underestimate the
scatter. The top-right, bottom-left, and bottom-right panels of
Fig.~\ref{f:dlbmg} show the residuals as a function of the offset
between the cluster X-ray flux-weighted center and BCG position,
luminosity fraction within $0.2r_{500}$, and central cooling time,
respectively. We do not observe as clearly evident correlations as for the
$L^{\rm in}_{\rm bol,500} - M_{\rm gas, 500}$ relation in
Fig.~\ref{f:dlbvin} in Appendix~\ref{a:lin}. This indicates that the
cluster core correction may sufficiently suppress the scatter caused
by the presence of cool cores.

The intrinsic scatter in logarithmic space of the $L - M_{\rm gas}$ relation
using $L^{\rm co}$ is similar to that using $L^{\rm ex}$
(Appendix~\ref{a:lex}), but is 0.05~dex lower than that using $L^{\rm in}$
(Appendix~\ref{a:lin}). The cluster core correction in deriving the X-ray
luminosity significantly reduces the intrinsic scatter in the $L-M_{\rm gas}$
relation. 

We note that both quantities are derived from the X-ray surface brightness
distribution in the soft band. If the gas is clumped, the emission measure can
be overestimated by $\langle n_{\rm e}^2 \rangle / \langle n_{\rm e}
\rangle^2$, which results in overestimation of both the X-ray luminosity and
the gas mass. Therefore, one possibly underestimates the scatter in the
relation.

\subsection{$r_{500} - \sigma$ relation}
\label{s:r500v}

In Fig.~\ref{f:r500v}, we present the relation between the velocity dispersion
and cluster radius, the latter being determined from the mass vs. gas-mass
relation (see Sect.~\ref{s:r500}). In Table~\ref{t:scaling}, we summarize the
best power-law fits. The slopes for the undisturbed clusters, disturbed
clusters, cool-core clusters, and non-cool-core clusters are statistically
indistinguishable.

Surprisingly the intrinsic scatter in logarithmic space
(Table~\ref{t:scaling}) for the cool-core clusters is about a factor of two
larger than for the non-cool-core clusters. Since most undisturbed clusters
are cool-core clusters, the undisturbed clusters exhibit significantly larger
intrinsic scatter than the disturbed clusters. This indicates that the
presence of cool cores is the main driver of the scatter instead of the
morphological substructure. We note that the scatter becomes increasingly
large toward the low-mass end, which is coincidentally again caused by the
systems that have fewer than 45 cluster members with spectroscopic redshifts.

The top-left panel of Fig.~\ref{f:dr500v} shows the histogram of the residuals
in logarithmic space from the $r_{500} - \sigma$ relation. The best Gaussian
fit gives ($0.063^{+0.010}_{-0.008}$) scatter in logarithmic space, comparable
to the intrinsic scatter. We note that the histogram has a strong asymmetric
shape, such that the Gaussian distribution slightly underestimates the
scatter. The top-right, bottom-left, and bottom-right panels of
Fig.~\ref{f:dr500v} show the residuals as a function of the offset between the
X-ray flux-weighted cluster center and BCG position, the luminosity fraction
within $0.2r_{500}$, and the central cooling time, respectively. The residuals
are not very tightly correlated with any of these three parameters. However,
20 of the 26 cool-core clusters are above the best fit, and two thirds of
the non-cool-core clusters are below the best fit. The undisturbed and
disturbed clusters display homogeneously distributed residuals. This confirms
that the presence of cool cores is the main cause of the intrinsic scatter in
the $r_{500} -\sigma$ relation.

\section{Simulated vs. observational samples}
\label{s:sim}

To understand the cluster physics behind the observed scaling relations, it is
crucial to compare observational samples to representative samples in
simulations. Our analysis of our sample shows that the presence of cool cores
is one of the main causes of the scatter. In addition, it has become
increasingly clear that active galactic nuclei (AGN) play an important role in
understanding the properties of clusters (e.g., McNamara \& Nulsen 2007) and
their scaling relations. We therefore investigated how well simulations can
explain the observed results by comparing our observational measurements to
those for a sample of 21 clusters and groups simulated at a very high
resolution both with and without AGN feedback (Puchwein et al. 2008). The AGN
feedback model that was employed resolves some of the long-standing problems
that hydro-dynamical simulations of galaxy clusters typically have, i.e.,
excessive overcooling within the densest cluster regions and too bright and
too blue central galaxies. The AGN feedback model also brings the simulated
X-ray luminosity -- temperature scaling relation into excellent agreement with
the observational one. We note that simulations with different cluster physics
(e.g., Borgani et al. 2004; Evrard et al. 2008) may give different predictions
about the normalization, slope, and scatter of the scaling relations.

\subsection{A sample drawn from simulations}

Puchwein et al. (2008) carried out a set of high-resolution hydrodynamical
re-simulations of clusters selected from the Millennium simulation with and
without AGN feedback (for the AGN feedback model see Sijacki et al. 2007), and
present the corresponding $L-T$ relation. In Puchwein et al. (2010), they also
show the properties of the stellar components and halo baryon fractions for
the same sample. Their simulations have high enough resolution to accurately
resolve galaxy populations down to the smallest galaxies that are expected to
contribute significantly to the stellar mass budget. We select a sample of 21
galaxy clusters from their simulations, whose gas masses span a similar range
as our observational sample, i.e., $(2.95 \times 10^{11} - 1.10 \times 10^{14})
M_{\odot}$ for the case without AGN feedback and $(0.74 \times 10^{11} - 1.23
\times 10^{14}) M_{\odot}$ for the case with AGN feedback. The gas masses for
the same sample of 21 simulated clusters but with AGN feedback expand to a
broader range due to the feedback.

\subsection{Analyzing the sample from simulations}

Both X-ray bolometric luminosity (also corrected for the cluster core) and
velocity dispersion are derived in the same manner as for the observational
sample of the 62 clusters except that we do not use the caustic method to
identify interlopers since they are known in simulations. In
determining the velocity dispersion, we exclude sub-halos that do not contain
any stellar components and should thus not be considered as galaxies.

In the simulations, the virialized region of every cluster is known. We derive
both the velocity dispersion using those galaxies within a projected radius of
1.2 Abell radii (hereafter $\sigma_{\rm dirty}$), and the velocity dispersion
using those galaxies not only within a projected radius of 1.2 Abell radii,
but also within the virialized region of the cluster (hereafter $\sigma_{\rm
  clean}$)\footnote{More precisely, we include only galaxies that are part of
  the friends-of-friends (FoF) group of the cluster in calculating
  $\sigma_{\rm clean}$. The FoF group is computed using a linking length of
  0.2 times the mean inter-particle distance, and extends to roughly $\sim 1.7
  r_{200}$ which corresponds to $\sim 1.2$ Abell radii for clusters with a
  virial mass of $2.7 \times 10^{14} M_{\odot} /h$. For more massive clusters,
  the FoF group is of course larger, while it is significantly smaller for
  poor groups.}. In the $\sigma_{\rm clean}$ case, nearby galaxies and
interlopers are completely removed. Therefore, $\sigma_{\rm clean}$ gives a
reliable estimate of the velocity dispersion. We note that even $\sigma_{\rm
  dirty}$ contains only interlopers that are very close to the cluster center
since they are all within the high resolution region of the cluster
re-simulation, which typically extends to five times the cluster virial radius
or somewhat farther depending on its exact geometry.

\subsection{$L^{\rm co}_{\rm bol}-\sigma$ relation}

The $L^{\rm co}_{\rm bol}-\sigma_{\rm dirty}$ relation is slightly shallower
than the $L^{\rm co}_{\rm bol}-\sigma_{\rm clean}$ relation for the simulated
sample (Fig.~\ref{f:lbv}). This may suggest that interlopers bias the velocity
dispersion estimates toward lower values, which might be the case for the six
clusters that have fewer than 45 cluster members with spectroscopic redshifts
in the observational sample. We discuss this further in
Sect.~\ref{s:bias}. The scatter in the $L-\sigma$ relation of the simulated
sample is larger for the run with AGN feedback.

The shape and scatter of the $L-\sigma$ relation of the simulated sample with
AGN feedback is comparable to that of the cool-core clusters in the
observational sample, disregarding the six clusters with $<45$ cluster members
with spectroscopic redshifts. AGN feedback suppresses cool cores, thus reduces
the X-ray luminosity in simulations. AGN feedback also significantly lowers
halo gas mass fractions in low-mass systems. We therefore find that the
scatter becomes larger toward the low-mass end, i.e., systems with $\sigma \le
500$~km~s$^{-1}$, which is also present in our observations. The observational
sample, disregarding the six clusters with $<45$ cluster members with
spectroscopic redshifts, gives a best fit of $\log_{10} \left (\frac{L^{\rm
      co}_{\rm bol,500}}{E(z)\;{\rm erg~s^{-1}}}\right)=(4.46\pm
0.23)\log_{10}\left (\frac{\sigma}{\rm km~s^{-1}}\right )+(31.40 \pm 0.66)$,
which closely follows the simulated sample with AGN feedback.

In the high-mass regime, the observational sample shows that non-cool-core
clusters are the main driver of the scatter. Since most non-cool-core clusters
are disturbed clusters, their substructures cause overestimations of the
$\sigma$ values (see Fig.~6 and Fig.~9 in Biviano et al. 2006). Different
fractions of substructures therefore translate into scatter in the $L-\sigma$
relation, which is exactly what we find in the observational sample. The
scatter in the $L-\sigma$ relation of the simulated sample is smaller than
that of the observational sample in the high-mass regime since the simulated
sample does not predominantly contain mergers. In part, this difference might
also be due to too few massive clusters in our simulations, i.e., six systems
with $\sigma_{\rm clean} > 500$~km~s$^{-1}$. Forthcoming simulations of much
larger cosmic volumes will be very useful in differentiating the scatter in
the $L-\sigma$ relation caused by measurement systematics from that
attributable to cluster physics and achieving a clearer understanding of the
cluster dynamics and gas physics in the high-mass regime.

\subsection{$L-M_{\rm gas}$ relation}

The intrinsic scatter in the $L-M_{\rm gas}$ relation is small for both the
observational sample and the simulated sample. AGN feedback mainly tends to
move clusters downward along the $L-M_{\rm gas}$ relation rather than strongly
changing the relation because removing gas from within $r_{500}$ also
significantly reduces the X-ray luminosity. Hence, AGN feedback does not
produce significant scatter in the $L-M_{\rm gas}$ relation found in
simulations. Nevertheless, the simulated sample with AGN feedback has a
slightly steeper slope than that of the case without AGN feedback. The
difference between the simulated samples without and with AGN feedback is
still small, and comparable to the intrinsic scatter. There is a good
agreement between the $L-M_{\rm gas}$ relations from observations and
simulations as shown in Fig.~\ref{f:lbmg}. Interestingly, the best fits of the
disturbed clusters and undisturbed clusters in the observational sample
closely match the simulated samples with and without AGN feedback,
respectively. This suggests that one of the main causes of the scatter could
be AGN activities providing feedback in different phases, e.g., during a
feedback cycle.

\subsection{$r_{500} - \sigma$ relation}

As shown in Fig.~\ref{f:r500v}, the slope of the $r_{500} - \sigma$ relation
for the observational sample is similar to that of the simulated sample with
AGN feedback, disregarding the clusters again that have fewer than 45 cluster
members with spectroscopic redshifts. For the simulated sample, the fraction
of gas removed by AGN feedback becomes significant toward low-mass systems.
As a consequence, their DM distributions expand slightly. Both the removal of
gas and the expansion of the DM distribution result in increasingly smaller
cluster radii with decreasing mass in the simulated sample with AGN feedback
compared to the simulated sample without AGN feedback. The simulated sample
with AGN feedback thus has a slightly steeper slope than that of the sample
without AGN feedback. For the observational sample, the subsample of the
undisturbed clusters exhibit a steeper slope than that of the subsample of the
disturbed clusters. This is consistent with the scenario that
  incorporates AGN activities in the undisturbed clusters (mostly cool-core
clusters).

There is some offset in the normalization between the simulated sample
and the observational sample, which is however still within the
scatter. X-ray masses are lower than the true masses in numerical
simulations (e.g., Evrard 1990; Lewis et al. 2000; Rasia et al. 2006;
Nagai et al. 2007; Piffaretti \& Valdarnini 2008; Jeltema et al. 2008;
Lau et al. 2009; Meneghetti et al. 2010). This may in part account for
the offset in the normalization, which relies on the total mass
vs. gas mass calibration. 
The galaxy selection is complete for the simulated cluster. However, we have
no homogeneous photometry data to constrain the completeness for the clusters
in the observational sample. Differences in the selection of galaxies used to
compute the $\sigma$ may also in part cause this offset.

\section{Systematic errors in estimates of $\sigma$}
\label{s:bias}

\subsection{Galaxy selection by projected radial distance}

Most velocity dispersion profiles of galaxy clusters become flat
beyond $1 h^{-1}$~Mpc which suggests that the measured velocity
dispersion within a larger radius is more representative of the total
kinetic energy of the cluster galaxies (e.g., Fadda et al. 1996; den
Hartog \& Katgert 1996; Biviano \& Girardi 2003; Boschin et al. 2010).

We test how the radial selection of cluster members affects the velocity
dispersion estimates of the observational sample as follows. We measure the
velocity dispersion within [0.5, 1.0, 1.5, 2.0, 2.5]$\times r_{500}$, and
normalize it to the value measured within 1.2 Abell radii (top-left panel of
Fig.~\ref{f:sigma_dist}). On average, the velocity dispersion measured within
small radii, i.e., [0.5,1.0]$\times r_{500}$, is $\sim 10$\% larger than the
one measured within larger radii. This is consistent with den Hartog \&
Katgert (1996) finding that more clusters with relatively large velocity
dispersion than small when measuring velocity dispersion close to the cluster
center. We also note that the scatter in the measured velocity dispersion
within small radii is $\sim 3 $ times that measured within $2.5r_{500}$.

In the top-right panel of Fig.~\ref{f:sigma_dist}, we show the normalized
velocity dispersion measurements within $r_{500}$ as a function of the
velocity dispersion measured within 1.2 Abell radii. For systems of velocity
dispersion greater than 500~km~s$^{-1}$, there is on average less than 10\%
difference between the velocity dispersion measurements within $r_{500}$ and
within 1.2 Abell radii. The difference becomes larger for low-mass systems,
and is up to $\sim 30$\% on average for our sample. The scatter in the ratio
of the velocity dispersion measurements within $r_{500}$ and to those within
1.2 Abell radii is almost independent of the absolute value of the velocity
dispersion, at $\sim 25$\%.

In the bottom panels of Fig.~\ref{f:sigma_dist}, we also show the normalized
velocity dispersion measurements within $r_{500}$ as a function of the offset
between the X-ray flux-weighted cluster center and BCG position and the
central cooling time. For clusters with a smaller offset between the X-ray
flux-weighted cluster center and BCG position or shorter central cooling time,
the velocity dispersion measurements within $r_{500}$ are significantly larger
than the values measured within 1.2 Abell radii.

We note that interlopers introduce uncertainties in the above tests,
particularly for the six systems with $<45$ cluster members with spectroscopic
redshifts. We therefore also carried out tests using the simulated sample
as shown in Figs.~\ref{f:sigma_dist_simu}--\ref{f:sigma_r500_sigma_simu}.

The trends of velocity dispersion decrease with increasing radius
agree between the simulated sample and the observational sample. In
Fig.~\ref{f:sigma_dist_simu}, the simulated sample shows that AGN feedback
does not clearly affect the velocity dispersion estimates. However,
interlopers increase both the amplitude and the scatter in the deviations of
the $\sigma$ estimates as a function of projected cluster-centric
distance. The average deviation for the simulated sample without interlopers
is similar to that of the observational sample. However, the scatter for the
simulated sample with interlopers is comparable to that of the observational
sample. This indicates interlopers may affect the velocity dispersion
estimates for a few but not the majority of systems in the observational
sample.

As shown in Fig.~\ref{f:sigma_dist_simu}, the velocity dispersion within
1.2~Abell radii for two groups in the simulated sample is $\sim 30$\% larger
than that within smaller cluster-centric radii. One group is in a
strongly clustered region with several group-size objects within 1.2 Abell
radius in projection. In particular, one of the group-size objects has a
similar mass to the group we analyzed. The other group is in the process of
merging, which biases the $\sigma$ estimate toward larger values (see also
Biviano et al. 2006).

In Fig.~\ref{f:sigma_r500_sigma_simu}, the simulated sample without
interlopers confirms that systems with velocity dispersions greater than
500~km~s$^{-1}$ have $<10$\% difference between the velocity dispersion
measurements within $r_{500}$ and within 1.2 Abell radii. The results for the
simulated sample also indicates that interlopers can boost the scatter in the
bias for the low-mass systems of velocity dispersion $<500$~km~s$^{-1}$. For
the low-mass systems, the uncertainties in the velocity dispersion estimates
may indeed be as large as 40\%, disregarding the radial selection. We have to
keep this in mind when we consider the six clusters with $<45$ cluster members
with spectroscopic redshifts in the observational sample.

\subsection{Mass selection} 

For the observational sample, we collected cluster galaxy redshifts from the
literature. This may introduce a bias in the velocity dispersion estimates
because brighter cluster galaxies may be more likely to have published
redshifts than fainter ones. This bias becomes less significant when many
cluster galaxies with spectroscopic redshifts are available. Since our
observed sample of cluster galaxies is incomplete, we test how the mass
selection of cluster members affects the velocity dispersion estimates using
the simulated sample, which is homogeneous in terms of cluster galaxies.  In
Fig.~\ref{f:sigma_mag_simu}, we display the velocity dispersion determined for
a fraction of cluster members at the massive end for the 21 simulated
clusters\footnote{Note that only galaxies above the resolution limit of the
  simulations are included in Fig.~\ref{f:sigma_mag_simu} and
  Figs.~\ref{f:sigma_45_simu}--\ref{f:sigma_45_sigma_simu}. The three figures
  are meant to illustrate trends instead of giving quantitative constraints.}.

AGN feedback does not have an obvious effect on the velocity dispersion
estimates.  For the simulated sample, the local regression non-parametric fit
illustrates that the velocity dispersion estimate tends to be increasingly
biased toward smaller values with decreasing fraction of cluster members. The
bias on average is within a few per cent as long as more than 10\% of the
cluster members at the massive end are used. The scatter in the velocity
dispersion also increases as a smaller fraction of cluster galaxies are used,
and is within 10\% when at least 50\% cluster members at the massive end are
used. The scatter is slightly smaller when there are no interlopers.  As shown
in Figs.~\ref{f:sigma_45_simu}--\ref{f:sigma_45_sigma_simu}, when we consider
only 45 of the most massive cluster members, the uncertainties in the velocity
dispersion estimates can be up to 40\% for some low-mass systems
($\sigma<500$~km~s$^{-1}$).

\subsection{Interlopers}

Except for one system in the simulated sample, interlopers always bias the
measurements of the velocity dispersion toward smaller values (see also
Biviano et al. 2006). A significant fraction of galaxies (up to $\sim 50$\% of
$n_{\rm gal}$) within 1.2 Abell radii are not in the virialized region for
poor systems. This is not the case for massive systems. As shown in
Fig.~\ref{f:caustic_simu}, a caustic cannot efficiently exclude
interlopers at larger radii, i.e., $[1-2.5]r_{500}$, and may significantly
bias the measurements of the velocity dispersion toward smaller values for
poor systems. The $\sigma_{\rm clean}$ is a far more robust indicator of the
cluster mass than the $\sigma_{\rm dirty}$ for poor systems.

We note that $\sigma_{\rm dirty}$ for the simulated sample only contains
interlopers very close to the cluster. In the observations, there may be more
distant interlopers for poor systems. As shown in Fig.~10 in Biviano et
al. (2006), unrecognized interlopers that are outside the virial radius but
dynamically linked to the host cluster and do not form major
substructures, bias the $\sigma$ estimate toward smaller values than cluster
galaxies.

\section{Conclusions}
\label{s:con}

We have presented the $L-\sigma$, $L-M_{\rm gas}$, and $r_{500}-\sigma$
relations for the 62 clusters in the HIFLUGCS, a purely X-ray flux-limited sample
selected to minimize bias toward any cluster morphology. The systems in this
sample span a broad range of morphological substructure, central cooling time,
and offset between the X-ray flux-weighted cluster center and BCG position,
respectively. Owing to our representative, statistically large sample, with
$\sim 1.3$~Ms of clean X-ray \emph{XMM-Newton} data and 13439
spectroscopically confirmed cluster members for 62 clusters, we have been able
to minimize our measurement uncertainties in both X-ray observables and
velocity dispersion. Our main results are as follows:

\begin{itemize}

\item The luminosity vs. velocity dispersion relation agrees with the
  self-similar prediction. The presence of cool cores is one of the major
  contributors to the scatter in the $L-\sigma$ relation. Correcting the
  central region in deriving the X-ray luminosity reduces the intrinsic
  scatter from 0.33~dex to 0.27~dex. Even after correcting the X-ray
  luminosity for the cool core, the scatter caused by cool cores becomes
  increasingly large toward the low-mass end. The scatter caused by the
  non-cool-core clusters does not strongly depend on the mass range, but
  becomes dominant for massive systems. The intrinsic scatter for the
  non-cool-core clusters, 0.25~dex, is statistically indistinguishable from
  that of the cool-core clusters, 0.28~dex, after correcting the central
  region when deriving the X-ray luminosity.

\item The presence of cool cores is also one of the major contributors to the
  scatter in the $L-M_{\rm gas}$ relation. Using the X-ray luminosity
  corrected for the cool core, the disturbed clusters with significant X-ray
  substructures exhibit similar scatter as the undisturbed clusters,
  partly because of the preponderance of cool-core clusters in the undisturbed
  subsample.

\item The shape of the $L^{\rm co}-\sigma$ relation in simulations with AGN
  feedback matches the observational sample, specifically the cool-core
  clusters, disregarding the clusters that have fewer than 45 cluster members
  with spectroscopic redshifts. A common trend in both observations and
  simulations is that the scatter becomes larger toward the low-mass end,
  i.e., systems with $\sigma \le 500$~km~s$^{-1}$.  The shape and intrinsic
  scatter in the $L^{\rm co}-\sigma$ relation of the observational sample
  closely matches that of the simulated sample for the low-mass clusters
  indicating that AGN feedback operates there. In the high-mass regime, the
  observational sample shows that non-cool-core clusters (their substructures)
  are the main driver of the scatter.  The scatter in the $L - \sigma$
  relation at the high-mass end is larger than the scatter in the simulated
  sample. This may be in part because there are too few massive clusters and
  no significantly disturbed clusters in the simulated sample.

\item Interestingly, the best fits of the luminosity vs. gas mass relations
  for the disturbed clusters and undisturbed clusters in the observational
  sample closely match those of the simulated samples with and without AGN
  feedback, respectively. This suggests that one of the main causes of the
  scatter could be AGN providing feedback in different phases, e.g., during a
  feedback cycle.

\item The $r_{500} - \sigma$ relation of the observational sample is similar
  to that of the simulated sample, disregarding the clusters with $<45$
  cluster members with spectroscopic redshifts. For the simulated sample, the
  fraction of gas removed by AGN feedback becomes significant toward low-mass
  systems, which makes their potential wells shallower. The slope for the
  simulated sample with AGN feedback is thus steeper than that for the sample
  without AGN feedback. For the observational sample, the subsample of the
  undisturbed clusters exhibits a steeper $r_{500} - \sigma$ relation than
  that of the subsample of the disturbed clusters. This suggests that there is
  AGN activity in the undisturbed clusters, which are mostly cool-core
  clusters.
 
\item Both the selections of the aperture and mass limit of the cluster
  members and interlopers cause systematic uncertainties in estimating the
  velocity dispersion. For the observational sample, the scatter in the
  velocity dispersion measured within small radii, i.e., [0.5,1.0]$\times
  r_{500}$ is $\sim 3 $ times that measured within $2.5r_{500}$. The analysis
  of the simulated sample indicates that interlopers bias the velocity
  dispersion estimates toward smaller values. The interlopers increase both
  the amplitude and the scatter in the bias, which is particularly significant
  for low-mass systems ($\sigma <500$~km~s$^{-1}$). The scatter in the bias of
  the velocity dispersion estimates increases as the fraction of cluster
  galaxies used decreases. The scatter is slightly smaller when there are no
  interlopers.

\end{itemize}

\bigskip 

\begin{acknowledgements}

  The \emph{XMM-Newton} project is an ESA Science Mission with instruments and
  contributions directly funded by ESA Member States and the USA (NASA). The
  \emph{XMM-Newton} project is supported by the Bundesministerium f\"ur
  Wirtschaft und Technologie/Deutsches Zentrum f\"ur Luft- und Raumfahrt
  (BMWI/DLR, FKZ 50 OX 0001) and the Max-Planck Society. We acknowledge the
  anonymous referee for suggestions that improved the disucssion of the
  results. Y.Y.Z., H.A., and T.H.R. acknowledge support by the DFG through
  Emmy Noether Research Grant RE\,1462/2 and by the German BMBF through the
  Verbundforschung under grant 50\,OR\,0601 and 50\,OR\,1005, and T.H.R. in
  addition acknowledges DFG Heisenberg Grant RE 1462/5. H.A. and
  C.A.C. acknowledge support from Mexican CONACyT grant 50921-F, and H.A. in
  addition benefited from CONACyT grants 81356 and 118295. Y.Y.Z. acknowledges
  A. Babul, A. Biviano, S. Borgani, and R. Piffaretti for useful discussions.

\end{acknowledgements}

\bibliography{Abell}


\clearpage

\begin{table*} { \begin{center} \footnotesize
      {\renewcommand{\arraystretch}{1.3} \caption[]{Offset between the X-ray
          flux-weighted cluster center and BCG position, velocity dispersion,
          and X-ray bolometric
          luminosity.}  \label{t:sigma}} \begin{tabular}{lccccrrccc} \hline
        Cluster& \multicolumn{2}{c}{Cluster center (J2000)}
        &\multicolumn{2}{c}{BCG position (J2000)} & \multicolumn{2}{c}{Offset}
        & $N_{\rm galaxy}$& $\sigma$
        & $L^{\rm co}_{\rm bol}$ \\
        &  R.A. & decl. & R.A. & decl. & kpc & $r_{500}$  & & (km~s$^{-1}$) & (erg~s$^{-1}$)\\
        \hline

A0085        & 00:41:50.306 & -09:18:11.11 & 00:41:50.48  & -09:18:11.2  &    2.8 & 0.0023 &  350 &$  963  \pm   39  $ &$(6.23 \pm0.43  )\times 10^{44  }$\\
A0119        & 00:56:17.119 & -01:15:11.98 & 00:56:16.20  & -01:15:20.0  &   13.8 & 0.0130 &  339 &$  797  \pm   38  $ &$(3.09 \pm0.14  )\times 10^{44  }$\\
A0133        & 01:02:43.141 & -21:52:47.04 & 01:02:41.70  & -21:52:55.0  &   23.8 & 0.0269 &  137 &$  725  \pm   44  $ &$(1.54 \pm0.11  )\times 10^{44  }$\\
NGC507       & 01:23:38.567 & +33:15:02.08 & 01:23:39.89  & +33:15:21.0  &    8.5 & 0.0152 &  110 &$  503  \pm   33  $ &$(1.49 \pm0.14  )\times 10^{43  }$\\
A0262        & 01:52:45.610 & +36:09:03.92 & 01:52:46.50  & +36:09:07.0  &    3.7 & 0.0049 &  138 &$  527  \pm   30  $ &$(5.04 \pm0.79  )\times 10^{43  }$\\
A0400        & 02:57:41.349 & +06:01:36.93 & 02:57:41.60  & +06:01:28.9  &    4.3 & 0.0060 &  114 &$  647  \pm   40  $ &$(4.16 \pm0.35  )\times 10^{43  }$\\
A0399        & 02:57:51.635 & +13:02:49.53 & 02:57:53.10  & +13:01:51.0  &   85.0 & 0.0801 &  101 &$ 1223  \pm   75  $ &$(4.20 \pm0.33  )\times 10^{44  }$\\
A0401        & 02:58:57.216 & +13:34:46.56 & 02:58:57.80  & +13:34:58.0  &   20.3 & 0.0161 &  116 &$ 1144  \pm   74  $ &$(1.16 \pm0.10  )\times 10^{45  }$\\
A3112        & 03:17:58.713 & -44:14:08.39 & 03:17:57.55  & -44:14:16.1  &   20.9 & 0.0216 &  111 &$  740  \pm   63  $ &$(2.81 \pm0.20  )\times 10^{44  }$\\
Fornax       & 03:38:28.791 & -35:27:04.50 & 03:38:29.00  & -35:27:01.0  &    0.4 & 0.0016 &  339 &$  366  \pm   13  $ &$(3.28 \pm0.11  )\times 10^{42  }$\\
IIIZw54      & 03:41:18.729 & +15:24:13.91 & 03:41:17.52  & +15:23:47.7  &   19.6 & 0.0270 &   45 &$  657  \pm   62  $ &$(5.52 \pm0.56  )\times 10^{43  }$\\
A3158        & 03:42:53.583 & -53:37:51.71 & 03:42:57.53  & -53:37:55.9  &   40.4 & 0.0399 &  258 &$ 1044  \pm   45  $ &$(3.22 \pm0.26  )\times 10^{44  }$\\
A0478        & 04:13:25.296 & +10:27:57.96 & 04:13:25.23  & +10:27:56.1  &    3.5 & 0.0029 &   13 &$  944  \pm  223  $ &$(1.14 \pm0.08  )\times 10^{45  }$\\
NGC1550      & 04:19:38.021 & +02:24:33.36 & 04:19:37.92  & +02:24:35.5  &    0.7 & 0.0012 &   22 &$  263  \pm   34  $ &$(1.33 \pm0.32  )\times 10^{43  }$\\
EXO0422      & 04:25:51.224 & -08:33:40.34 & 04:25:51.30  & -08:33:38.6  &    1.6 & 0.0021 &   42 &$  298  \pm   59  $ &$(8.03 \pm0.71  )\times 10^{43  }$\\
A3266        & 04:31:14.909 & -61:26:54.13 & 04:31:13.22  & -61:27:12.0  &   24.8 & 0.0196 &  559 &$ 1174  \pm   41  $ &$(7.92 \pm0.34  )\times 10^{44  }$\\
A0496        & 04:33:37.818 & -13:15:38.55 & 04:33:37.80  & -13:15:43.0  &    2.9 & 0.0030 &  360 &$  687  \pm   28  $ &$(1.97 \pm0.13  )\times 10^{44  }$\\
A3376        & 06:02:10.108 & -39:57:35.75 & 06:00:41.09  & -40:02:40.4  &  955.1 & 0.9982 &  165 &$  798  \pm   46  $ &$(1.49 \pm0.12  )\times 10^{44  }$\\
A3391        & 06:26:24.222 & -53:41:24.02 & 06:26:20.40  & -53:41:35.0  &   36.9 & 0.0380 &   71 &$  716  \pm   62  $ &$(2.43 \pm0.09  )\times 10^{44  }$\\
A3395s       & 06:26:46.080 & -54:32:43.08 & 06:27:36.29  & -54:26:57.9  &  543.0 & 0.5714 &  215 &$  841  \pm   39  $ &$(2.07 \pm0.32  )\times 10^{44  }$\\
A0576        & 07:21:26.115 & +55:45:34.22 & 07:21:32.52  & +55:45:27.2  &   41.2 & 0.0474 &  237 &$  837  \pm   39  $ &$(1.18 \pm0.14  )\times 10^{44  }$\\
A0754        & 09:09:18.187 & -09:41:15.95 & 09:08:32.50  & -09:37:48.0  &  727.7 & 0.6884 &  470 &$  928  \pm   34  $ &$(4.84 \pm0.51  )\times 10^{44  }$\\
HydraA       & 09:18:05.988 & -12:05:36.15 & 09:18:05.60  & -12:05:44.0  &   10.2 & 0.0110 &   37 &$  687  \pm   82  $ &$(1.97 \pm0.13  )\times 10^{44  }$\\
A1060        & 10:36:42.859 & -27:31:42.10 & 10:36:42.71  & -27:31:42.9  &    0.5 & 0.0007 &  389 &$  652  \pm   21  $ &$(3.95 \pm0.70  )\times 10^{43  }$\\
A1367        & 11:44:44.501 & +19:43:55.82 & 11:44:02.20  & +19:57:00.0  &  430.8 & 0.4821 &  343 &$  639  \pm   24  $ &$(1.02 \pm0.06  )\times 10^{44  }$\\
MKW4         & 12:04:27.660 & +01:53:41.50 & 12:04:27.08  & +01:53:45.3  &    3.8 & 0.0066 &  145 &$  417  \pm   37  $ &$(1.77 \pm0.23  )\times 10^{43  }$\\
ZwCl1215     & 12:17:40.637 & +03:39:29.66 & 12:17:41.13  & +03:39:21.0  &   16.2 & 0.0153 &  154 &$  889  \pm   51  $ &$(4.75 \pm0.27  )\times 10^{44  }$\\
NGC4636      & 12:42:50.265 & +02:41:30.64 & 12:42:49.67  & +02:41:15.4  &    1.4 & 0.0055 &  115 &$  224  \pm   12  $ &$(4.56 \pm1.97  )\times 10^{41  }$\\
A3526        & 12:48:50.643 & -41:18:15.28 & 12:48:48.94  & -41:18:42.0  &    7.0 & 0.0092 &  235 &$  486  \pm   24  $ &$(5.92 \pm0.99  )\times 10^{43  }$\\
A1644        & 12:57:10.735 & -17:24:10.28 & 12:57:11.59  & -17:24:34.4  &   25.2 & 0.0234 &  307 &$  980  \pm   48  $ &$(2.79 \pm0.27  )\times 10^{44  }$\\
A1650        & 12:58:41.885 & -01:45:32.91 & 12:58:41.50  & -01:45:42.4  &   17.6 & 0.0168 &  220 &$  794  \pm   43  $ &$(4.99 \pm0.60  )\times 10^{44  }$\\
A1651        & 12:59:22.352 & -04:11:46.60 & 12:59:22.48  & -04:11:46.2  &    3.2 & 0.0030 &  222 &$  896  \pm   36  $ &$(5.31 \pm0.43  )\times 10^{44  }$\\
Coma         & 12:59:45.341 & +27:57:05.63 & 12:59:35.67  & +27:57:33.6  &   61.5 & 0.0480 &  972 &$  970  \pm   22  $ &$(7.26 \pm0.64  )\times 10^{44  }$\\
NGC5044      & 13:15:23.782 & -16:23:11.68 & 13:15:23.97  & -16:23:07.9  &    0.9 & 0.0019 &  156 &$  308  \pm   20  $ &$(5.81 \pm0.49  )\times 10^{42  }$\\
A1736        & 13:26:53.712 & -27:10:35.40 & 13:27:27.90  & -27:19:30.0  &  636.6 & 0.6467 &  148 &$  832  \pm   43  $ &$(1.93 \pm0.57  )\times 10^{44  }$\\
A3558        & 13:28:00.410 & -31:30:00.78 & 13:27:56.80  & -31:29:44.0  &   46.2 & 0.0389 &  509 &$  902  \pm   27  $ &$(5.56 \pm0.18  )\times 10^{44  }$\\
A3562        & 13:33:36.487 & -31:40:25.54 & 13:33:34.73  & -31:40:20.3  &   22.5 & 0.0249 &  265 &$ 1029  \pm   41  $ &$(1.96 \pm0.19  )\times 10^{44  }$\\
A3571        & 13:47:27.868 & -32:51:37.65 & 13:47:28.30  & -32:51:53.0  &   12.8 & 0.0113 &  172 &$  853  \pm   45  $ &$(4.91 \pm0.25  )\times 10^{44  }$\\
A1795        & 13:48:52.790 & +26:35:34.36 & 13:48:52.47  & +26:35:34.0  &    5.1 & 0.0047 &  179 &$  791  \pm   41  $ &$(4.49 \pm0.19  )\times 10^{44  }$\\
A3581        & 14:07:30.627 & -27:00:47.33 & 14:07:29.60  & -27:01:05.0  &    9.7 & 0.0160 &   83 &$  439  \pm   41  $ &$(2.08 \pm0.38  )\times 10^{43  }$\\
MKW8         & 14:40:42.150 & +03:28:17.87 & 14:40:42.81  & +03:27:55.3  &   13.4 & 0.0188 &  183 &$  450  \pm   25  $ &$(4.56 \pm0.93  )\times 10^{43  }$\\
A2029        & 15:10:55.990 & +05:44:33.64 & 15:10:56.07  & +05:44:41.5  &   11.6 & 0.0093 &  202 &$ 1247  \pm   61  $ &$(1.03 \pm0.06  )\times 10^{45  }$\\
A2052        & 15:16:44.411 & +07:01:12.57 & 15:16:44.55  & +07:01:18.3  &    4.2 & 0.0049 &  168 &$  590  \pm   35  $ &$(1.14 \pm0.08  )\times 10^{44  }$\\
MKW3S        & 15:21:50.277 & +07:42:11.77 & 15:21:51.86  & +07:42:32.0  &   27.5 & 0.0308 &   94 &$  599  \pm   42  $ &$(1.43 \pm0.10  )\times 10^{44  }$\\
A2065        & 15:22:29.082 & +27:43:14.39 & 15:22:29.05  & +27:42:35.0  &   54.1 & 0.0556 &  204 &$ 1146  \pm   47  $ &$(3.34 \pm0.32  )\times 10^{44  }$\\
A2063        & 15:23:05.772 & +08:36:25.37 & 15:23:05.20  & +08:36:32.0  &    7.6 & 0.0087 &  224 &$  646  \pm   33  $ &$(1.28 \pm0.08  )\times 10^{44  }$\\
A2142        & 15:58:19.776 & +27:14:00.96 & 15:58:19.97  & +27:13:59.7  &    4.8 & 0.0035 &  233 &$ 1008  \pm   46  $ &$(1.84 \pm0.17  )\times 10^{45  }$\\
A2147        & 16:02:16.305 & +15:58:18.46 & 16:02:17.00  & +15:58:27.0  &    9.2 & 0.0087 &  397 &$  859  \pm   32  $ &$(2.94 \pm0.57  )\times 10^{44  }$\\
A2163        & 16:15:46.392 & -06:08:36.96 & 16:15:48.98  & -06:08:41.5  &  128.8 & 0.0915 &  311 &$ 1498  \pm   61  $ &$(6.41 \pm0.53  )\times 10^{45  }$\\
A2199        & 16:28:37.126 & +39:32:53.29 & 16:28:38.25  & +39:33:04.3  &   10.3 & 0.0108 &  374 &$  733  \pm   29  $ &$(1.90 \pm0.16  )\times 10^{44  }$\\
A2204        & 16:32:47.059 & +05:34:32.03 & 16:32:46.90  & +05:34:33.0  &    6.8 & 0.0060 &  111 &$  917  \pm   99  $ &$(1.24 \pm0.09  )\times 10^{45  }$\\
A2244        & 17:02:41.976 & +34:03:28.08 & 17:02:42.49  & +34:03:36.0  &   18.3 & 0.0170 &  106 &$ 1116  \pm   63  $ &$(6.05 \pm1.11  )\times 10^{44  }$\\
A2256        & 17:03:52.468 & +78:40:19.14 & 17:04:00.81  & +78:38:06.2  &  157.1 & 0.1244 &  296 &$ 1216  \pm   45  $ &$(8.17 \pm0.43  )\times 10^{44  }$\\
A2255        & 17:12:54.538 & +64:03:51.46 & 17:12:28.79  & +64:03:38.8  &  256.0 & 0.2488 &  189 &$  998  \pm   55  $ &$(3.88 \pm0.36  )\times 10^{44  }$\\
A3667        & 20:12:40.708 & -56:50:27.06 & 20:12:27.29  & -56:49:36.4  &  131.8 & 0.1006 &  580 &$ 1073  \pm   37  $ &$(9.02 \pm0.28  )\times 10^{44  }$\\
S1101        & 23:13:58.312 & -42:43:36.11 & 23:13:58.60  & -42:43:39.0  &    4.8 & 0.0058 &   20 &$  422  \pm   55  $ &$(1.17 \pm0.10  )\times 10^{44  }$\\
A2589        & 23:23:56.772 & +16:46:33.19 & 23:23:57.41  & +16:46:35.0  &    7.7 & 0.0092 &   94 &$  762  \pm   57  $ &$(1.15 \pm0.08  )\times 10^{44  }$\\
A2597        & 23:25:20.009 & -12:07:27.18 & 23:25:19.71  & -12:07:27.0  &    7.0 & 0.0074 &   44 &$  525  \pm   54  $ &$(2.49 \pm0.18  )\times 10^{44  }$\\
A2634        & 23:38:29.045 & +27:01:51.66 & 23:38:29.40  & +27:01:51.0  &    3.0 & 0.0038 &  192 &$  721  \pm   38  $ &$(6.92 \pm0.70  )\times 10^{43  }$\\
A2657        & 23:44:56.743 & +09:11:52.93 & 23:44:57.45  & +09:11:35.3  &   16.4 & 0.0200 &   64 &$  764  \pm   92  $ &$(1.26 \pm0.10  )\times 10^{44  }$\\
A4038        & 23:47:44.652 & -28:08:42.45 & 23:47:45.04  & -28:08:26.2  &    9.7 & 0.0113 &  202 &$  764  \pm   37  $ &$(9.99 \pm1.03  )\times 10^{43  }$\\
A4059        & 23:57:01.698 & -34:45:29.13 & 23:57:00.71  & -34:45:33.0  &   11.6 & 0.0130 &  188 &$  674  \pm   43  $ &$(1.56 \pm0.13  )\times 10^{44  }$\\
\hline
  \end{tabular}
  \end{center}
  \hspace*{0.3cm}{\footnotesize Note: The clusters
    are sorted by R.A.. }} 
\end{table*}

\begin{table*} { \begin{center} \footnotesize  
      {\renewcommand{\arraystretch}{1.3} \caption[]{
          \emph{XMM-Newton} observations and cluster properties.}
        \label{t:xmm}}
\begin{tabular}{lrrrrccccc}
\hline
Cluster&OBS-ID & \multicolumn{3}{c}{Net exposure (ks)} & $z$& $N_{\rm H} $ & $M_{\rm gas,500}$ & Undisturbed & Cool core$^{*}$ \\
       &       &MOS1 & MOS2 & pn   &     & $10^{22}{\rm cm}^{-2}$& $M_{\odot}$ \\
\hline
A0085 & 0065140101 & 12.1 & 11.7 &  8.6 & 0.0556 & 0.0283 &$ ( 6.67 \pm 0.32 ) \times 10^{ 13 } $& Y & S \\
A0119 & 0505211001 &  8.2 &  8.0 &  7.6 & 0.0440 & 0.0328 &$ ( 4.27 \pm 0.20 ) \times 10^{ 13 } $& N & N \\
A0133 & 0144310101 & 18.6 & 17.2 & 15.3 & 0.0569 & 0.0158 &$ ( 2.31 \pm 0.10 ) \times 10^{ 13 } $& Y & S \\
NGC507& 0080540101 & 16.7 & 16.4 & 28.6 & 0.0165 & 0.0556 &$ ( 3.68 \pm 0.15 ) \times 10^{ 12 } $& N & S \\
A0262 & 0109980101 & 21.1 & 20.7 & 17.6 & 0.0161 & 0.0638 &$ ( 1.08 \pm 0.12 ) \times 10^{ 13 } $& Y & S \\
A0400 & 0404010101 & 17.0 & 13.8 & 20.7 & 0.0240 & 0.0833 &$ ( 9.12 \pm 0.36 ) \times 10^{ 12 } $& N & N \\
A0399 & 0112260101 & 11.5 & 12.8 &  5.9 & 0.0715 & 0.1050 &$ ( 4.67 \pm 0.14 ) \times 10^{ 13 } $& N & N \\
A0401 & ---        &  0.0 &  0.0 &  0.0 & 0.0748 & 0.0988 &$ ( 8.81 \pm 0.52 ) \times 10^{ 13 } $& Y & N \\
A3112 & 0105660101 & 21.4 & 21.5 & 17.1 & 0.0750 & 0.0394 &$ ( 3.41 \pm 0.06 ) \times 10^{ 13 } $& Y & S \\
Fornax& 0400620101 & 67.6 & 62.1 & 67.7 & 0.0046 & 0.0138 &$ ( 5.30 \pm 0.07 ) \times 10^{ 11 } $& Y & S \\
IIIZw54& 0505230401& 23.3 & 22.3 & 30.3 & 0.0311 & 0.1470 &$ ( 1.01 \pm 0.20 ) \times 10^{ 13 } $& Y & W \\
A3158 & 0300211301 &  8.4 &  8.2 &  4.9 & 0.0590 & 0.0121 &$ ( 3.75 \pm 0.33 ) \times 10^{ 13 } $& Y & N \\
A0478 & ---        &  0.0 &  0.0 &  0.0 & 0.0900 & 0.1350 &$ ( 8.19 \pm 0.38 ) \times 10^{ 13 } $& Y & S \\
NGC1550& 0152150101& 16.4 & 13.8 & 16.5 & 0.0123 & 0.0981 &$ ( 3.82 \pm 0.85 ) \times 10^{ 12 } $& Y & S \\
EXO0422& 0300210401& 31.5 & 32.2 & 32.7 & 0.0390 & 0.0808 &$ ( 1.32 \pm 0.30 ) \times 10^{ 13 } $& Y & S \\
A3266 & 0105260901 & 22.9 & 22.2 & 17.1 & 0.0594 & 0.0184 &$ ( 8.39 \pm 0.39 ) \times 10^{ 13 } $& N & W \\
A0496 & 0135120201 & 20.4 & 21.4 & 11.8 & 0.0328 & 0.0398 &$ ( 2.79 \pm 0.12 ) \times 10^{ 13 } $& Y & S \\
A3376 & 0151900101 & 16.5 & 16.5 & 19.4 & 0.0455 & 0.0442 &$ ( 2.90 \pm 0.15 ) \times 10^{ 13 } $& N & N \\
A3391 & 0505210401 & 23.3 & 24.6 & 18.2 & 0.0531 & 0.0559 &$ ( 3.16 \pm 0.19 ) \times 10^{ 13 } $& Y & N \\
A3395s& 0400010301 & 28.2 & 28.8 & 24.2 & 0.0498 & 0.0736 &$ ( 2.88 \pm 0.31 ) \times 10^{ 13 } $& N & N \\
A0576 & 0205070301 &  9.8 & 10.1 &  7.2 & 0.0381 & 0.0548 &$ ( 2.00 \pm 0.71 ) \times 10^{ 13 } $& Y & W \\
A0754 & 0136740101 & 12.8 & 13.0 & 11.9 & 0.0528 & 0.0479 &$ ( 4.28 \pm 0.39 ) \times 10^{ 13 } $& N & N \\
HydraA& 0109980301 & 20.6 & 20.1 & 14.3 & 0.0538 & 0.0425 &$ ( 2.67 \pm 0.12 ) \times 10^{ 13 } $& Y & S \\
A1060 & 0206230101 & 35.4 & 36.3 & 26.1 & 0.0114 & 0.0503 &$ ( 8.55 \pm 0.96 ) \times 10^{ 12 } $& Y & W \\
A1367 & 0061740101 & 28.3 & 28.4 & 23.6 & 0.0216 & 0.0189 &$ ( 2.07 \pm 0.08 ) \times 10^{ 13 } $& N & N \\
MKW4  & 0093060101 &  8.0 &  6.8 &  9.4 & 0.0200 & 0.0171 &$ ( 4.30 \pm 0.21 ) \times 10^{ 12 } $& Y & S \\
ZwCl1215& 0300211401&22.8 & 21.0 & 18.5 & 0.0750 & 0.0177 &$ ( 4.75 \pm 0.24 ) \times 10^{ 13 } $& Y & N \\
NGC4636& 0111190701& 32.7 & 32.0 & 51.4 & 0.0037 & 0.0185 &$ ( 1.74 \pm 0.23 ) \times 10^{ 11 } $& Y & S \\
A3526 & 0406200101 &106.1 & 107.2& 88.7 & 0.0103 & 0.0854 &$ ( 1.09 \pm 0.10 ) \times 10^{ 13 } $& N & S \\
A1644 & 0010420201 & 13.4 & 13.5 & 11.7 & 0.0474 & 0.0401 &$ ( 4.23 \pm 0.61 ) \times 10^{ 13 } $& N & S \\
A1650 & 0093200101 & 32.3 & 31.8 & 31.6 & 0.0845 & 0.0130 &$ ( 4.68 \pm 0.73 ) \times 10^{ 13 } $& Y & W \\
A1651 & 0203020101 & 8.7  & 8.6  & 5.4  & 0.0860 & 0.0146 &$ ( 4.93 \pm 0.32 ) \times 10^{ 13 } $& Y & W \\
Coma  & 0124711401 & 16.8 & 14.9 & 14.4 & 0.0232 & 0.0085 &$ ( 7.63 \pm 0.62 ) \times 10^{ 13 } $& N & N \\
NGC5044& 0037950101& 14.4 & 14.5 & 12.9 & 0.0090 & 0.0507 &$ ( 1.85 \pm 0.04 ) \times 10^{ 12 } $& Y & S \\
A1736 & ---        &  0.0 &  0.0 &  0.0 & 0.0461 & 0.0455 &$ ( 3.22 \pm 0.52 ) \times 10^{ 13 } $& N & N \\
A3558 & 0107260101 & 41.7 & 40.7 & 36.3 & 0.0480 & 0.0400 &$ ( 6.39 \pm 0.19 ) \times 10^{ 13 } $& N & W \\
A3562 & 0105261801 & 10.4 & 11.0 & 4.3  & 0.0499 & 0.0389 &$ ( 2.38 \pm 0.09 ) \times 10^{ 13 } $& Y & W \\
A3571 & 0086950201 & 25.2 & 25.3 & 8.4  & 0.0397 & 0.0420 &$ ( 5.17 \pm 0.28 ) \times 10^{ 13 } $& Y & W \\
A1795 & 0097820101 & 38.3 & 37.0 & 24.8 & 0.0616 & 0.0102 &$ ( 4.96 \pm 0.14 ) \times 10^{ 13 } $& Y & S \\
A3581 & 0205990101 & 31.3 & 30.6 & 30.5 & 0.0214 & 0.0431 &$ ( 5.11 \pm 0.59 ) \times 10^{ 12 } $& Y & S \\
MKW8  & 0300210701 & 13.6 & 11.9 & 17.8 & 0.0270 & 0.0234 &$ ( 9.31 \pm 1.33 ) \times 10^{ 12 } $& N & N \\
A2029 & 0111270201 & 11.1 & 11.4 & 9.5  & 0.0767 & 0.0330 &$ ( 8.25 \pm 0.29 ) \times 10^{ 13 } $& Y & S \\
A2052 & 0109920101 & 28.3 & 27.0 & 24.1 & 0.0348 & 0.0268 &$ ( 1.95 \pm 0.11 ) \times 10^{ 13 } $& Y & S \\
MKW3S & 0109930101 & 33.8 & 33.6 & 28.4 & 0.0450 & 0.0286 &$ ( 2.25 \pm 0.10 ) \times 10^{ 13 } $& Y & S \\
A2065 & 0112240201 & 10.2 & 11.0 & 1.5  & 0.0721 & 0.0308 &$ ( 3.43 \pm 0.41 ) \times 10^{ 13 } $& N & W \\
A2063 & 0550360101 & 24.3 & 24.8 & 16.8 & 0.0354 & 0.0273 &$ ( 2.00 \pm 0.09 ) \times 10^{ 13 } $& Y & W \\
A2142 & 0111870301 & 8.7  & 8.5  & 3.3  & 0.0899 & 0.0383 &$ ( 1.34 \pm 0.08 ) \times 10^{ 14 } $& Y & W \\
A2147 & 0505210601 & 6.9  & 8.5  & 2.3  & 0.0351 & 0.0275 &$ ( 3.91 \pm 0.40 ) \times 10^{ 13 } $& N & N \\
A2163 & ---        & 0.0  & 0.0  & 0.0  & 0.2010 & 0.1090 &$ ( 2.12 \pm 0.15 ) \times 10^{ 14 } $& N & N \\
A2199 & 0008030201 & 13.8 & 13.9 & 11.9 & 0.0302 & 0.0089 &$ ( 2.70 \pm 0.24 ) \times 10^{ 13 } $& Y & S \\
A2204 & 0112230301 & 18.2 & 18.4 & 14.2 & 0.1523 & 0.0607 &$ ( 7.90 \pm 0.48 ) \times 10^{ 13 } $& Y & S \\
A2244 & ---        & 0.0  & 0.0  & 0.0  & 0.0970 & 0.0188 &$ ( 5.39 \pm 0.67 ) \times 10^{ 13 } $& Y & W \\
A2256 & 0141380201 & 10.2 & 11.0 & 8.8  & 0.0601 & 0.0433 &$ ( 8.34 \pm 0.32 ) \times 10^{ 13 } $& N & N \\
A2255 & 0112260801 & 8.2  & 8.2  & 3.4  & 0.0800 & 0.0235 &$ ( 4.31 \pm 0.19 ) \times 10^{ 13 } $& N & N \\
A3667 & 0206850101 & 55.4 & 53.0 & 47.8 & 0.0560 & 0.0459 &$ ( 9.39 \pm 0.33 ) \times 10^{ 13 } $& N & W \\
S1101 & 0123900101 & 25.7 & 28.1 & 28.7 & 0.0580 & 0.0105 &$ ( 1.90 \pm 0.13 ) \times 10^{ 13 } $& Y & S \\
A2589 & 0204180101 & 21.2 & 23.2 & 19.6 & 0.0416 & 0.0287 &$ ( 1.77 \pm 0.12 ) \times 10^{ 13 } $& Y & W \\
A2597 & 0147330101 & 37.9 & 37.5 & 41.9 & 0.0852 & 0.0246 &$ ( 3.35 \pm 0.29 ) \times 10^{ 13 } $& Y & S \\
A2634 & 0002960101 & 7.9  & 8.2  & 4.8  & 0.0312 & 0.0514 &$ ( 1.38 \pm 0.09 ) \times 10^{ 13 } $& Y & W \\
A2657 & 0402190301 & 17.6 & 20.3 &  2.2 & 0.0404 & 0.0605 &$ ( 1.64 \pm 0.07 ) \times 10^{ 13 } $& Y & W \\
A4038 & 0204460101 & 28.0 & 27.1 & 26.1 & 0.0283 & 0.0154 &$ ( 1.79 \pm 0.14 ) \times 10^{ 13 } $& Y & W \\
A4059 & 0109950201 & 21.3 & 20.4 & 17.8 & 0.0460 & 0.0119 &$ ( 2.24 \pm 0.17 ) \times 10^{ 13 } $& Y & S \\
\hline
  \end{tabular}
  \end{center}
  \hspace*{0.3cm}{\footnotesize $^{*}$ ``S'' denotes cool-core clusters, and ``W''
    and ``N'' denote non-cool-core clusters. }} 
\end{table*}

\begin{table*} { \begin{center} \footnotesize
      {\renewcommand{\arraystretch}{1.3} \caption[]{Power-law fit,
          $\log_{10}(Y)=A + B \log_{10}(X)$, to the scaling
          relations of the observational sample.}
        \label{t:scaling}}
\begin{tabular}{llllccc}
\hline
$Y$                       & $X$       &   Method & Sample     & $A$               & $B$
& $\sigma_{\rm int}$ (dex)\\
\hline
$\frac{L^{\rm co}_{\rm bol,500}}{E(z)\;{\rm erg~s}^{-1}}$ & $\frac{\sigma}{1000\;{\rm
    km~s}^{-1}}$ &BCES bisector
    & Whole          & $    44.76 \pm 0.97  $ & $4.02\pm 0.33$ & $ 0.27\pm  0.03  $\\
 && & Undisturbed    & $    44.88 \pm 1.27  $ & $4.16\pm 0.44$ & $ 0.29\pm  0.04  $\\
 && & Disturbed      & $    44.66 \pm 1.68  $ & $4.12\pm 0.57$ & $ 0.22\pm  0.03  $\\
 && & Cool core      & $    45.00 \pm 1.48  $ & $4.40\pm 0.52$ & $ 0.28\pm  0.06  $\\
 && & Non-cool core  & $    44.69 \pm 1.59  $ & $4.73\pm 0.54$ & $ 0.25\pm  0.03  $\\
&&BCES orthogonal
    & Whole          & $    44.81 \pm 0.88  $ & $4.47\pm 0.30$ & $ 0.28\pm  0.03  $\\
 && & Undisturbed    & $    44.98 \pm 1.16  $ & $4.66\pm 0.40$ & $ 0.30\pm  0.05  $\\
 && & Disturbed      & $    44.69 \pm 1.71  $ & $4.43\pm 0.58$ & $ 0.23\pm  0.03  $\\
 && & Cool core      & $    45.09 \pm 1.45  $ & $4.83\pm 0.52$ & $ 0.29\pm  0.07  $\\
 && & Non-cool core  & $    44.75 \pm 1.84  $ & $5.45\pm 0.62$ & $ 0.28\pm  0.03  $\\
$\frac{L^{\rm co}_{\rm bol,500}}{E(z)\;{\rm erg~s}^{-1}}$ & $\frac{M_{\rm
    gas,500}E(z)}{10^{14} M_{\odot}}$  &BCES bisector
    & Whole          & $    45.06 \pm 0.68  $ & $1.29\pm 0.05$ & $ 0.09\pm  0.01  $\\
 && & Undisturbed    & $    45.08 \pm 0.77  $ & $1.27\pm 0.06$ & $ 0.09\pm  0.01  $\\
 && & Disturbed      & $    45.06 \pm 0.99  $ & $1.39\pm 0.07$ & $ 0.07\pm  0.01  $\\
 && & Cool core      & $    45.06 \pm 0.77  $ & $1.24\pm 0.06$ & $ 0.10\pm  0.01  $\\
 && & Non-cool core  & $    45.08 \pm 0.62  $ & $1.42\pm 0.05$ & $ 0.06\pm  0.01  $\\
&&BCES orthogonal
    & Whole          & $    45.06 \pm 0.68  $ & $1.29\pm 0.05$ & $ 0.09\pm  0.01  $\\
 && & Undisturbed    & $    45.08 \pm 0.75  $ & $1.27\pm 0.06$ & $ 0.09\pm  0.01  $\\
 && & Disturbed      & $    45.06 \pm 1.01  $ & $1.39\pm 0.07$ & $ 0.07\pm  0.01  $\\
 && & Cool core      & $    44.96 \pm 0.76  $ & $1.24\pm 0.06$ & $ 0.10\pm  0.02  $\\
 && & Non-cool core  & $    45.08 \pm 0.63  $ & $1.42\pm 0.05$ & $ 0.06\pm  0.01  $\\
$\frac{L^{\rm co}_{\rm 0.5-2keV,500}}{E(z)\;{\rm erg~s}^{-1}}$ & $\frac{\sigma}{1000\;{\rm
    km~s}^{-1}}$ &BCES bisector
    & Whole          & $    44.28 \pm 0.89  $ & $3.46\pm 0.30$ & $ 0.24\pm  0.03  $\\
 && & Undisturbed    & $    44.33 \pm 1.17  $ & $3.61\pm 0.41$ & $ 0.27\pm  0.04  $\\
 && & Disturbed      & $    44.09 \pm 1.31  $ & $3.43\pm 0.45$ & $ 0.18\pm  0.03  $\\
 && & Cool core      & $    44.41 \pm 1.38  $ & $3.87\pm 0.49$ & $ 0.26\pm  0.05  $\\
 && & Non-cool core  & $    44.10 \pm 1.31  $ & $4.00\pm 0.45$ & $ 0.22\pm  0.02  $\\
&&BCES orthogonal
    & Whole          & $    44.24 \pm 0.82  $ & $3.88\pm 0.28$ & $ 0.26\pm  0.03  $\\
 && & Undisturbed    & $    44.40 \pm 1.12  $ & $4.10\pm 0.39$ & $ 0.29\pm  0.04  $\\
 && & Disturbed      & $    44.17 \pm 1.35  $ & $3.69\pm 0.46$ & $ 0.19\pm  0.03  $\\
 && & Cool core      & $    44.61 \pm 1.40  $ & $4.27\pm 0.50$ & $ 0.27\pm  0.06  $\\
 && & Non-cool core  & $    44.19 \pm 1.62  $ & $4.63\pm 0.55$ & $ 0.24\pm  0.03  $\\
$\frac{L^{\rm co}_{\rm 0.5-2keV,500}}{E(z)\;{\rm erg~s}^{-1}}$ & $\frac{M_{\rm
    gas,500}E(z)}{10^{14} M_{\odot}}$  &BCES bisector
    & Whole          & $    44.44 \pm 0.44  $ & $1.11\pm 0.03$ & $ 0.06\pm  0.01  $\\
 && & Undisturbed    & $    44.40 \pm 0.54  $ & $1.10\pm 0.04$ & $ 0.06\pm  0.01  $\\
 && & Disturbed      & $    44.44 \pm 0.61  $ & $1.16\pm 0.04$ & $ 0.02\pm  0.01  $\\
 && & Cool core      & $    44.46 \pm 0.53  $ & $1.09\pm 0.04$ & $ 0.07\pm  0.01  $\\
 && & Non-cool core  & $    44.40 \pm 0.33  $ & $1.20\pm 0.02$ & $ 0.01\pm  0.01  $\\
&&BCES orthogonal
    & Whole          & $    44.44 \pm 0.44  $ & $1.11\pm 0.03$ & $ 0.06\pm  0.01  $\\
 && & Undisturbed    & $    44.40 \pm 0.54  $ & $1.10\pm 0.04$ & $ 0.06\pm  0.01  $\\
 && & Disturbed      & $    44.44 \pm 0.60  $ & $1.16\pm 0.04$ & $ 0.02\pm  0.01  $\\
 && & Cool core      & $    44.46 \pm 0.53  $ & $1.09\pm 0.04$ & $ 0.07\pm  0.01  $\\
 && & Non-cool core  & $    44.50 \pm 0.33  $ & $1.20\pm 0.02$ & $ 0.01\pm  0.01  $\\
$\frac{r_{500}}{\rm kpc}$ & $\frac{\sigma}{1000\;{\rm km~s}^{-1}}$ &BCES bisector
    & Whole          & $     3.05 \pm 0.30  $ & $0.85\pm 0.10$ & $ 0.07\pm  0.01  $\\
 && & Undisturbed    & $     3.07 \pm 0.40  $ & $0.90\pm 0.14$ & $ 0.08\pm  0.01  $\\
 && & Disturbed      & $     3.03 \pm 0.28  $ & $0.78\pm 0.09$ & $ 0.05\pm  0.01  $\\
 && & Cool core      & $     3.13 \pm 0.47  $ & $1.01\pm 0.17$ & $ 0.09\pm  0.02  $\\
 && & Non-cool core  & $     3.03 \pm 0.29  $ & $0.82\pm 0.10$ & $ 0.05\pm  0.01  $\\
&&BCES orthogonal
    & Whole          & $     3.05 \pm 0.34  $ & $0.82\pm 0.12$ & $ 0.07\pm  0.01  $\\
 && & Undisturbed    & $     3.07 \pm 0.47  $ & $0.89\pm 0.16$ & $ 0.08\pm  0.01  $\\
 && & Disturbed      & $     3.03 \pm 0.31  $ & $0.76\pm 0.10$ & $ 0.05\pm  0.01  $\\
 && & Cool core      & $     3.12 \pm 0.53  $ & $1.01\pm 0.19$ & $ 0.09\pm  0.02  $\\
 && & Non-cool core  & $     3.02 \pm 0.34  $ & $0.79\pm 0.12$ & $ 0.05\pm  0.01  $\\
\hline
  \end{tabular}
  \end{center}
  \hspace*{0.3cm}{\footnotesize}} 
\end{table*}

\begin{figure*}
\begin{center}
\includegraphics[angle=270,width=8cm]{plots/15830f4a.ps}
\includegraphics[angle=270,width=8cm]{plots/15830f4b.ps}

\includegraphics[angle=270,width=8cm]{plots/15830f4c.ps}
\includegraphics[angle=270,width=8cm]{plots/15830f4d.ps}
\end{center}
\caption{{\it Top-left:} Histogram of residuals in logarithmic space from
  the best-fit $L^{\rm co}_{\rm bol}-\sigma$ relation for the 62 clusters
  using the BCES bisector method. {\it Top-right:} Residual vs.  offset
  between the X-ray flux-weighted center and BCG position. {\it Bottom-left:}
  Residual vs. fraction of the X-ray luminosity within $0.2r_{500}$. {\it
    Bottom-right:} Residual vs. central cooling time. The colors and symbols
  have the same meaning as those in Fig.~\ref{f:lbv}.
  \label{f:dlbv}}
\end{figure*}

\begin{figure*}
\begin{center}
\includegraphics[angle=270,width=16cm]{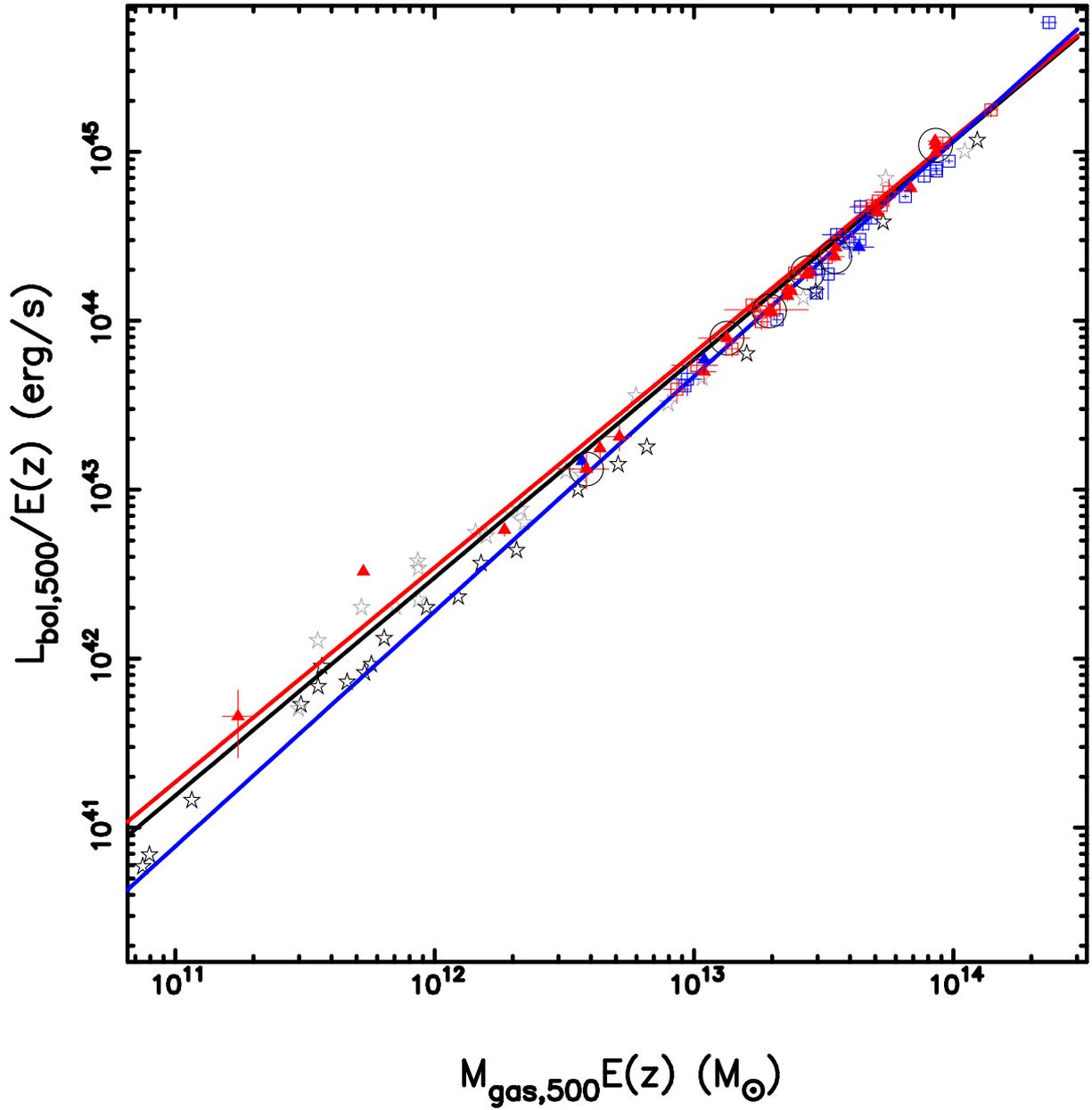}
\end{center}
\caption{X-ray bolometric luminosity vs. gas mass with 
  luminosity corrected for the cluster core ($L^{\rm co}_{\rm bol}$).
  The colors, lines, and symbols have the same meaning as those in
  Fig.~\ref{f:lbv}.
  \label{f:lbmg}}
\end{figure*}

\begin{figure*}
\begin{center}
\includegraphics[angle=270,width=8cm]{plots/15830f6a.ps}
\includegraphics[angle=270,width=8cm]{plots/15830f6b.ps}

\includegraphics[angle=270,width=8cm]{plots/15830f6c.ps}
\includegraphics[angle=270,width=8cm]{plots/15830f6d.ps}
\end{center}
\caption{{\it Top-left:} Histogram of residuals in logarithmic space from
  the best-fit $L^{\rm co}_{\rm
    bol}-M_{\rm gas}$ relation for the 62 clusters
  using the BCES bisector method. {\it Top-right:} Residual vs.  offset
  between the X-ray flux-weighted center and BCG position. {\it Bottom-left:}
  Residual vs. fraction of the X-ray luminosity within $0.2r_{500}$. {\it
    Bottom-right:} Residual vs. central cooling time. The colors and symbols
  have the same meaning as those in Fig.~\ref{f:lbv}.
  \label{f:dlbmg}}
\end{figure*}

\begin{figure*}
\begin{center}
\includegraphics[angle=270,width=16cm]{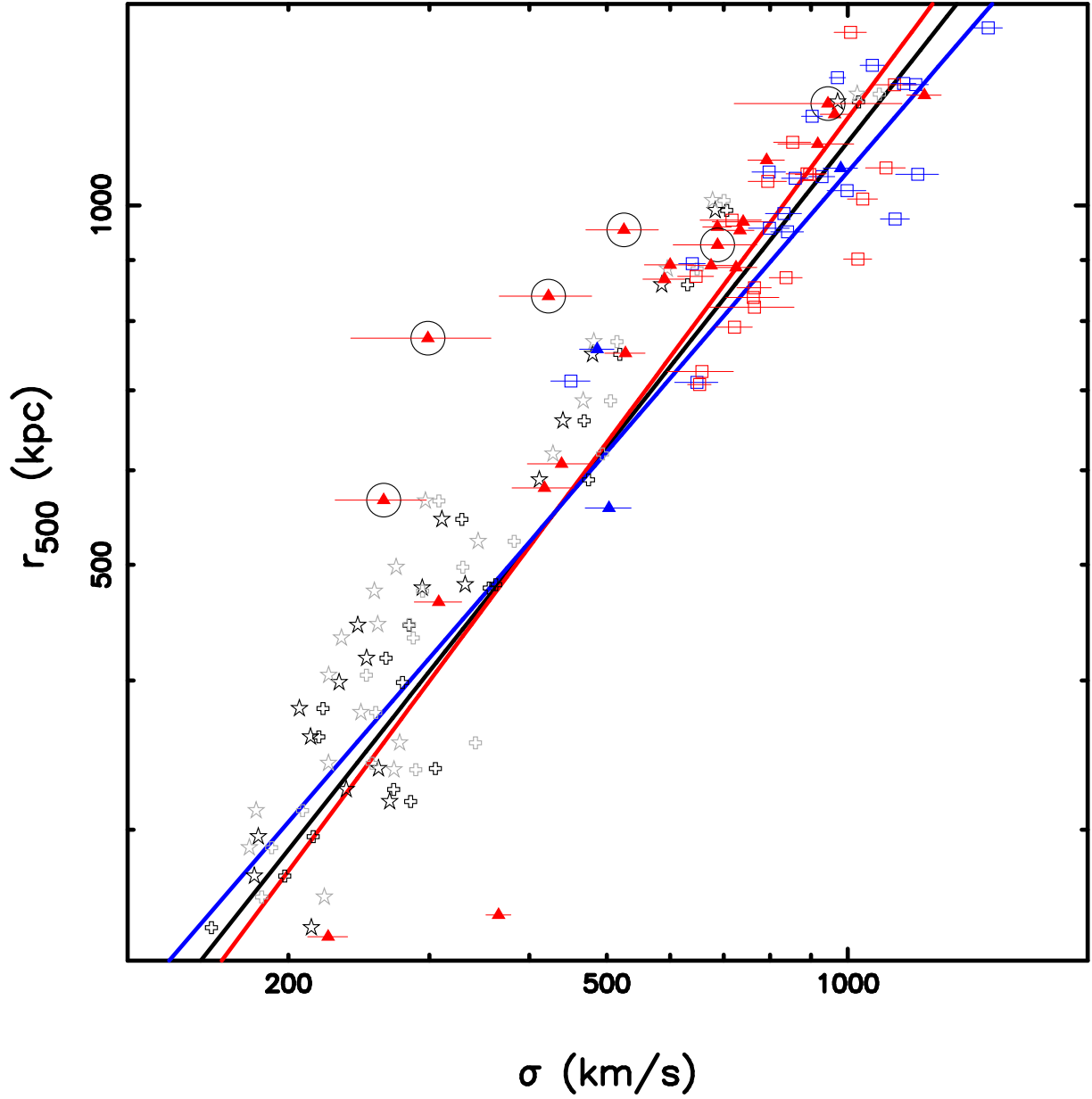}
\end{center}
\caption{Gas-mass-determined cluster radius vs. velocity dispersion.  The
  colors, lines, and symbols have the same meaning as those in
  Fig.~\ref{f:lbv}. 
  \label{f:r500v}}
\end{figure*}

\begin{figure*}
\begin{center}
\includegraphics[angle=270,width=8cm]{plots/15830f8a.ps}
\includegraphics[angle=270,width=8cm]{plots/15830f8b.ps}

\includegraphics[angle=270,width=8cm]{plots/15830f8c.ps}
\includegraphics[angle=270,width=8cm]{plots/15830f8d.ps}
\end{center}
\caption{{\it Top-left:} Histogram of residuals in logarithmic space from
  the best-fit $r_{500}-\sigma$ relation for the 62 clusters
  using the BCES bisector method. {\it Top-right:} Residual vs.  offset
  between the X-ray flux-weighted center and BCG position. {\it Bottom-left:}
  Residual vs. fraction of the X-ray luminosity within $0.2r_{500}$. {\it
    Bottom-right:} Residual vs. central cooling time. The colors and symbols
  have the same meaning as those in Fig.~\ref{f:lbv}.
  \label{f:dr500v}}
\end{figure*}

\begin{figure*}
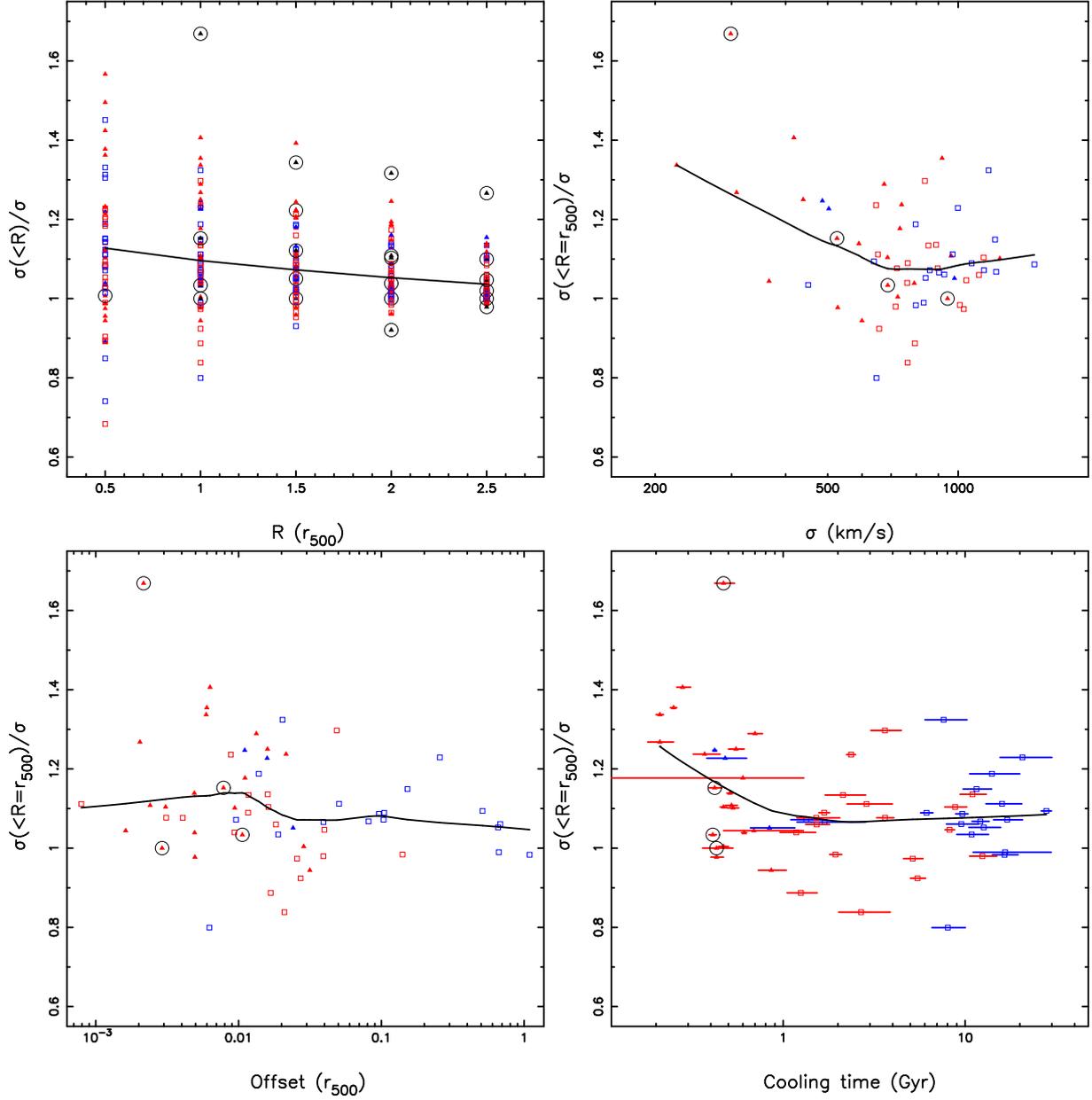

\begin{center}
\includegraphics[angle=270,width=8cm]{plots/15830f9a.ps}
\includegraphics[angle=270,width=8cm]{plots/15830f9b.ps}

\includegraphics[angle=270,width=8cm]{plots/15830f9c.ps}
\includegraphics[angle=270,width=8cm]{plots/15830f9d.ps}
\end{center}
\caption{{\it Top-left:} Velocity dispersion within a projected radius of R
  normalized by the velocity dispersion within 1.2~Abell radii as a
  function of the projected radius. We do not include the values for
  the clusters having fewer than 10 members within the projected
  radius we are interested. {\it Top-right:} Normalized velocity
  dispersion within a projected radius of $r_{500}$ vs. velocity
  dispersion. {\it Bottom-left:} Normalized velocity dispersion within
  a projected radius of $r_{500}$ vs. offset between the X-ray
  flux-weighted center and BCG position. {\it Bottom-right:}
  Normalized velocity dispersion within a projected radius of
  $r_{500}$ vs. central cooling time. The results are only based on
  the observational sample of the 62 clusters. The colors and symbols
  have the same meaning as those in Fig.~\ref{f:lbv}. The curves are
  the local regression non-parametric fits.
\label{f:sigma_dist}}
\end{figure*}

\begin{figure*}
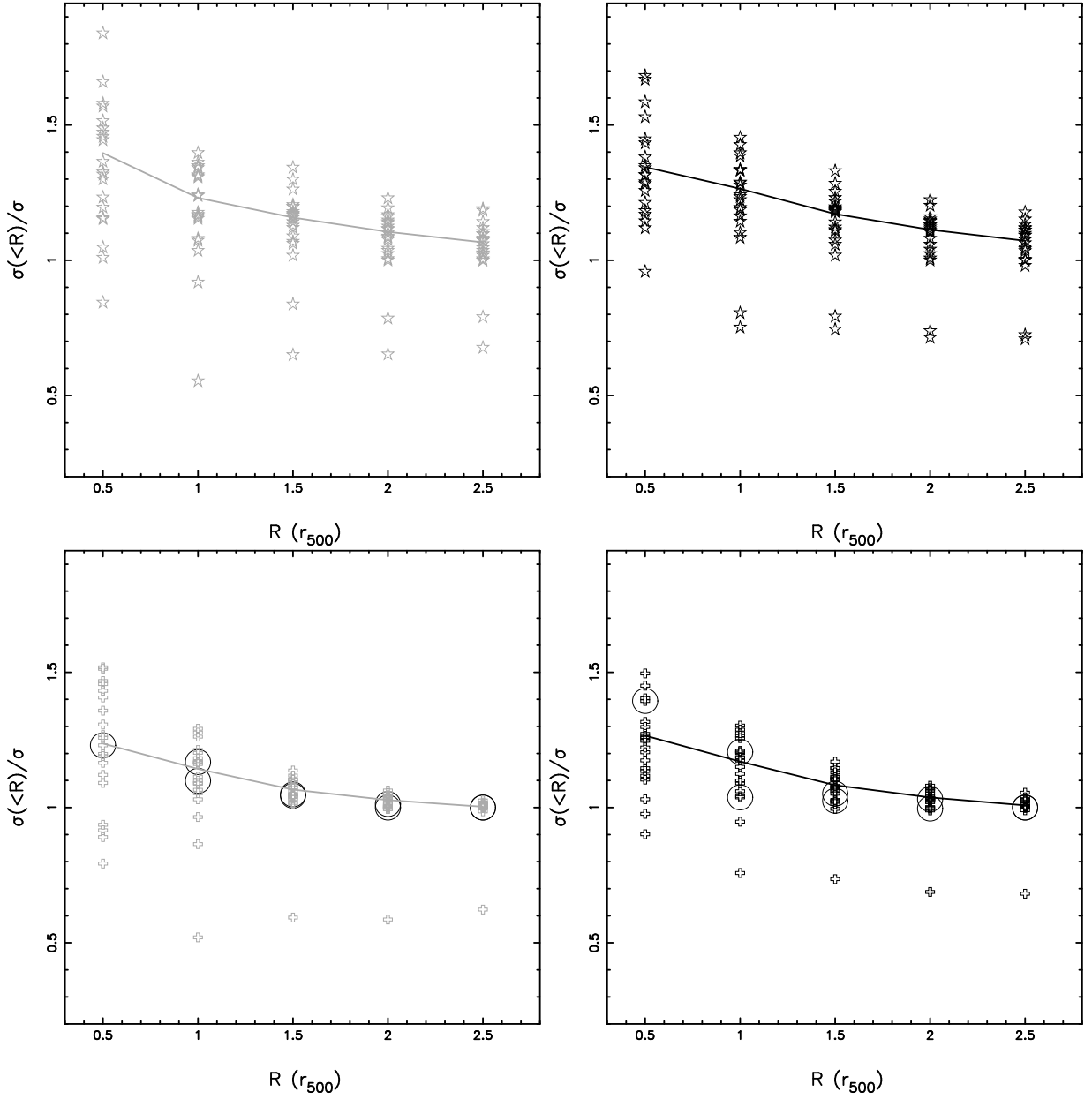

\begin{center}
\includegraphics[angle=270,width=8cm]{plots/15830f10a.ps}
\includegraphics[angle=270,width=8cm]{plots/15830f10b.ps}

\includegraphics[angle=270,width=8cm]{plots/15830f10c.ps}
\includegraphics[angle=270,width=8cm]{plots/15830f10d.ps}
\end{center}
\caption{Velocity dispersion measured by the galaxies within a projected
  radius of R normalized by the velocity dispersion within 1.2~Abell radii as
  a function of the projected radius for the simulated sample. The results are
  only based on the simulated sample of the 21 clusters. The colors and
  symbols have the same meaning as those in Fig.~\ref{f:lbv}. The curves are
  the local regression non-parametric fits. We do not include the values for
  the clusters having fewer than 10 members within the projected radius we are
  interested. The black circles highlight the derived velocity dispersion with
  $<45$ cluster members. \label{f:sigma_dist_simu}}
\end{figure*}

\begin{figure*}
\begin{center}
\includegraphics[angle=270,width=8cm]{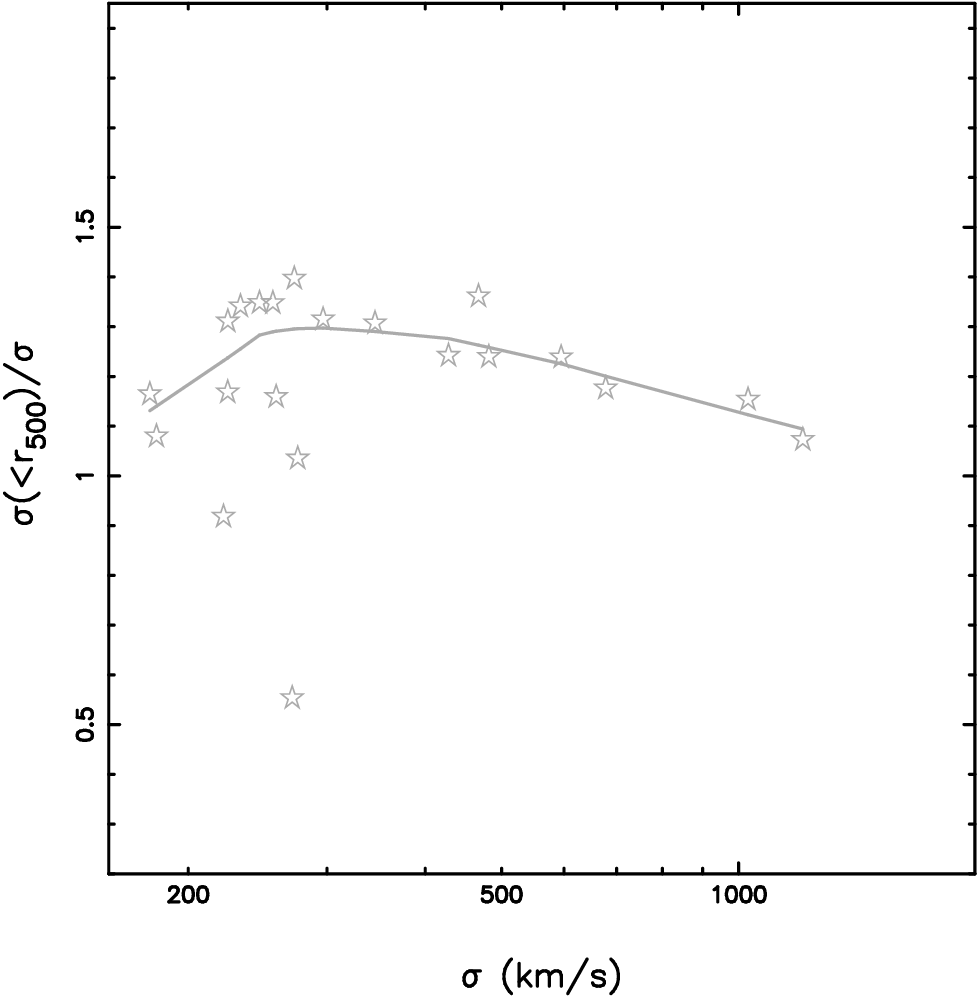}
\includegraphics[angle=270,width=8cm]{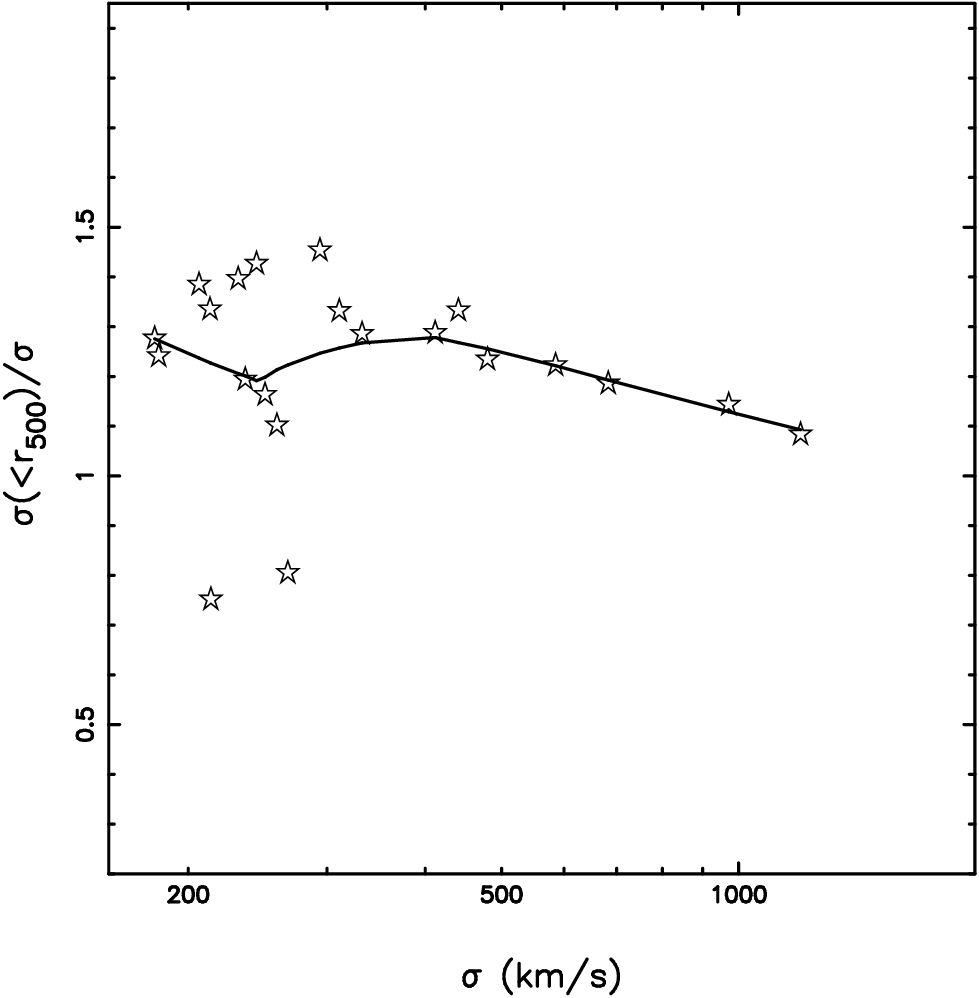}

\includegraphics[angle=270,width=8cm]{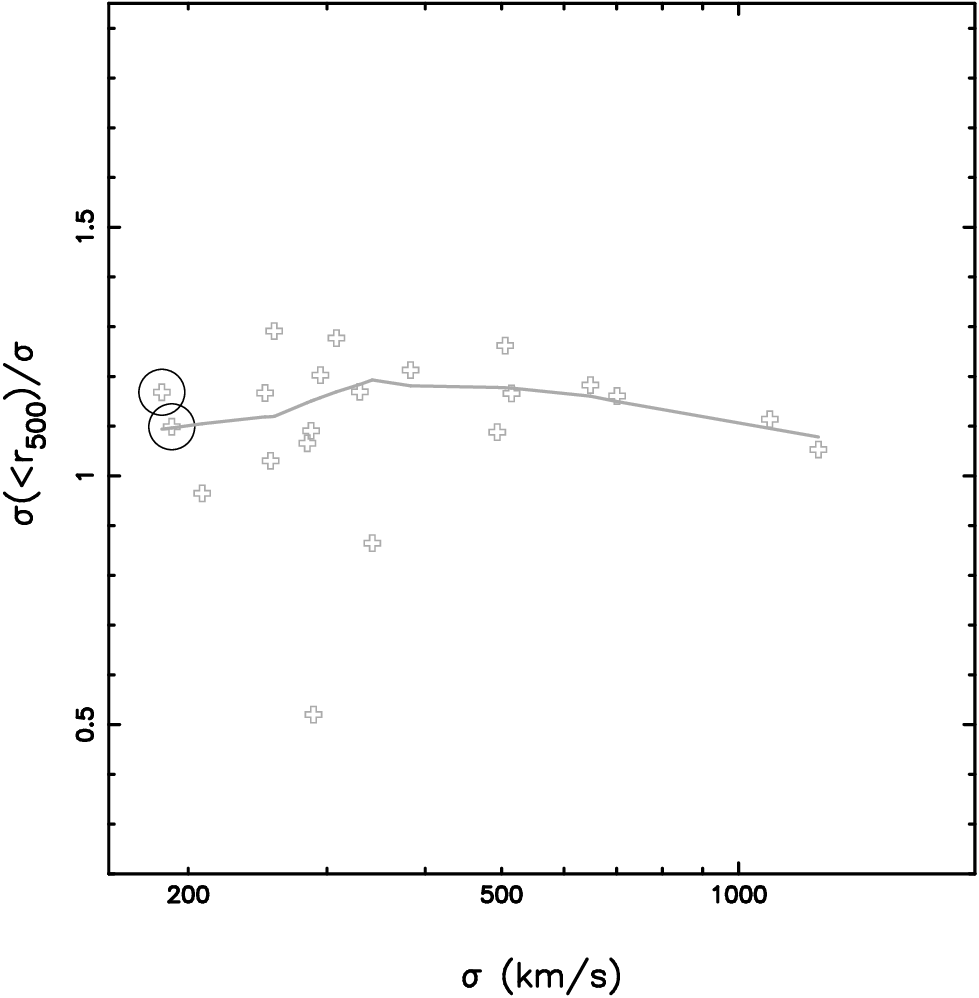}
\includegraphics[angle=270,width=8cm]{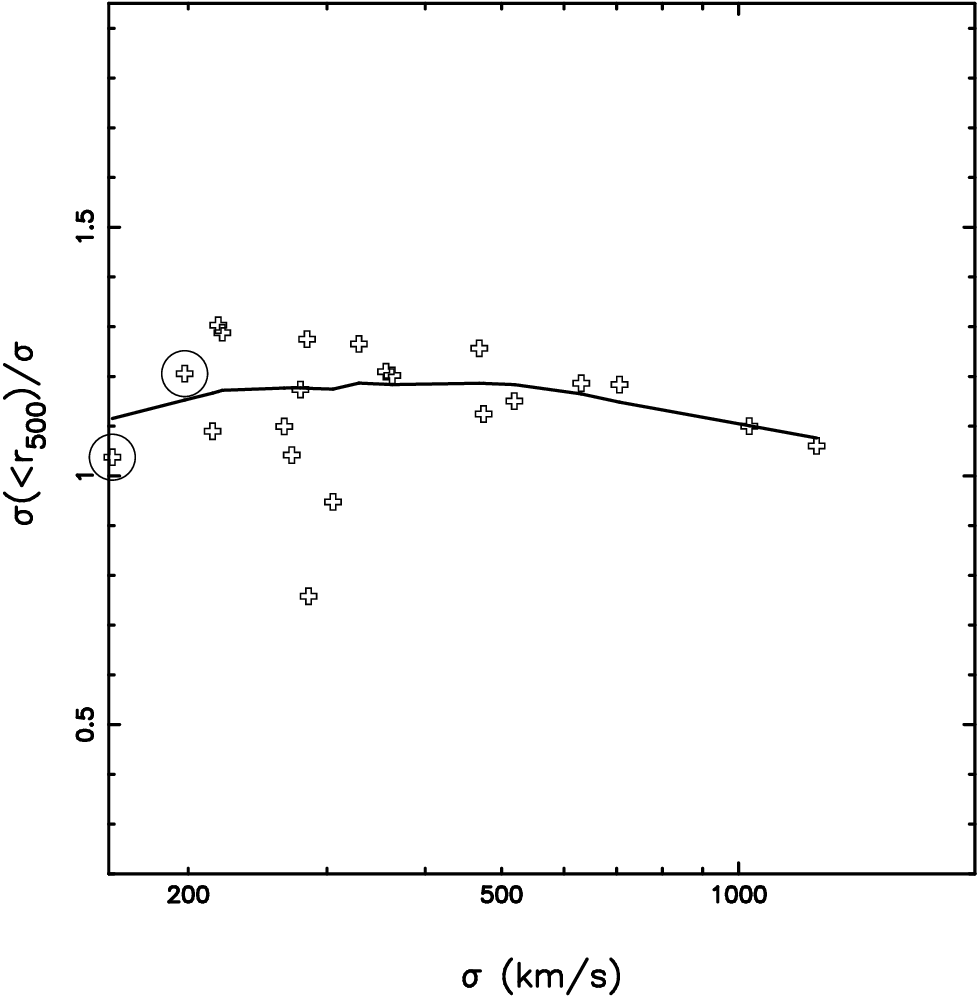}
\end{center}
\caption{Velocity dispersion measured by the galaxies within a projected
  radius of $r_{500}$ normalized by the
  velocity dispersion within 1.2~Abell radii as a function of velocity
  dispersion for the simulated sample. The results are
  only based on the simulated sample of the 21 clusters. The 
colors and symbols have the same meaning as those in
Fig.~\ref{f:sigma_dist_simu}. The curves are the local regression non-parametric
fits.
\label{f:sigma_r500_sigma_simu}}
\end{figure*}

\begin{figure*}
\begin{center}
\includegraphics[angle=270,width=8cm]{plots/15830f12a.ps}
\includegraphics[angle=270,width=8cm]{plots/15830f12b.ps}

\includegraphics[angle=270,width=8cm]{plots/15830f12c.ps}
\includegraphics[angle=270,width=8cm]{plots/15830f12d.ps}
\end{center}
\caption{Velocity dispersion measured by the $n$ most massive galaxies 
  normalized by the velocity dispersion within 1.2~Abell radii as a
  function of the fraction of galaxies for the simulated sample. The
  results are only based on the simulated sample of the 21
  clusters. The colors and symbols have the same meaning as those in
  Fig.~\ref{f:sigma_dist_simu}. The curves are the local regression
  non-parametric fits.
\label{f:sigma_mag_simu}}
\end{figure*}

\begin{figure*}
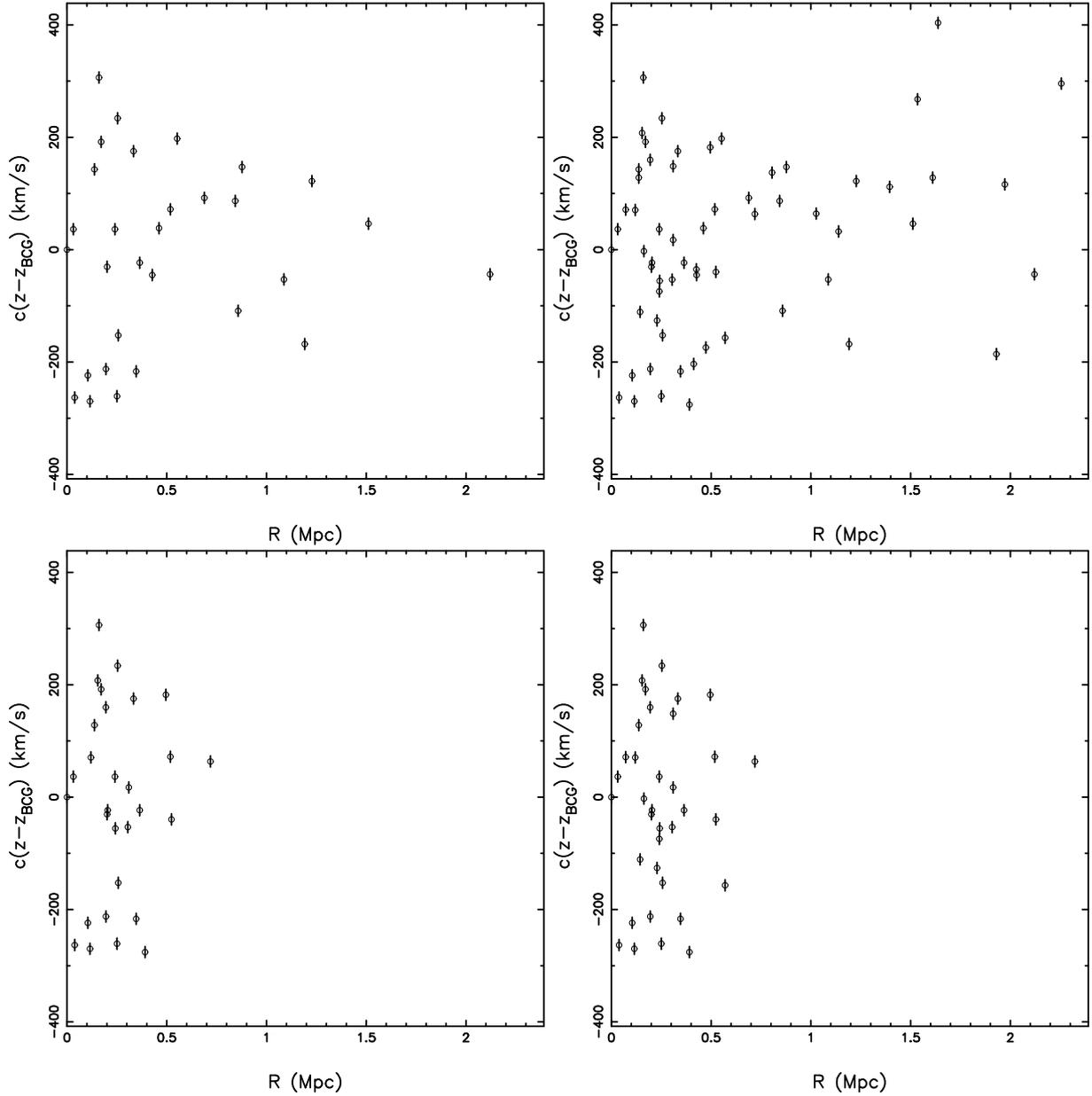

\begin{center}
\includegraphics[angle=270,width=8cm]{plots/15830f13a.ps}
\includegraphics[angle=270,width=8cm]{plots/15830f13b.ps}

\includegraphics[angle=270,width=8cm]{plots/15830f13c.ps}
\includegraphics[angle=270,width=8cm]{plots/15830f13d.ps}
\end{center}
\caption{Line-of-sight velocity vs. projected radius of the 30 
brightest member galaxies (left panels) and all members within
1.2~Abell radii (right panels), respectively, for a simulated cluster
(without AGN feedback) having 40 cluster galaxies when excluding
interlopers. The top panels correspond to $\sigma_{\rm dirty}$, and
the bottom panels correspond to $\sigma_{\rm clean}$.
\label{f:caustic_simu}}
\end{figure*}

\clearpage
\appendix

\section{Luminosity cross-calibration}
\label{a:lxlr}

\begin{figure*}
\begin{center}
\includegraphics[angle=270,width=16cm]{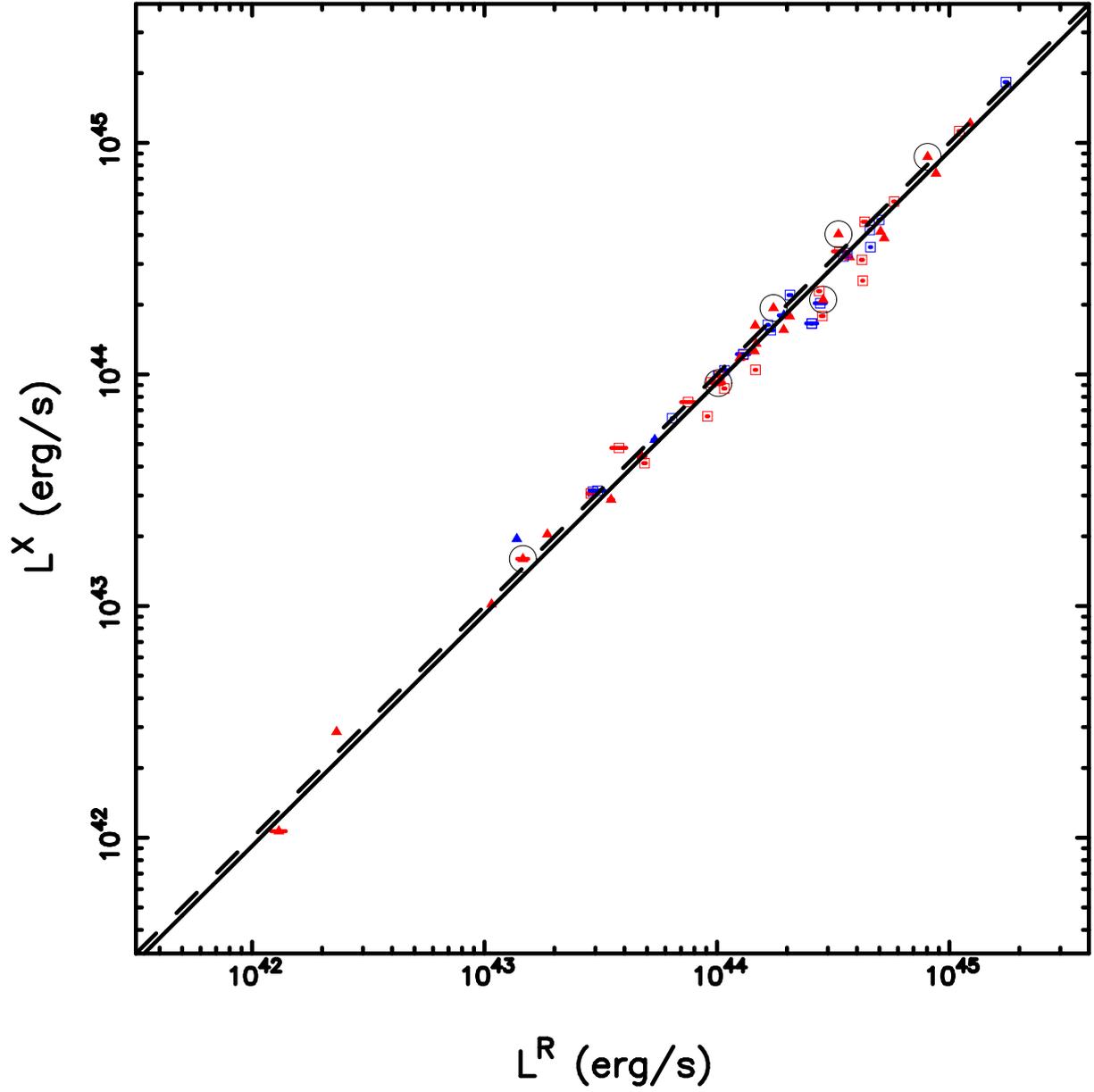}
\end{center}
\caption{\emph{XMM-Newton}-\emph{ROSAT} vs. \emph{ROSAT}-only 
measured luminosity in the
  0.1--2.4~keV band within $r_{500}$. The dashed line denotes 1:1. With a
  fixed slope to 1, the best-fit
  normalization of the \emph{XMM-Newton}-\emph{ROSAT} vs. 
\emph{ROSAT}-only measured
  luminosity for the 62 clusters is 0.92 shown in solid line. The
  colors and symbols have the same meaning as those in
  Fig.~\ref{f:lbv}.  \label{f:lxlr}}
\end{figure*}

To cross-calibrate the \emph{XMM-Newton}-\emph{ROSAT} with the
\emph{ROSAT}-only measured X-ray luminosity, we re-derived the X-ray
luminosity from \emph{ROSAT} within $r_{500}$ given in Sect.~\ref{s:r500} by
using the gas mass from the current work and the mass vs. gas mass relation in
Pratt et al. (2009). The same spectral model was used to derive the X-ray
luminosity using both \emph{ROSAT} data alone and a combination of
\emph{XMM-Newton} and \emph{ROSAT} data. The comparison between the
\emph{XMM-Newton}-\emph{ROSAT} and \emph{ROSAT}-only measured luminosity in
the 0.1--2.4~keV band is shown in Fig.~\ref{f:lxlr}.

The \emph{XMM-Newton}-\emph{ROSAT} to \emph{ROSAT}-only measured luminosity
ratio is $(92\pm2)$\%. The intrinsic scatter is $(0.07\pm0.01)$~dex. This was
found for the REFLEX-DXL sample of 14 massive galaxy clusters at $z\sim 0.3$
in Zhang et al. (2006) and the REXCESS sample of 31 nearby galaxy clusters in
Pratt et al. (2009, $\frac{L^{R}}{\rm erg~s^{-1}}=1.15\times \left
  (\frac{L^{X}}{\rm erg~s^{-1}}\right )^{0.94}$). The difference between the
\emph{XMM-Newton}-\emph{ROSAT} and \emph{ROSAT}-only measured luminosity is
well within the intrinsic scatter.

\section{Iron abundance vs. temperature}

\label{a:TZ}
\begin{figure*}
\begin{center}
\includegraphics[angle=270,width=8.0cm]{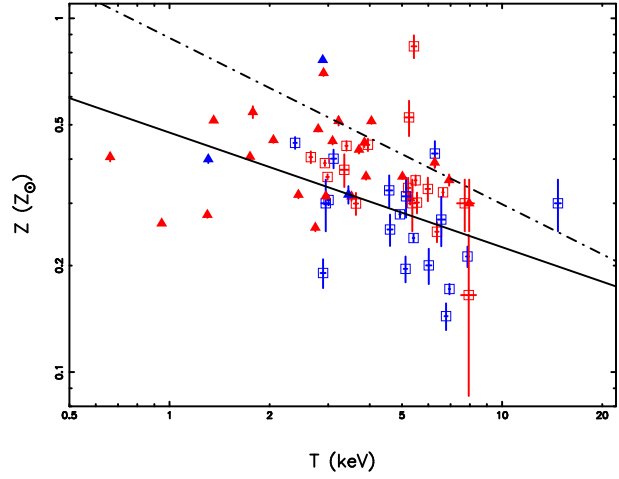}
\end{center}
\caption{Iron abundance vs. temperature for the 62 clusters. The 
black line denotes our best power-law fit using the bisector
method. The dot-dashed line is the best fit in Balestra et al. (2007)
for clusters at higher redshifts ($z \ge 0.3$) and in a higher
temperature range (3--15~keV).  The colors and symbols have the same
meaning as those in Fig.~\ref{f:lbv}.
\label{f:TZ}}
\end{figure*}

\section{Scaling relations using $L^{\rm in}$}
\label{a:lin}

\begin{figure*}
\begin{center}
\includegraphics[angle=270,width=16cm]{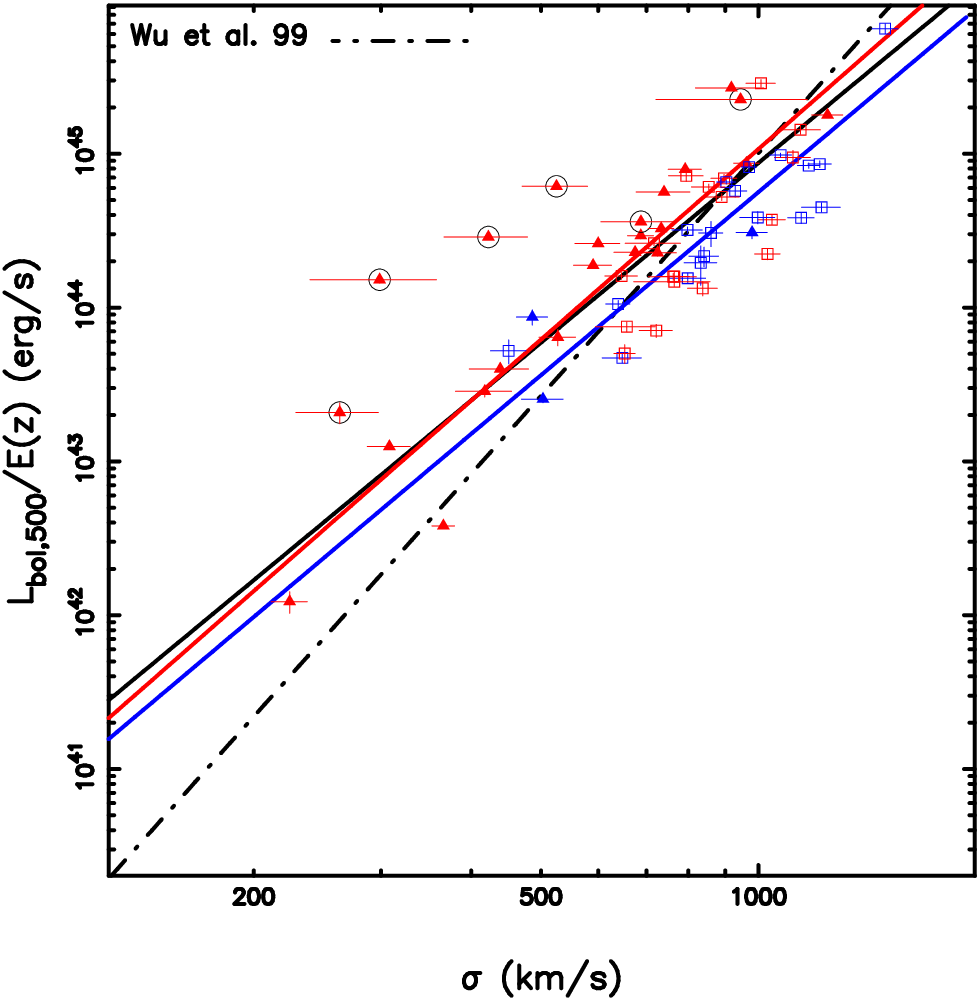}

\includegraphics[angle=270,width=4cm]{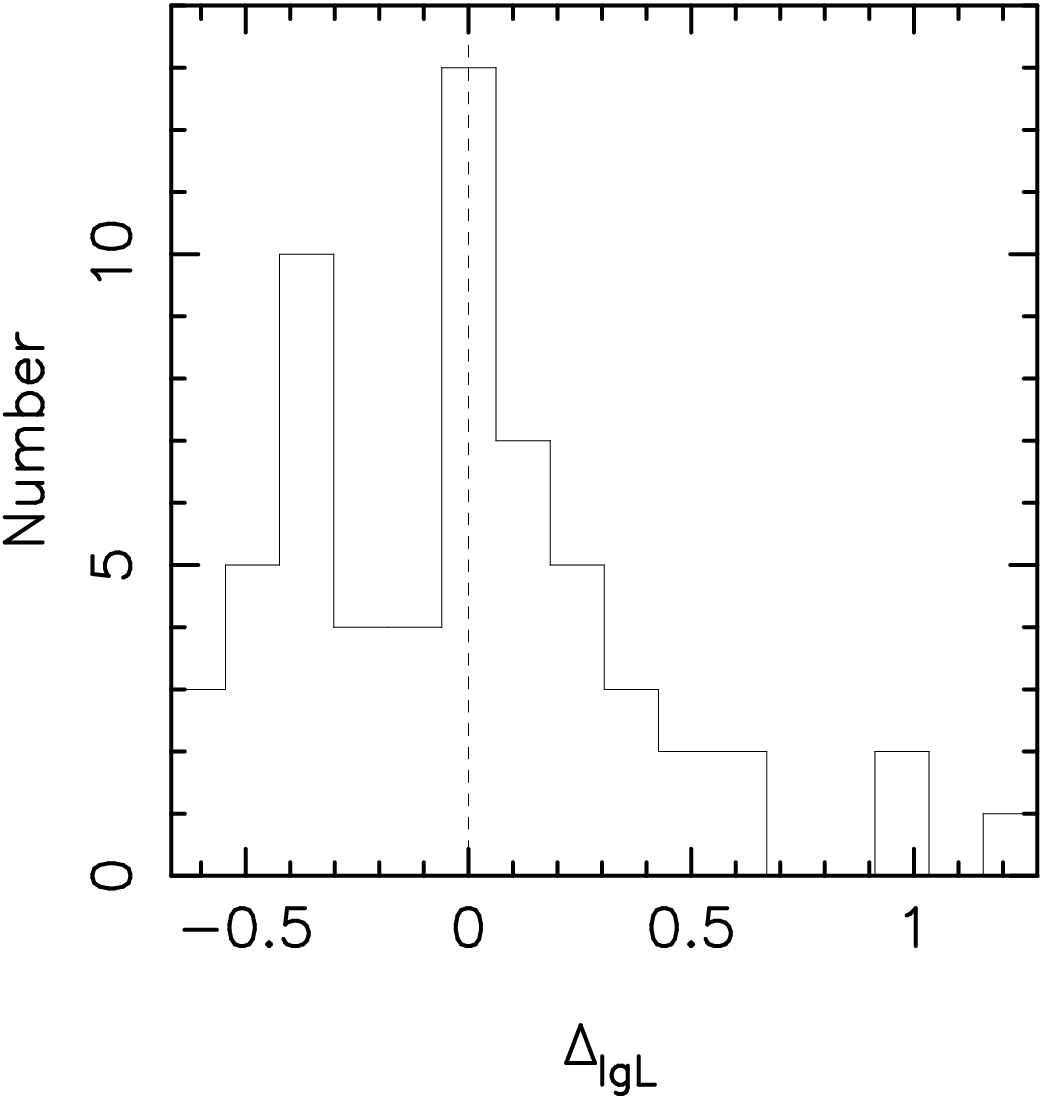}
\includegraphics[angle=270,width=4cm]{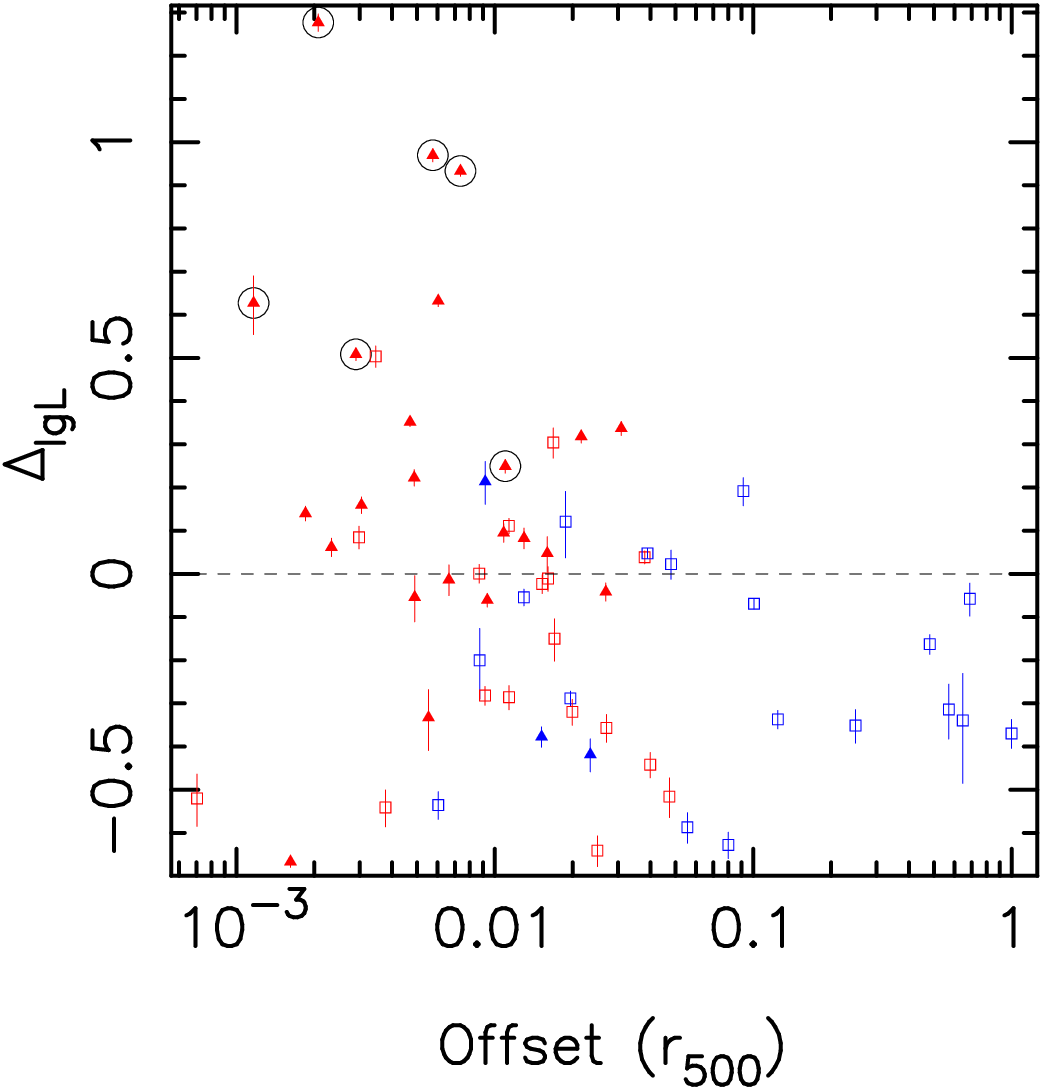}
\includegraphics[angle=270,width=4cm]{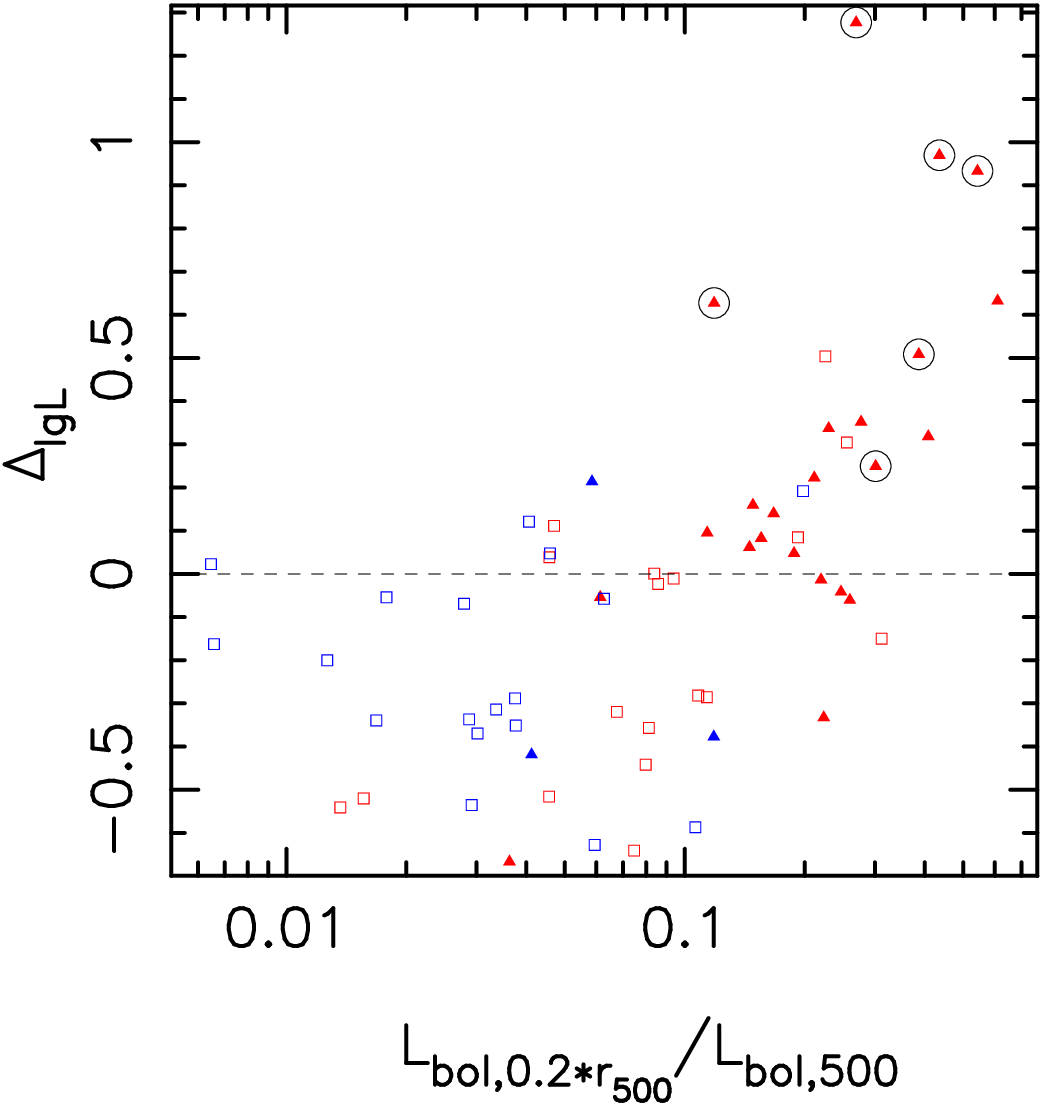}
\includegraphics[angle=270,width=4cm]{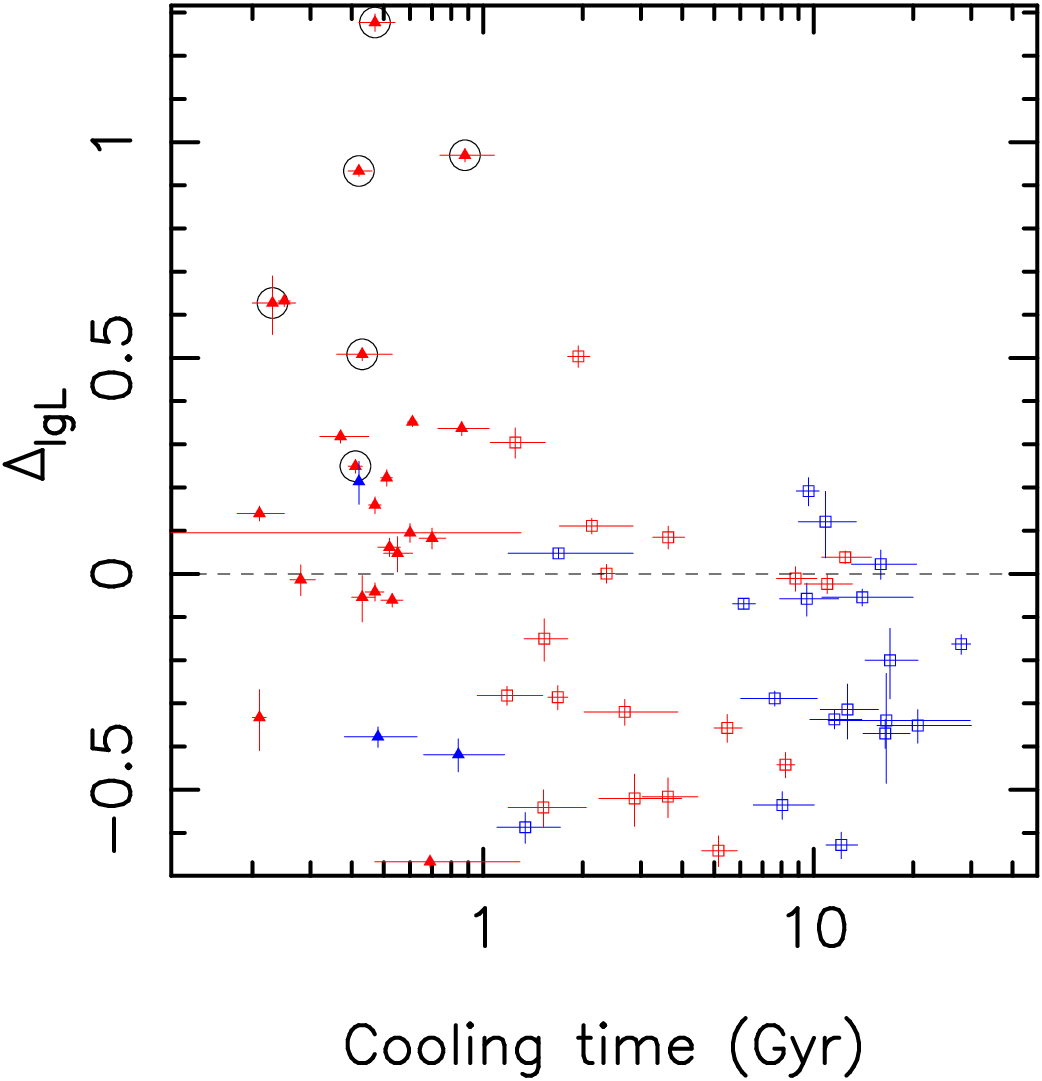}
\end{center}
\caption{{\it Upper panel:} X-ray bolometric luminosity vs. velocity
  dispersion with luminosity derived from all emission interior to $r_{500}$
  ($L^{\rm in}_{\rm bol}$). {\it Lower left panel:} Histogram of residuals in
  logarithmic space from the best-fit $L^{\rm in}_{\rm bol}-\sigma$ relation
  for the 62 clusters using the BCES bisector method.  {\it Lower 2nd panel:}
  Residual vs. offset between the X-ray flux-weighted center and BCG
  position. {\it Lower 3rd panel:} Residual vs. fraction of the X-ray
  luminosity within $0.2r_{500}$. {\it Lower right panel:} Residual
  vs. central cooling time. The colors, lines, and
  symbols have the same meaning as those in Fig.~\ref{f:lbv}.
  \label{f:dlbvin}}
\end{figure*}

\begin{figure*}
\begin{center}
\includegraphics[angle=270,width=16cm]{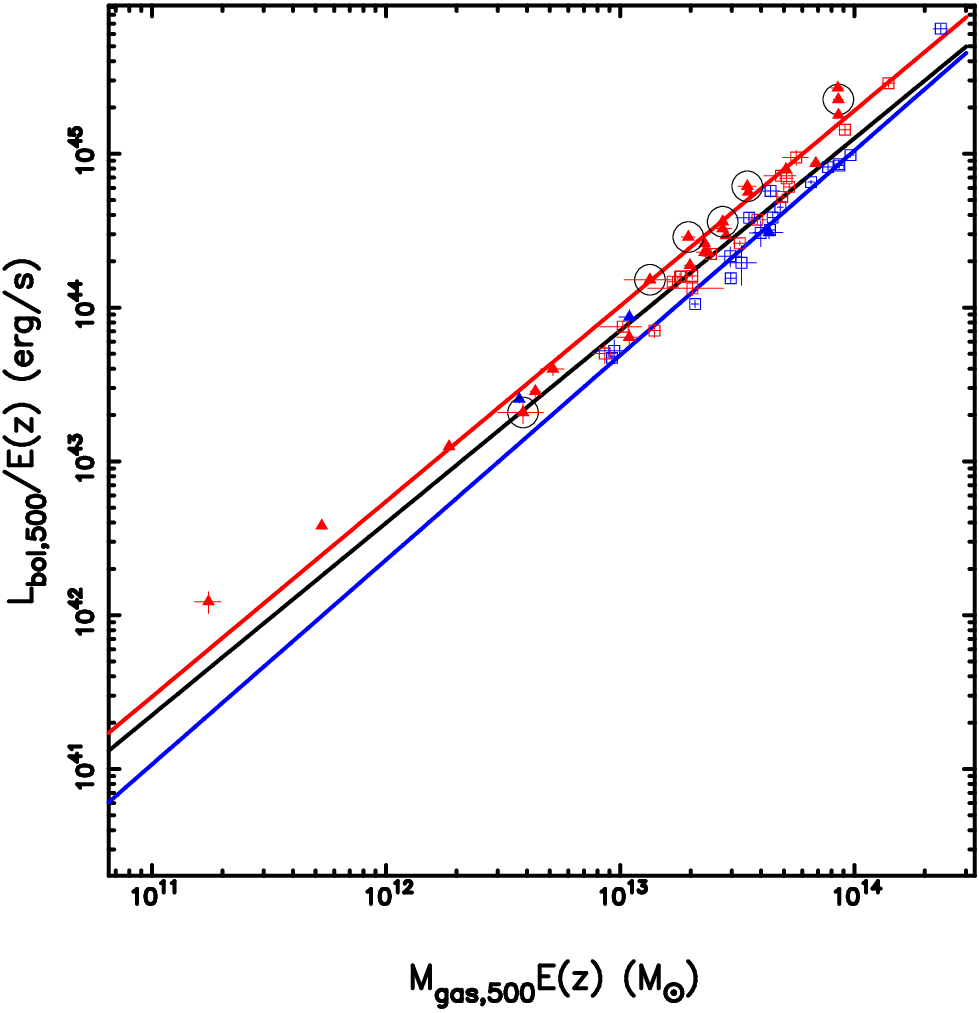}

\includegraphics[angle=270,width=4cm]{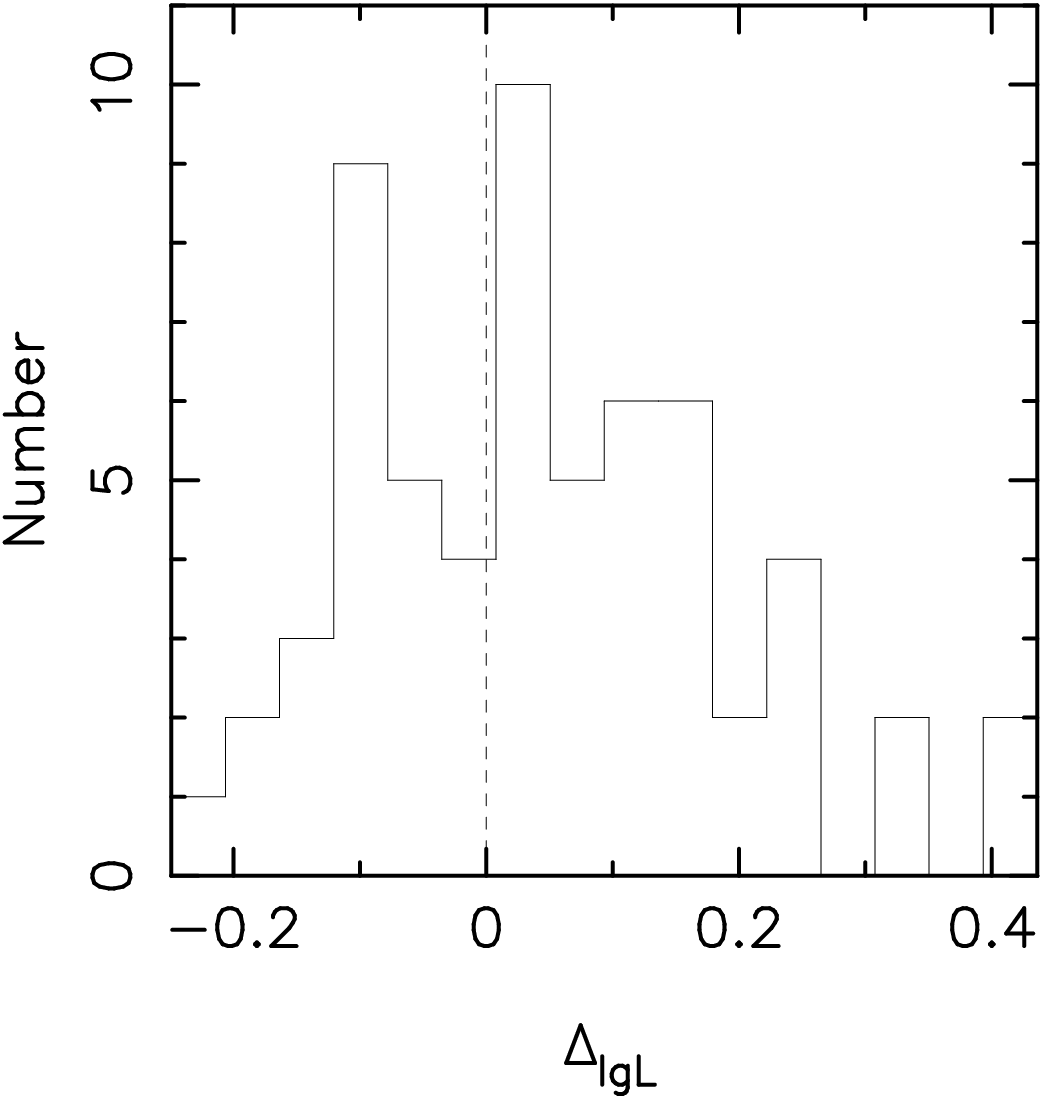}
\includegraphics[angle=270,width=4cm]{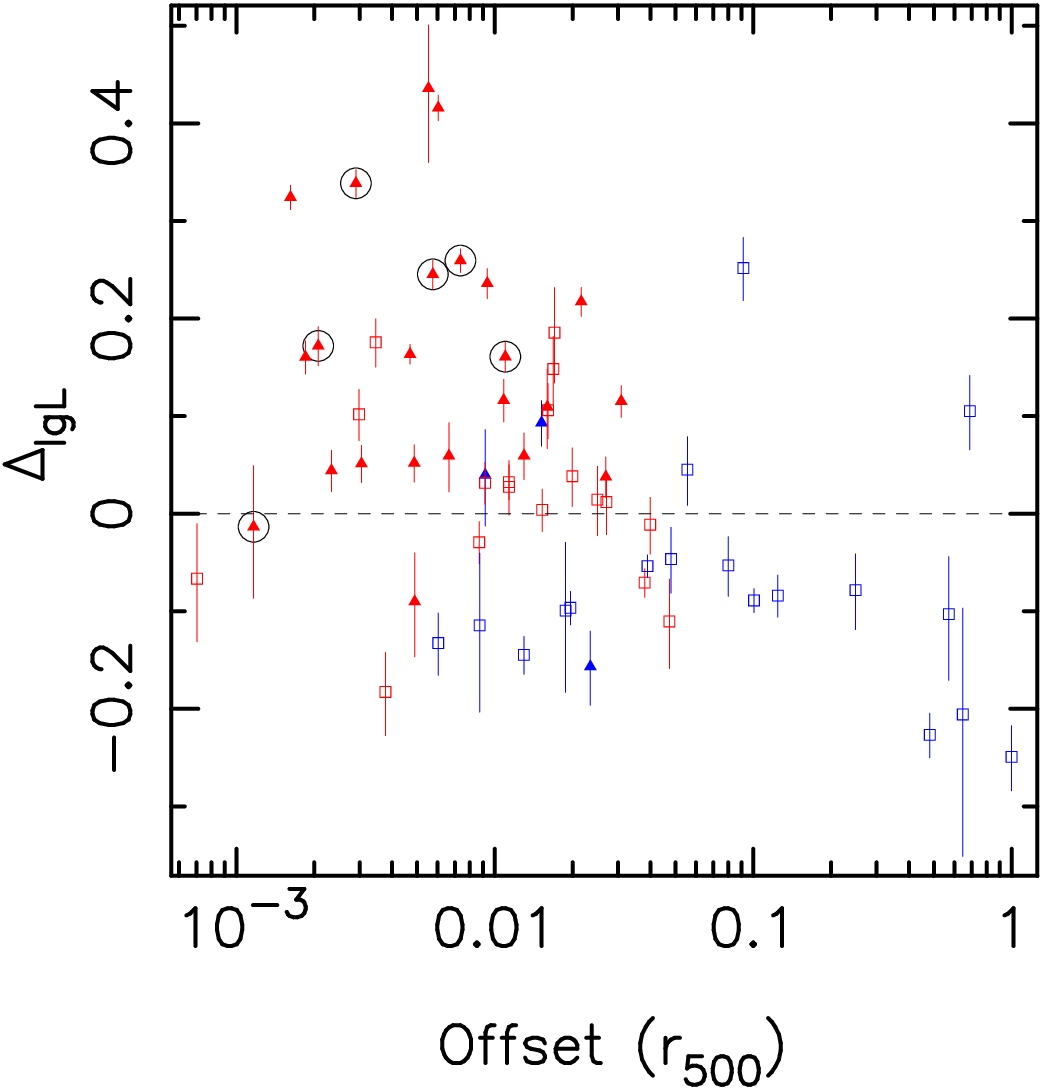}
\includegraphics[angle=270,width=4cm]{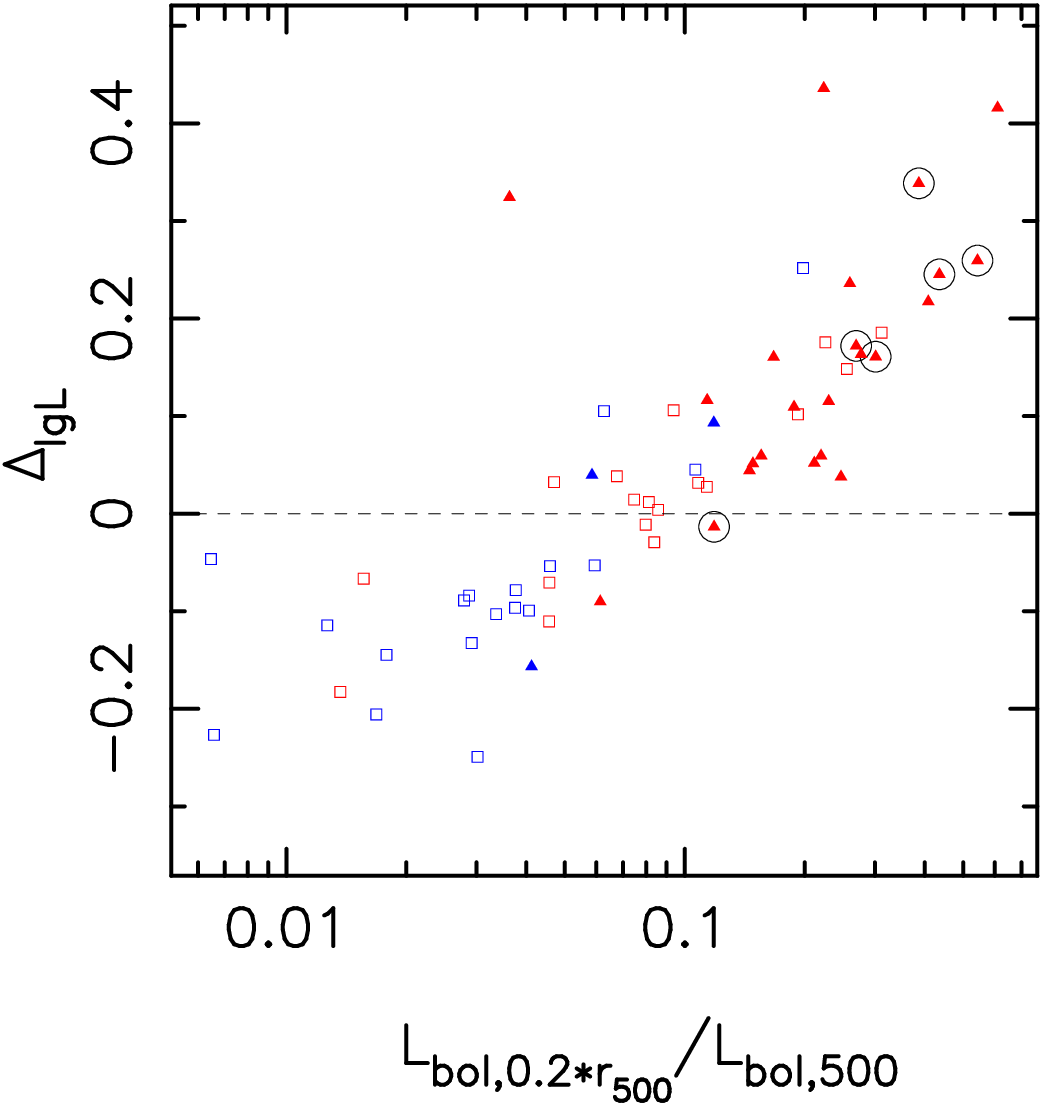}
\includegraphics[angle=270,width=4cm]{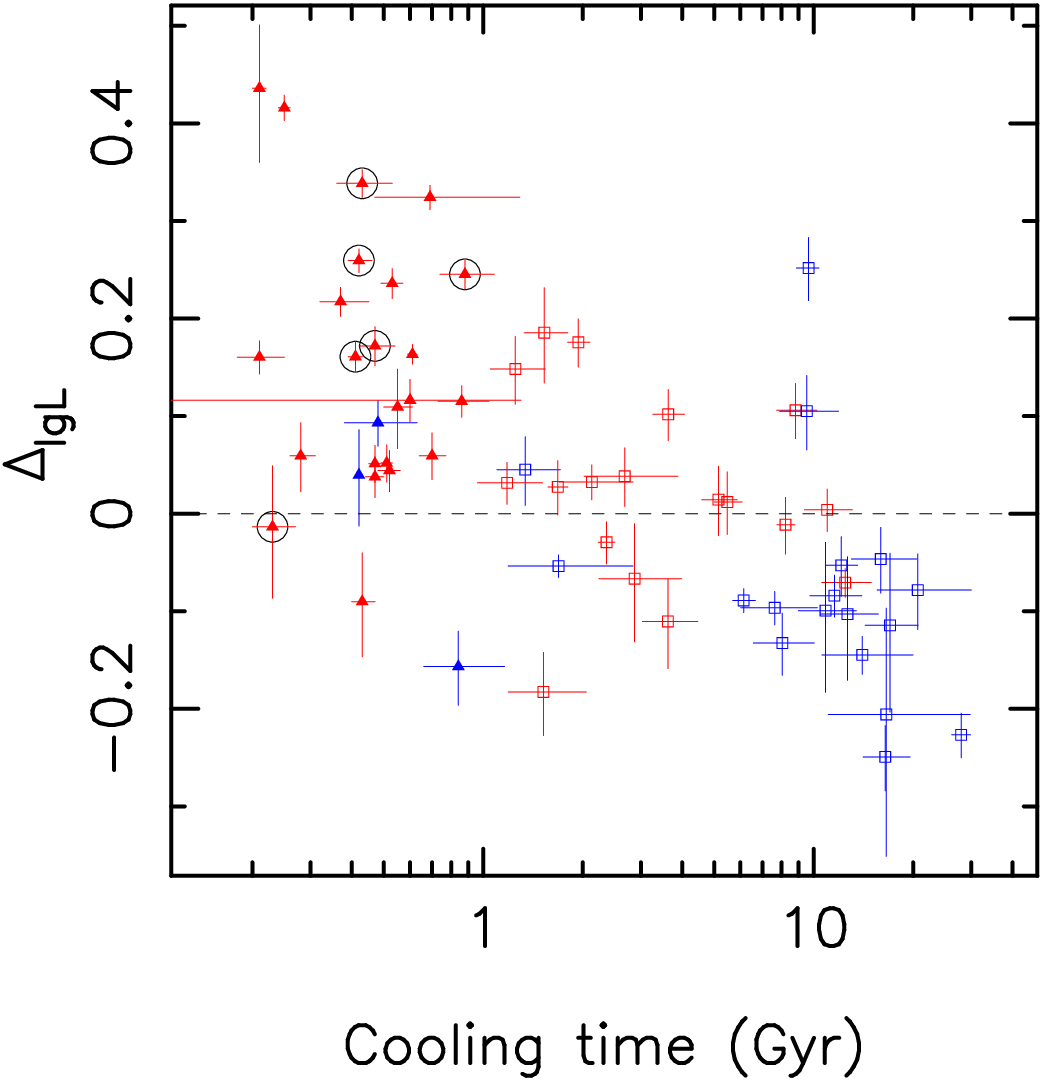}
\end{center}
\caption{{\it Upper panel:} X-ray bolometric luminosity vs. gas mass with
  luminosity derived from all emission interior to $r_{500}$ ($L^{\rm in}_{\rm
    bol}$). {\it Lower left panel:} Histogram of residuals in logarithmic
  space from the best-fit $L^{\rm in}_{\rm bol}-M_{\rm gas}$ relation for the
  62 clusters using the BCES bisector method.  {\it Lower 2nd panel:} Residual
  vs. offset between the X-ray flux-weighted center and BCG position. {\it
    Lower 3rd panel:} Residual vs. fraction of the X-ray luminosity within
  $0.2r_{500}$. {\it Lower right panel:} Residual vs. central cooling
  time. The colors, lines, and symbols have the same meaning as those in
  Fig.~\ref{f:lbv}.
\label{f:dlbmgin}}
\end{figure*}

In Table~\ref{t:lx_in_ex}, we present the X-ray bolometric luminosity within
$r_{500}$ ($L^{\rm in}_{\rm bol}$). We list the best fits to the corresponding
scaling relations using the bolometric and 0.5--2~keV band luminosity in
Table~\ref{t:scaling_lin}, and show those plots using the bolometric
luminosity in Figs.~\ref{f:dlbvin}--\ref{f:dlbmgin}, which helps us to
understand the scatter driven by the presence of cool cores.

\begin{table*} { \begin{center} \footnotesize
      {\renewcommand{\arraystretch}{1.3} \caption[]{X-ray bolometric
          luminosity within $r_{500}$, $L^{\rm in}$, and in the
          $[0.2-1]r_{500}$ annulus, $L^{\rm ex}$.}
        \label{t:lx_in_ex}}
\begin{tabular}{lcc}
\hline
Cluster         & $L^{\rm in}_{\rm bol}$ (erg~s$^{-1}$) & $L^{\rm ex}_{\rm bol}$ (erg~s$^{-1}$)\\
\hline
A0085         &$(8.91 \pm0.43  )\times 10^{44}$     &$(4.73 \pm0.28  )\times 10^{44  }$\\
A0119         &$(3.26 \pm0.14  )\times 10^{44}$     &$(2.57 \pm0.12  )\times 10^{44  }$\\
A0133         &$(2.33 \pm0.11  )\times 10^{44}$     &$(1.19 \pm0.07  )\times 10^{44  }$\\
NGC507        &$(2.56 \pm0.14  )\times 10^{43}$     &$(1.09 \pm0.08  )\times 10^{43  }$\\
A0262         &$(6.47 \pm0.79  )\times 10^{43}$     &$(4.14 \pm0.71  )\times 10^{43  }$\\
A0400         &$(4.76 \pm0.35  )\times 10^{43}$     &$(3.38 \pm0.25  )\times 10^{43  }$\\
A0399         &$(4.64 \pm0.33  )\times 10^{44}$     &$(3.44 \pm0.25  )\times 10^{44  }$\\
A0401         &$(1.48 \pm0.10  )\times 10^{45}$     &$(8.50 \pm0.66  )\times 10^{44  }$\\
A3112         &$(5.85 \pm0.20  )\times 10^{44}$     &$(1.99 \pm0.11  )\times 10^{44  }$\\
Fornax        &$(3.81 \pm0.11  )\times 10^{42}$     &$(2.74 \pm0.08  )\times 10^{42  }$\\
IIIZw54       &$(7.62 \pm0.56  )\times 10^{43}$     &$(3.97 \pm0.45  )\times 10^{43  }$\\
A3158         &$(3.83 \pm0.26  )\times 10^{44}$     &$(2.48 \pm0.17  )\times 10^{44  }$\\
A0478         &$(2.35 \pm0.08  )\times 10^{45}$     &$(7.94 \pm0.44  )\times 10^{44  }$\\
NGC1550       &$(2.09 \pm0.32  )\times 10^{43}$     &$(1.04 \pm0.28  )\times 10^{43  }$\\
EXO0422       &$(1.54 \pm0.07  )\times 10^{44}$     &$(5.64 \pm0.46  )\times 10^{43  }$\\
A3266         &$(8.61 \pm0.34  )\times 10^{44}$     &$(6.55 \pm0.26  )\times 10^{44  }$\\
A0496         &$(2.98 \pm0.13  )\times 10^{44}$     &$(1.55 \pm0.09  )\times 10^{44  }$\\
A3376         &$(1.58 \pm0.12  )\times 10^{44}$     &$(1.26 \pm0.10  )\times 10^{44  }$\\
A3391         &$(2.68 \pm0.09  )\times 10^{44}$     &$(2.02 \pm0.07  )\times 10^{44  }$\\
A3395s        &$(2.21 \pm0.32  )\times 10^{44}$     &$(1.79 \pm0.30  )\times 10^{44  }$\\
A0576         &$(1.36 \pm0.14  )\times 10^{44}$     &$(9.46 \pm1.13  )\times 10^{43  }$\\
A0754         &$(5.87 \pm0.51  )\times 10^{44}$     &$(3.45 \pm0.41  )\times 10^{44  }$\\
HydraA        &$(3.71 \pm0.13  )\times 10^{44}$     &$(1.37 \pm0.07  )\times 10^{44  }$\\
A1060         &$(5.05 \pm0.70  )\times 10^{43}$     &$(3.17 \pm0.65  )\times 10^{43  }$\\
A1367         &$(1.07 \pm0.06  )\times 10^{44}$     &$(8.55 \pm0.47  )\times 10^{43  }$\\
MKW4          &$(2.88 \pm0.23  )\times 10^{43}$     &$(1.38 \pm0.13  )\times 10^{43  }$\\
ZwCl1215      &$(5.42 \pm0.27  )\times 10^{44}$     &$(3.69 \pm0.20  )\times 10^{44  }$\\
NGC4636       &$(1.23 \pm0.20  )\times 10^{42}$     &$(3.37 \pm1.31  )\times 10^{41  }$\\
A3526         &$(8.72 \pm0.99  )\times 10^{43}$     &$(4.63 \pm0.91  )\times 10^{43  }$\\
A1644         &$(3.14 \pm0.27  )\times 10^{44}$     &$(2.41 \pm0.23  )\times 10^{44  }$\\
A1650         &$(7.49 \pm0.60  )\times 10^{44}$     &$(3.55 \pm0.50  )\times 10^{44  }$\\
A1651         &$(7.21 \pm0.43  )\times 10^{44}$     &$(3.93 \pm0.27  )\times 10^{44  }$\\
Coma          &$(8.27 \pm0.64  )\times 10^{44}$     &$(5.58 \pm0.58  )\times 10^{44  }$\\
NGC5044       &$(1.25 \pm0.05  )\times 10^{43}$     &$(4.37 \pm0.22  )\times 10^{42  }$\\
A1736         &$(2.00 \pm0.57  )\times 10^{44}$     &$(1.64 \pm0.46  )\times 10^{44  }$\\
A3558         &$(6.69 \pm0.18  )\times 10^{44}$     &$(4.31 \pm0.13  )\times 10^{44  }$\\
A3562         &$(2.28 \pm0.19  )\times 10^{44}$     &$(1.65 \pm0.14  )\times 10^{44  }$\\
A3571         &$(6.20 \pm0.25  )\times 10^{44}$     &$(3.81 \pm0.20  )\times 10^{44  }$\\
A1795         &$(8.15 \pm0.19  )\times 10^{44}$     &$(3.30 \pm0.11  )\times 10^{44  }$\\
A3581         &$(4.02 \pm0.38  )\times 10^{43}$     &$(1.53 \pm0.31  )\times 10^{43  }$\\
MKW8          &$(5.30 \pm0.93  )\times 10^{43}$     &$(3.64 \pm0.81  )\times 10^{43  }$\\
A2029         &$(1.85 \pm0.06  )\times 10^{45}$     &$(7.24 \pm0.34  )\times 10^{44  }$\\
A2052         &$(1.91 \pm0.08  )\times 10^{44}$     &$(8.63 \pm0.61  )\times 10^{43  }$\\
MKW3S         &$(2.66 \pm0.10  )\times 10^{44}$     &$(9.98 \pm0.62  )\times 10^{43  }$\\
A2065         &$(3.96 \pm0.32  )\times 10^{44}$     &$(2.65 \pm0.23  )\times 10^{44  }$\\
A2063         &$(1.63 \pm0.08  )\times 10^{44}$     &$(1.00 \pm0.06  )\times 10^{44  }$\\
A2142         &$(2.99 \pm0.17  )\times 10^{45}$     &$(1.35 \pm0.11  )\times 10^{45  }$\\
A2147         &$(3.10 \pm0.57  )\times 10^{44}$     &$(2.54 \pm0.54  )\times 10^{44  }$\\
A2163         &$(7.19 \pm0.53  )\times 10^{45}$     &$(4.80 \pm0.39  )\times 10^{45  }$\\
A2199         &$(3.30 \pm0.16  )\times 10^{44}$     &$(1.33 \pm0.11  )\times 10^{44  }$\\
A2204         &$(2.88 \pm0.09  )\times 10^{45}$     &$(8.62 \pm0.44  )\times 10^{44  }$\\
A2244         &$(9.89 \pm1.11  )\times 10^{44}$     &$(4.22 \pm0.67  )\times 10^{44  }$\\
A2256         &$(8.81 \pm0.43  )\times 10^{44}$     &$(6.25 \pm0.31  )\times 10^{44  }$\\
A2255         &$(4.00 \pm0.36  )\times 10^{44}$     &$(3.33 \pm0.29  )\times 10^{44  }$\\
A3667         &$(1.01 \pm0.03  )\times 10^{45}$     &$(7.32 \pm0.23  )\times 10^{44  }$\\
S1101         &$(2.95 \pm0.10  )\times 10^{44}$     &$(7.83 \pm0.53  )\times 10^{43  }$\\
A2589         &$(1.62 \pm0.08  )\times 10^{44}$     &$(8.39 \pm0.52  )\times 10^{43  }$\\
A2597         &$(6.39 \pm0.18  )\times 10^{44}$     &$(1.70 \pm0.10  )\times 10^{44  }$\\
A2634         &$(7.18 \pm0.70  )\times 10^{43}$     &$(6.08 \pm0.57  )\times 10^{43  }$\\
A2657         &$(1.50 \pm0.10  )\times 10^{44}$     &$(1.04 \pm0.08  )\times 10^{44  }$\\
A4038         &$(1.61 \pm0.10  )\times 10^{44}$     &$(7.54 \pm0.83  )\times 10^{43  }$\\
A4059         &$(2.34 \pm0.13  )\times 10^{44}$     &$(1.16 \pm0.09  )\times 10^{44  }$\\
\hline
  \end{tabular}
  \end{center}
  \hspace*{0.3cm}{\footnotesize Note: The bolometric X-ray luminosity within
    the cluster core can be derived by $L_{\rm bol,0.2r_{500}}=L^{\rm in}_{\rm bol}-L^{\rm ex}_{\rm bol}$. }} 
\end{table*}

\begin{table*} { \begin{center} \footnotesize
      {\renewcommand{\arraystretch}{1.3} \caption[]{Power-law fit,
          $\log_{10}(Y)=A + B \log_{10}(X)$, to the scaling
          relations for the observational sample using $L^{\rm in}$.}
        \label{t:scaling_lin}}
\begin{tabular}{llllccc}
\hline
$Y$                       & $X$       & Method & Sample     & $A$               & $B$
& $\sigma_{\rm int}$ (dex)\\
\hline
$\frac{L_{\rm bol,500}}{E(z)\;{\rm erg~s}^{-1}}$ & $\frac{\sigma}{1000\;{\rm km~s}^{-1}}$&BCES bisector
    & Whole          & $    44.94 \pm 0.99  $ & $3.88\pm 0.34$ & $ 0.33\pm  0.03  $\\
 && & Undisturbed    & $    45.03 \pm 1.31  $ & $4.11\pm 0.45$ & $ 0.37\pm  0.05  $\\
 && & Disturbed      & $    44.75 \pm 1.65  $ & $3.95\pm 0.56$ & $ 0.22\pm  0.03  $\\
 && & Cool core      & $    45.26 \pm 1.53  $ & $4.42\pm 0.54$ & $ 0.35\pm  0.06  $\\
 && & Non-cool core  & $    44.77 \pm 1.65  $ & $4.79\pm 0.56$ & $ 0.27\pm  0.03  $\\
&& BCES orthogonal
    & Whole          & $    45.08 \pm 1.11  $ & $4.66\pm 0.39$ & $ 0.37\pm  0.04  $\\
 && & Undisturbed    & $    45.15 \pm 1.47  $ & $4.95\pm 0.51$ & $ 0.40\pm  0.05  $\\
 && & Disturbed      & $    44.77 \pm 1.71  $ & $4.29\pm 0.58$ & $ 0.24\pm  0.03  $\\
 && & Cool core      & $    45.38 \pm 1.66  $ & $5.06\pm 0.60$ & $ 0.38\pm  0.08  $\\
 && & Non-cool core  & $    44.88 \pm 2.19  $ & $5.66\pm 0.74$ & $ 0.30\pm  0.04  $\\
$\frac{L_{\rm bol,500}}{E(z)\;{\rm erg~s}^{-1}}$ & $\frac{M_{\rm gas,500}E(z)}{10^{14} M_{\odot}}$ &BCES bisector
    & Whole          & $    45.10 \pm 0.86  $ & $1.25\pm 0.06$ & $ 0.14\pm  0.01  $\\
 && & Undisturbed    & $    45.28 \pm 0.99  $ & $1.27\pm 0.07$ & $ 0.12\pm  0.01  $\\
 && & Disturbed      & $    45.02 \pm 1.42  $ & $1.33\pm 0.10$ & $ 0.11\pm  0.02  $\\
 && & Cool core      & $    45.20 \pm 0.93  $ & $1.25\pm 0.07$ & $ 0.12\pm  0.02  $\\
 && & Non-cool core  & $    45.16 \pm 0.75  $ & $1.44\pm 0.06$ & $ 0.09\pm  0.01  $\\
&& BCES orthogonal
    & Whole          & $    45.14 \pm 0.90  $ & $1.26\pm 0.07$ & $ 0.14\pm  0.01  $\\
 && & Undisturbed    & $    45.18 \pm 1.01  $ & $1.27\pm 0.08$ & $ 0.12\pm  0.01  $\\
 && & Disturbed      & $    45.06 \pm 1.44  $ & $1.34\pm 0.11$ & $ 0.11\pm  0.02  $\\
 && & Cool core      & $    45.20 \pm 0.94  $ & $1.25\pm 0.07$ & $ 0.12\pm  0.02  $\\
 && & Non-cool core  & $    45.20 \pm 0.76  $ & $1.45\pm 0.06$ & $ 0.09\pm  0.01  $\\
$\frac{L_{\rm 0.5-2keV,500}}{E(z)\;{\rm erg~s}^{-1}}$ & $\frac{\sigma}{1000\;{\rm km~s}^{-1}}$ &BCES bisector
    & Whole          & $    44.39 \pm 0.90  $ & $3.33\pm 0.31$ & $ 0.31\pm  0.03  $\\
 && & Undisturbed    & $    44.54 \pm 1.20  $ & $3.58\pm 0.41$ & $ 0.35\pm  0.04  $\\
 && & Disturbed      & $    44.18 \pm 1.28  $ & $3.26\pm 0.43$ & $ 0.19\pm  0.03  $\\
 && & Cool core      & $    44.70 \pm 1.42  $ & $3.90\pm 0.50$ & $ 0.34\pm  0.06  $\\
 && & Non-cool core  & $    44.21 \pm 1.39  $ & $4.07\pm 0.47$ & $ 0.24\pm  0.03  $\\
&& BCES orthogonal
    & Whole          & $    44.49 \pm 1.13  $ & $4.13\pm 0.39$ & $ 0.36\pm  0.04  $\\
 && & Undisturbed    & $    44.61 \pm 1.51  $ & $4.47\pm 0.53$ & $ 0.40\pm  0.05  $\\
 && & Disturbed      & $    44.22 \pm 1.35  $ & $3.54\pm 0.46$ & $ 0.20\pm  0.03  $\\
 && & Cool core      & $    44.86 \pm 1.63  $ & $4.52\pm 0.59$ & $ 0.36\pm  0.07  $\\
 && & Non-cool core  & $    44.28 \pm 2.03  $ & $4.86\pm 0.69$ & $ 0.27\pm  0.03  $\\
$\frac{L_{\rm 0.5-2keV,500}}{E(z)\;{\rm erg~s}^{-1}}$ & $\frac{M_{\rm gas,500}E(z)}{10^{14} M_{\odot}}$ &BCES bisector
    & Whole          & $    44.52 \pm 0.68  $ & $1.08\pm 0.05$ & $ 0.14\pm  0.01  $\\
 && & Undisturbed    & $    44.64 \pm 0.83  $ & $1.11\pm 0.06$ & $ 0.13\pm  0.01  $\\
 && & Disturbed      & $    44.40 \pm 1.08  $ & $1.10\pm 0.08$ & $ 0.08\pm  0.02  $\\
 && & Cool core      & $    44.60 \pm 0.70  $ & $1.10\pm 0.05$ & $ 0.10\pm  0.02  $\\
 && & Non-cool core  & $    44.62 \pm 0.59  $ & $1.23\pm 0.04$ & $ 0.07\pm  0.01  $\\
&& BCES orthogonal
    & Whole          & $    44.66 \pm 0.71  $ & $1.09\pm 0.05$ & $ 0.14\pm  0.01  $\\
 && & Undisturbed    & $    44.64 \pm 0.86  $ & $1.11\pm 0.06$ & $ 0.13\pm  0.01  $\\
 && & Disturbed      & $    44.40 \pm 1.08  $ & $1.10\pm 0.08$ & $ 0.08\pm  0.02  $\\
 && & Cool core      & $    44.74 \pm 0.71  $ & $1.11\pm 0.05$ & $ 0.10\pm  0.01  $\\
 && & Non-cool core  & $    44.52 \pm 0.59  $ & $1.23\pm 0.04$ & $ 0.07\pm  0.01  $\\
\hline
  \end{tabular}
  \end{center}
  \hspace*{0.3cm}{\footnotesize}} 
\end{table*}

\section{Scaling relations using $L^{\rm ex}$}
\label{a:lex}

\begin{figure*}
\begin{center}
\includegraphics[angle=270,width=16cm]{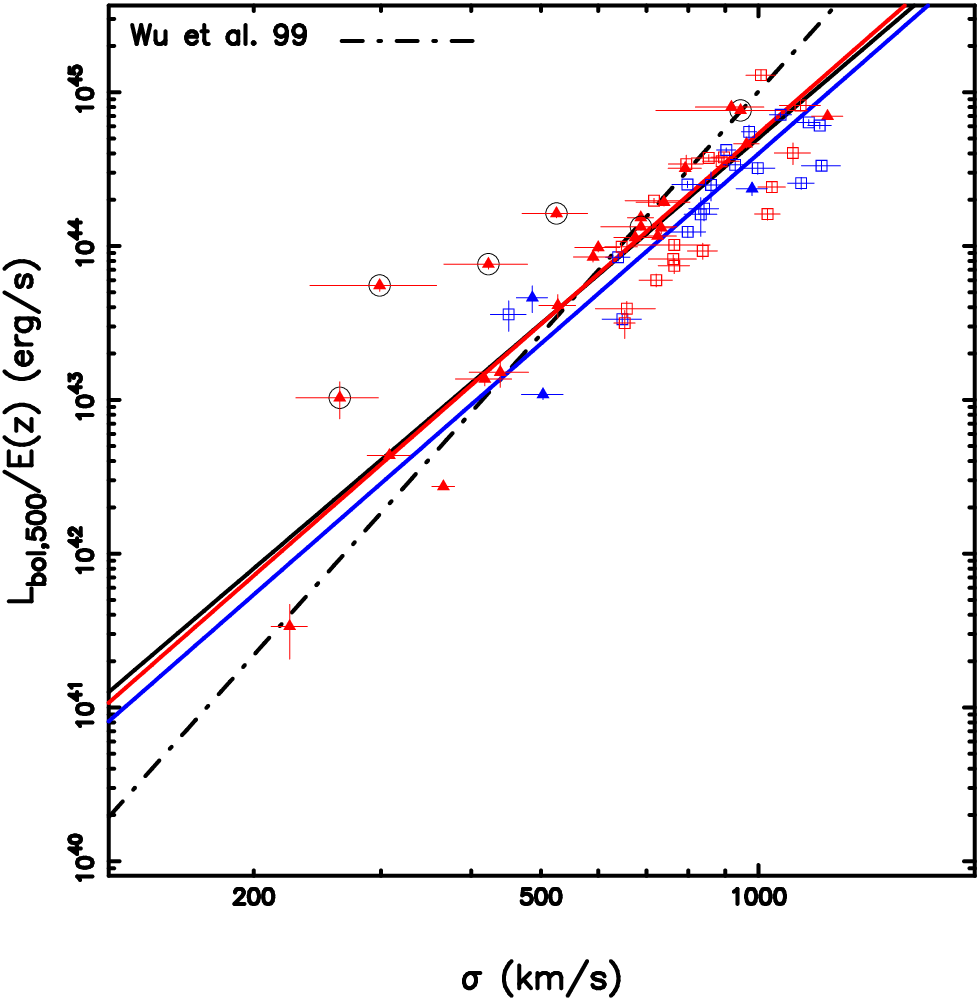}

\includegraphics[angle=270,width=4cm]{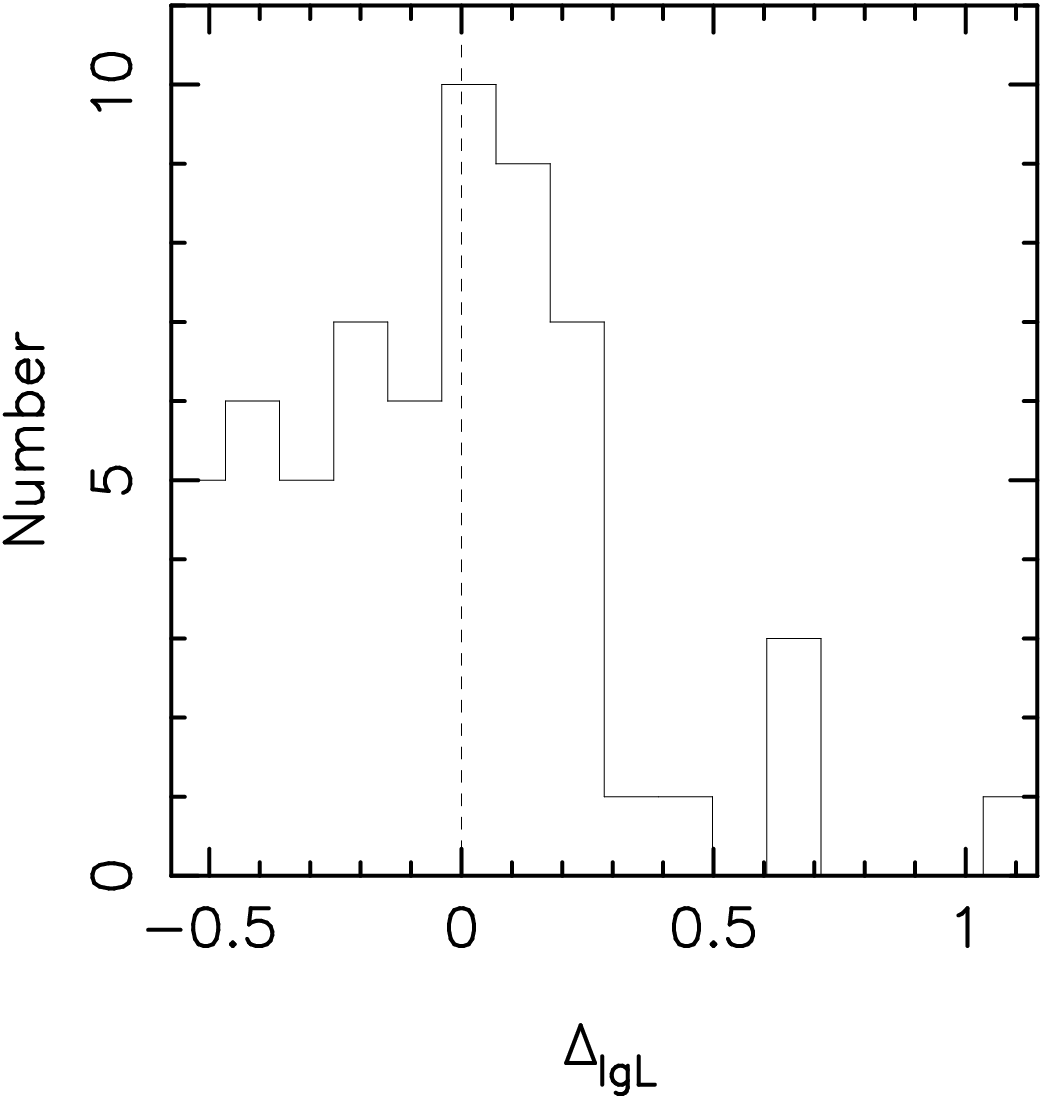}
\includegraphics[angle=270,width=4cm]{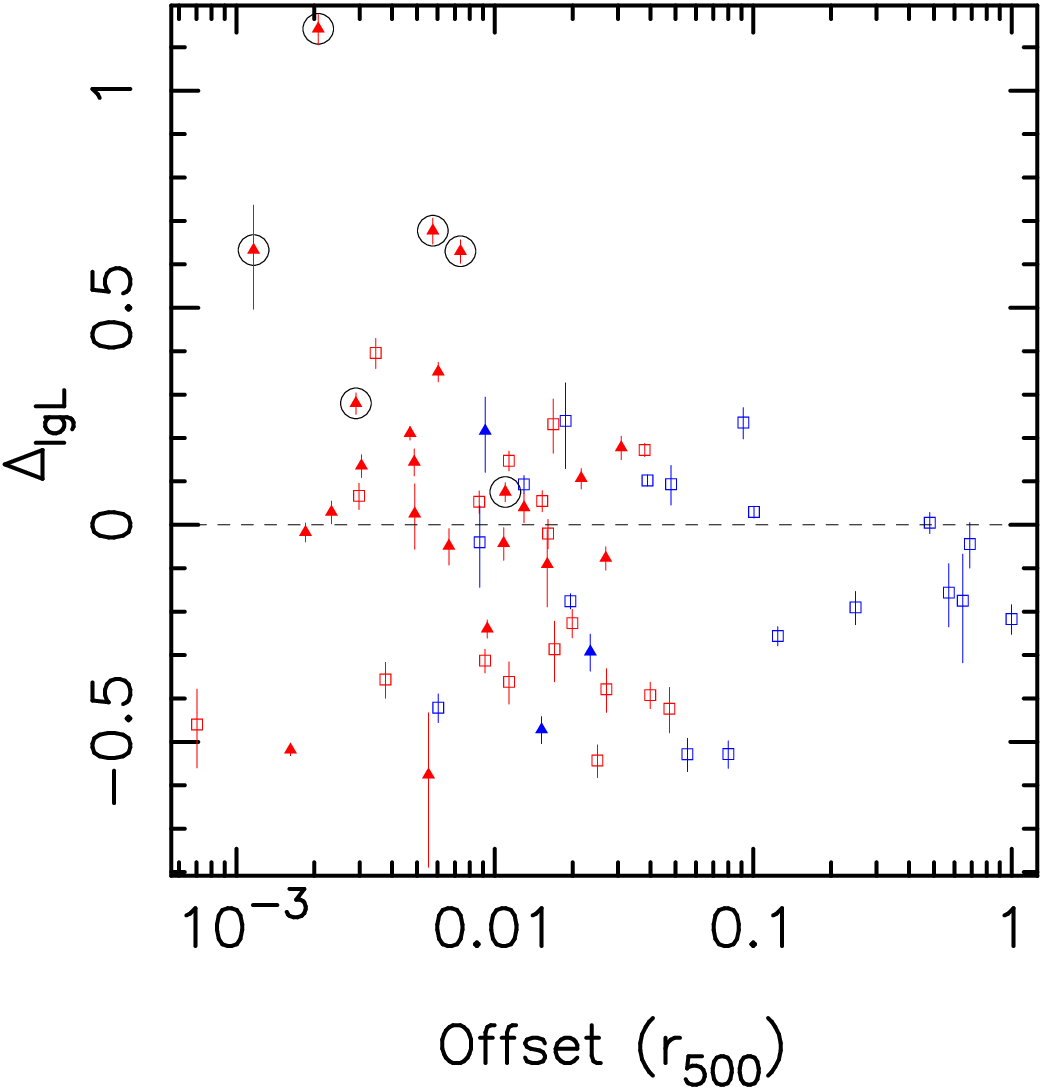}
\includegraphics[angle=270,width=4cm]{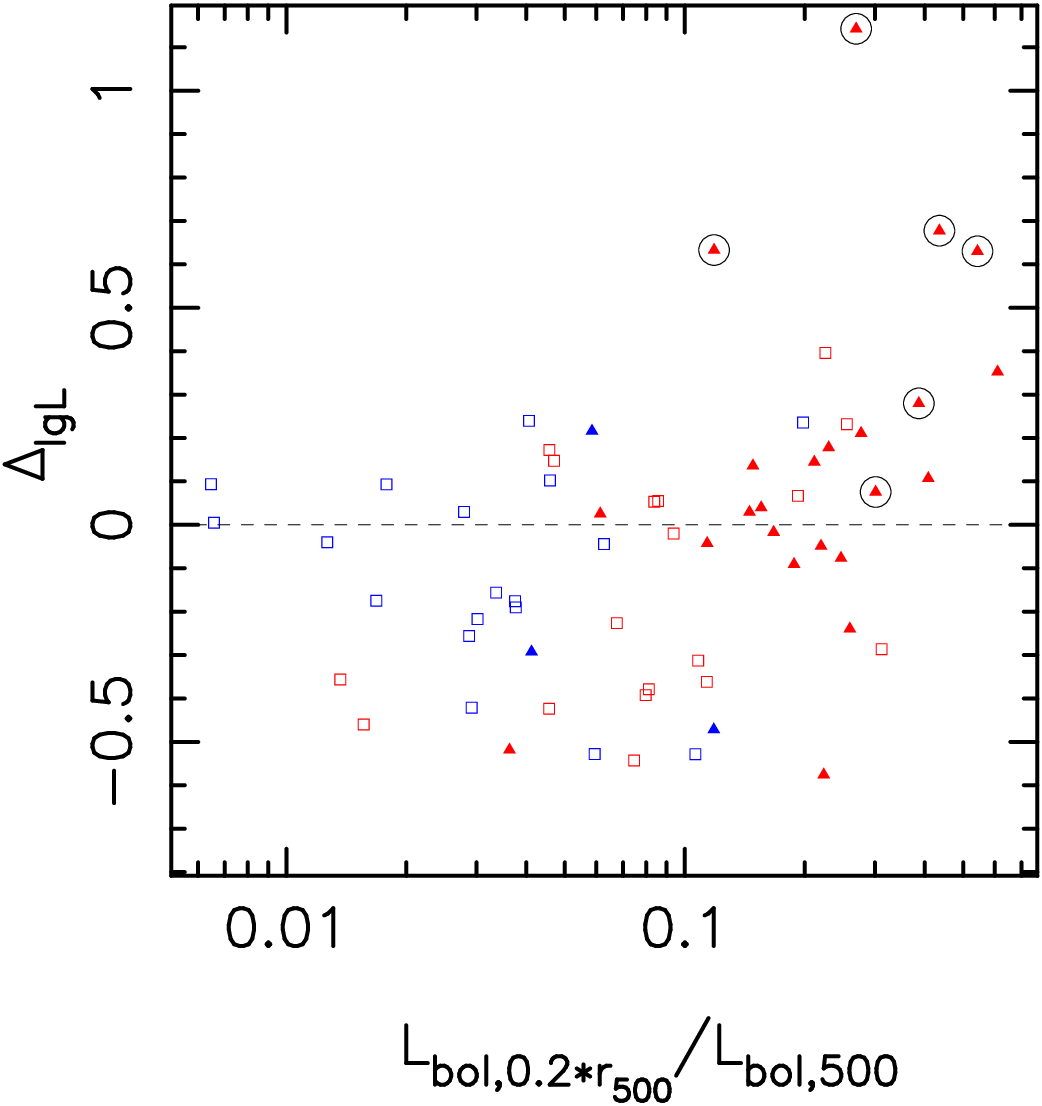}
\includegraphics[angle=270,width=4cm]{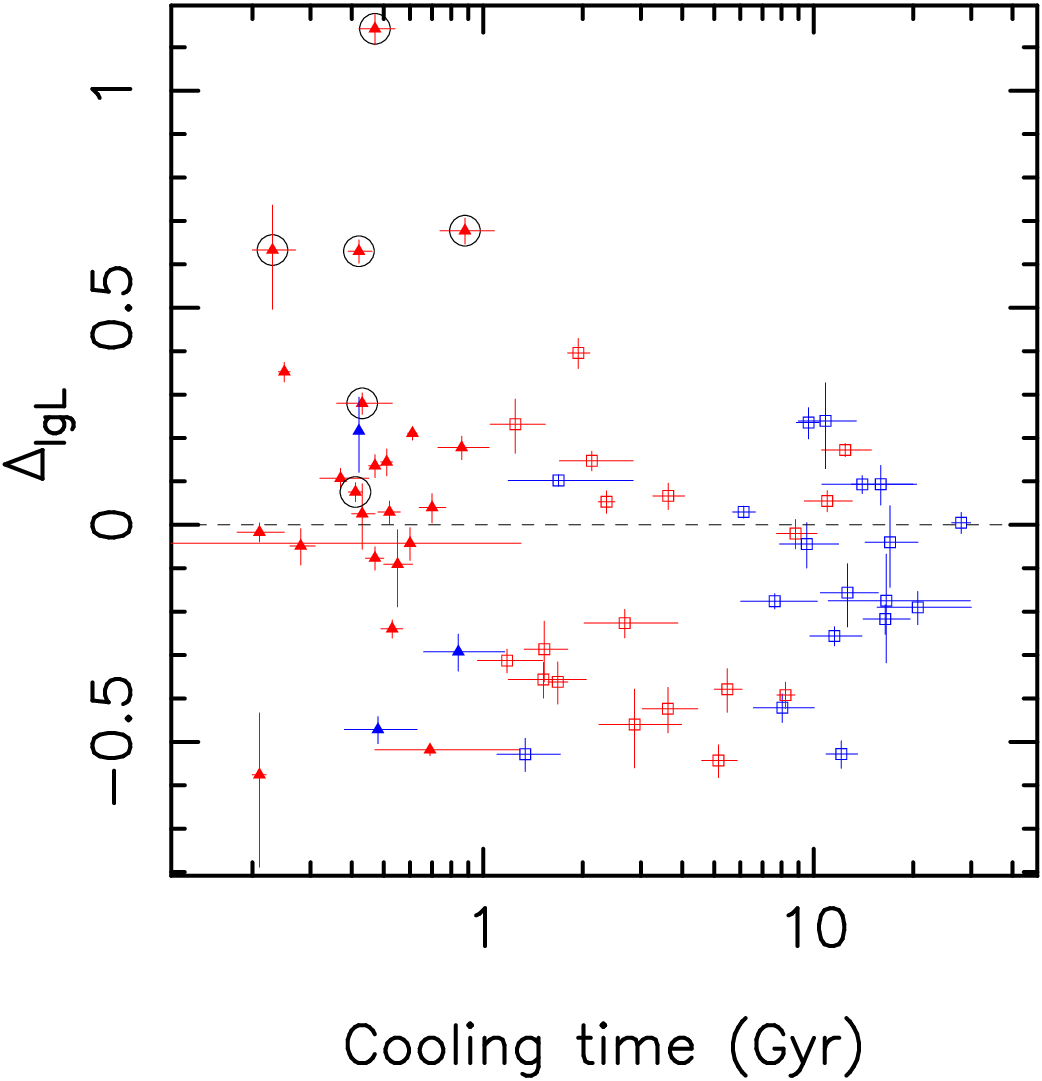}
\end{center}
\caption{{\it Upper panel:} X-ray bolometric luminosity vs. velocity
  dispersion with luminosity derived from emission in the $[0.2-1.0]r_{500}$
  aperture ($L^{\rm ex}_{\rm bol}$). {\it Lower left panel:} Histogram of
  residuals in logarithmic space from the best-fit $L^{\rm ex}_{\rm
    bol}-\sigma$ relation for the 62 clusters using the BCES bisector method.
  {\it Lower 2nd panel:} Residual vs. offset between the X-ray flux-weighted
  center and BCG position. {\it Lower 3rd panel:} Residual vs. fraction of the
  X-ray luminosity within $0.2r_{500}$. {\it Lower right panel:} Residual
  vs. central cooling time. The colors, lines, and symbols have the same
  meaning as those in Fig.~\ref{f:lbv}.
  \label{f:dlbvex}}
\end{figure*}

\begin{figure*}
\begin{center}
\includegraphics[angle=270,width=16cm]{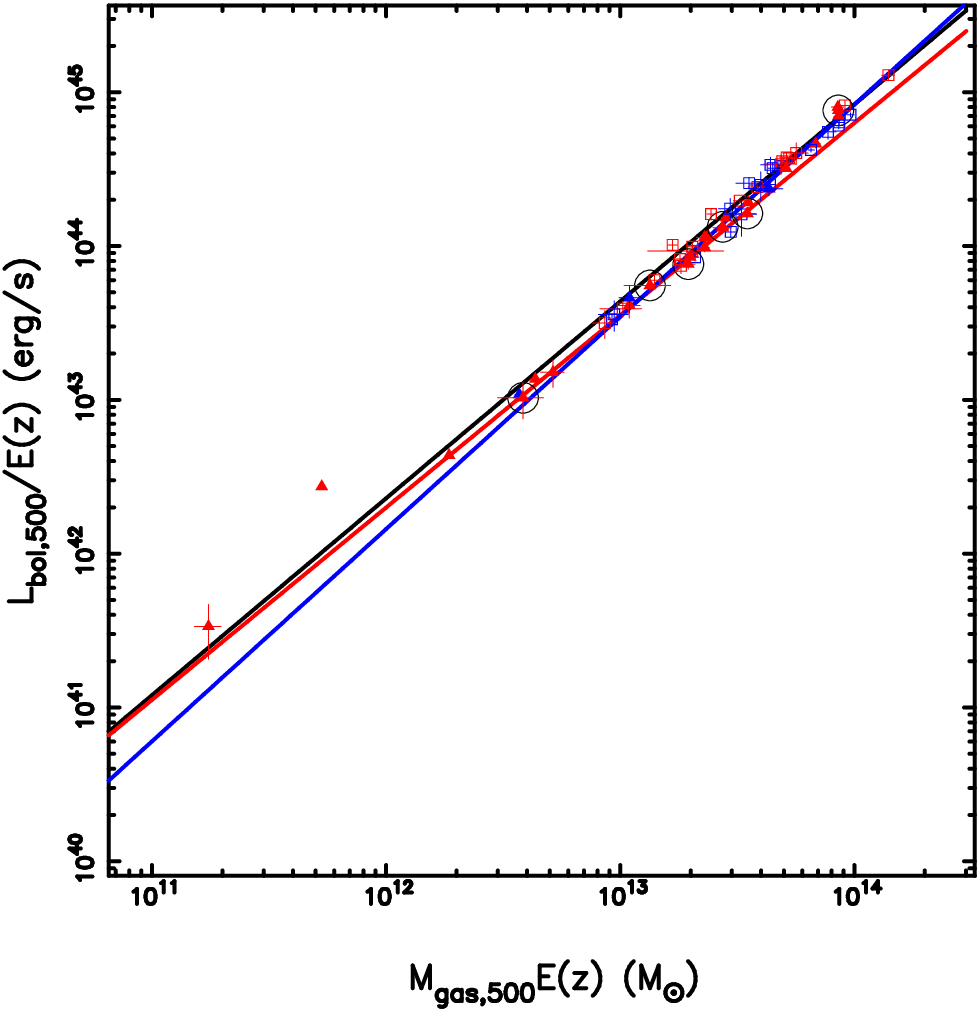}

\includegraphics[angle=270,width=4cm]{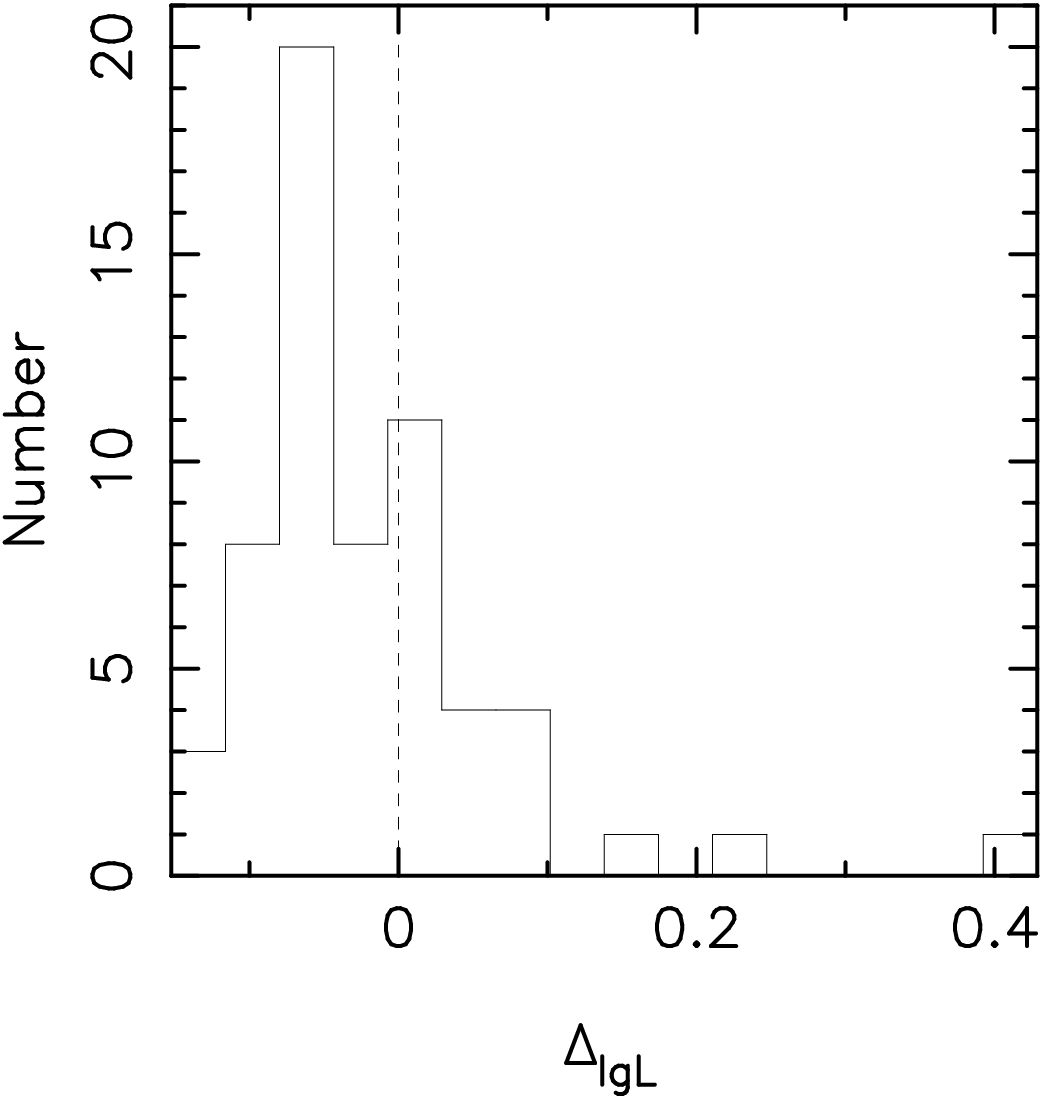}
\includegraphics[angle=270,width=4cm]{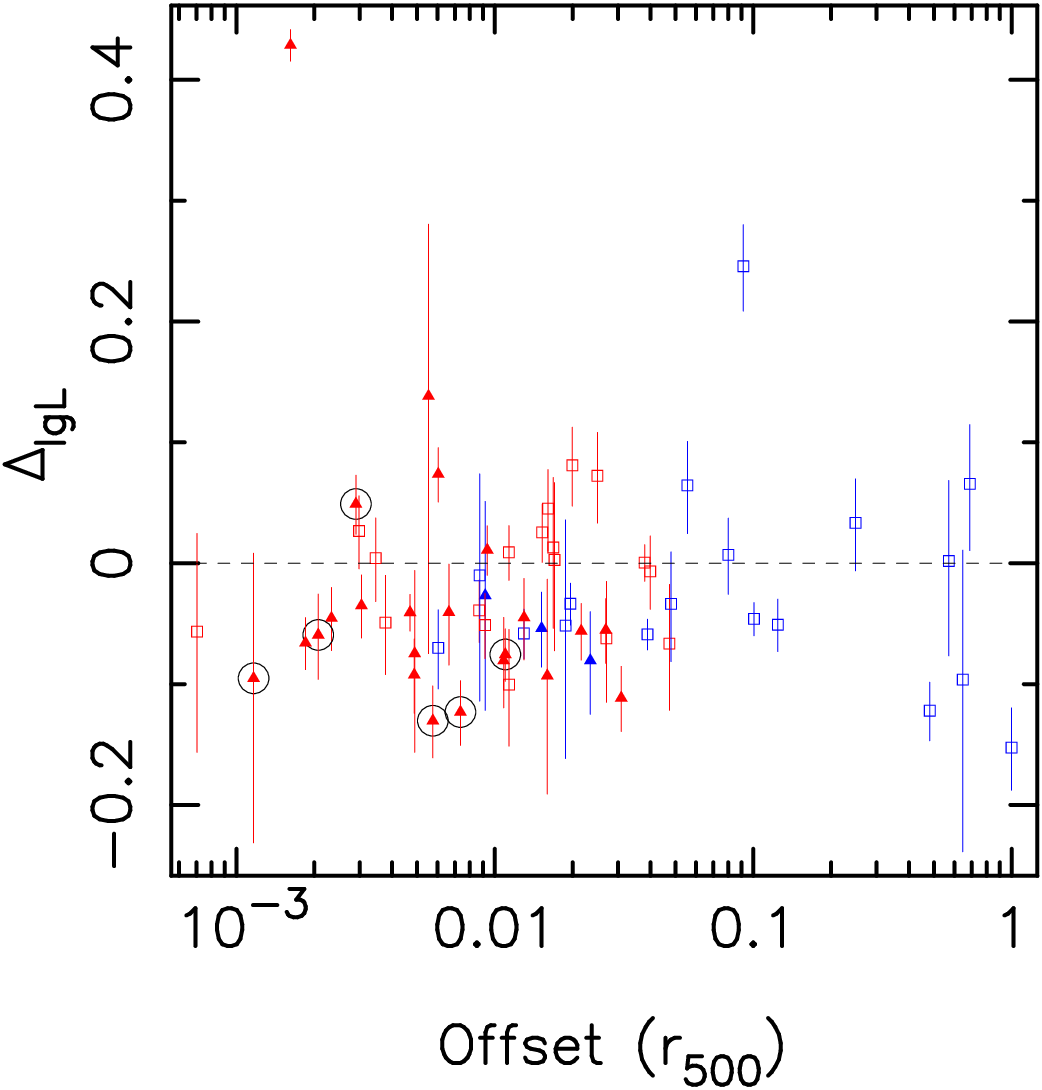}
\includegraphics[angle=270,width=4cm]{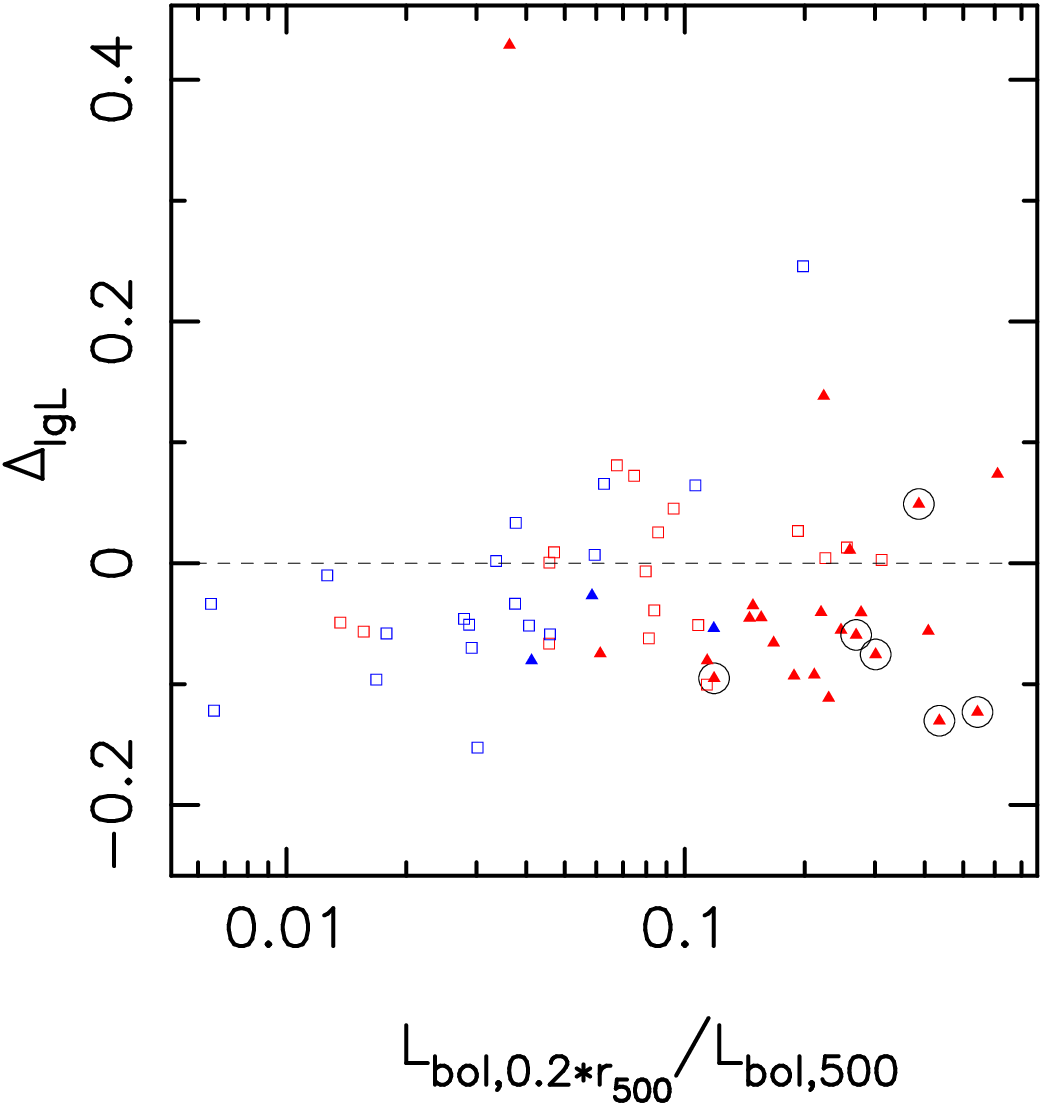}
\includegraphics[angle=270,width=4cm]{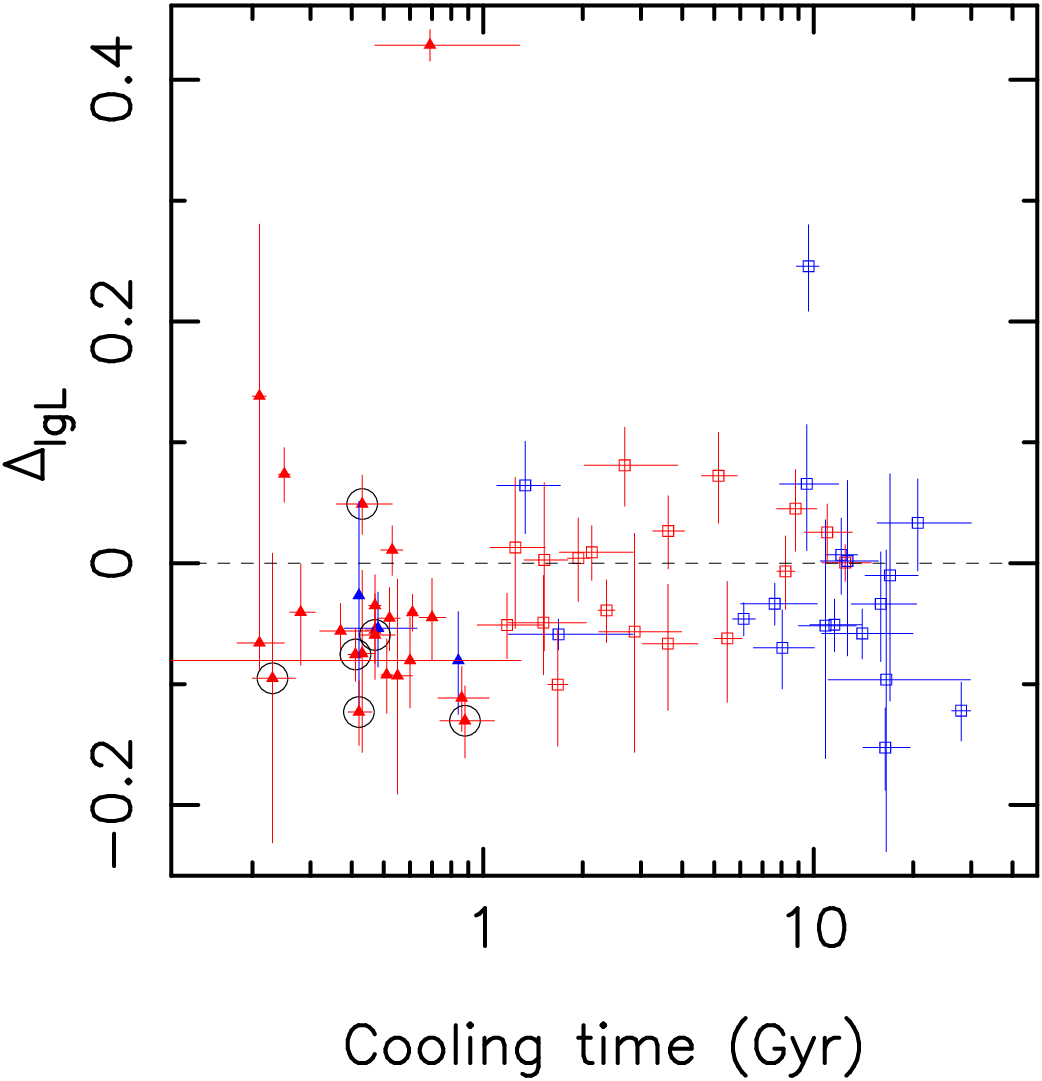}
\end{center}
\caption{{\it Upper panel:} X-ray bolometric luminosity vs. gas mass with
  luminosity derived from emission in the $[0.2-1.0]r_{500}$ aperture ($L^{\rm
    ex}_{\rm bol}$). {\it Lower left panel:} Histogram of residuals in
  logarithmic space from the best-fit $L^{\rm ex}_{\rm bol}-M_{\rm gas}$
  relation for the 62 clusters using the BCES bisector method.  {\it Lower 2nd
    panel:} Residual vs. offset between the X-ray flux-weighted center and BCG
  position. {\it Lower 3rd panel:} Residual vs. fraction of the X-ray
  luminosity within $0.2r_{500}$. {\it Lower right panel:} Residual
  vs. central cooling time. The colors, lines, and symbols have the same
  meaning as those in Fig.~\ref{f:lbv}.
  \label{f:dlbmgex}}
\end{figure*}

Since the luminosity derived in the [0.2--1]~$r_{500}$ radial range is widely
used to reduce the scatter caused by the presence of cool cores, we also
present the X-ray bolometric luminosity in the [0.2--1]~$r_{500}$ radial range
($L^{\rm ex}_{\rm bol}$) in Table~\ref{t:lx_in_ex}. We also list the best fits
to the corresponding scaling relations using the bolometric and 0.5--2~keV
band luminosity derived in the [0.2--1]~$r_{500}$ radial range in
Table~\ref{t:scaling_lex}, and show the plots using $L^{\rm ex}_{\rm bol}$ in
Figs.~\ref{f:dlbvex}--\ref{f:dlbmgex}.

\begin{table*} { \begin{center} \footnotesize
      {\renewcommand{\arraystretch}{1.3} \caption[]{Power-law fit,
          $\log_{10}(Y)=A + B \log_{10}(X)$, to the scaling
          relations for the observational sample using $L^{\rm ex}$.}
        \label{t:scaling_lex}}
\begin{tabular}{llllccc}
\hline
$Y$                       & $X$       &   Method & Sample     & $A$               & $B$
& $\sigma_{\rm int}$ (dex)\\
\hline
$\frac{L_{\rm bol,500}}{E(z)\;{\rm erg~s}^{-1}}$ & $\frac{\sigma}{1000\;{\rm km~s}^{-1}}$&BCES bisector
    & Whole          & $    44.70 \pm 0.97  $ & $4.00\pm 0.33$ & $ 0.25\pm  0.03  $\\
 && & Undisturbed    & $    44.73 \pm 1.26  $ & $4.11\pm 0.43$ & $ 0.28\pm  0.04  $\\
 && & Disturbed      & $    44.60 \pm 1.67  $ & $4.10\pm 0.57$ & $ 0.21\pm  0.03  $\\
 && & Cool core      & $    44.85 \pm 1.47  $ & $4.35\pm 0.52$ & $ 0.26\pm  0.06  $\\
 && & Non-cool core  & $    44.65 \pm 1.55  $ & $4.65\pm 0.53$ & $ 0.24\pm  0.02  $\\
    && BCES orthogonal 
    & Whole          & $    44.70 \pm 0.87  $ & $4.40\pm 0.30$ & $ 0.26\pm  0.03  $\\
 && & Undisturbed    & $    44.75 \pm 1.15  $ & $4.55\pm 0.40$ & $ 0.28\pm  0.04  $\\
 && & Disturbed      & $    44.57 \pm 1.70  $ & $4.39\pm 0.57$ & $ 0.22\pm  0.03  $\\
 && & Cool core      & $    45.02 \pm 1.43  $ & $4.74\pm 0.51$ & $ 0.27\pm  0.06  $\\
 && & Non-cool core  & $    44.66 \pm 1.75  $ & $5.32\pm 0.59$ & $ 0.26\pm  0.03  $\\
    $\frac{L_{\rm bol,500}}{E(z)\;{\rm erg~s}^{-1}}$ & $\frac{M_{\rm gas,500}E(z)}{10^{14} M_{\odot}}$ &BCES bisector
    & Whole          & $    44.92 \pm 0.67  $ & $1.28\pm 0.05$ & $ 0.09\pm  0.01  $\\
 && & Undisturbed    & $    44.80 \pm 0.75  $ & $1.25\pm 0.06$ & $ 0.09\pm  0.01  $\\
 && & Disturbed      & $    44.92 \pm 0.81  $ & $1.38\pm 0.06$ & $ 0.05\pm  0.01  $\\
 && & Cool core      & $    44.78 \pm 0.78  $ & $1.22\pm 0.06$ & $ 0.10\pm  0.02  $\\
 && & Non-cool core  & $    44.96 \pm 0.58  $ & $1.39\pm 0.04$ & $ 0.05\pm  0.01  $\\
    && BCES orthogonal
    & Whole          & $    44.92 \pm 0.66  $ & $1.28\pm 0.05$ & $ 0.09\pm  0.01  $\\
 && & Undisturbed    & $    44.80 \pm 0.74  $ & $1.25\pm 0.05$ & $ 0.09\pm  0.01  $\\
 && & Disturbed      & $    44.92 \pm 0.82  $ & $1.38\pm 0.06$ & $ 0.05\pm  0.01  $\\
 && & Cool core      & $    44.78 \pm 0.77  $ & $1.22\pm 0.06$ & $ 0.10\pm  0.02  $\\
 && & Non-cool core  & $    44.96 \pm 0.59  $ & $1.39\pm 0.04$ & $ 0.05\pm  0.01  $\\
$\frac{L_{\rm 0.5-2keV,500}}{E(z)\;{\rm erg~s}^{-1}}$ & $\frac{\sigma}{1000\;{\rm km~s}^{-1}}$ &BCES bisector
    & Whole          & $    44.12 \pm 0.89  $ & $3.44\pm 0.30$ & $ 0.23\pm  0.02  $\\
 && & Undisturbed    & $    44.18 \pm 1.16  $ & $3.56\pm 0.40$ & $ 0.26\pm  0.04  $\\
 && & Disturbed      & $    44.06 \pm 1.32  $ & $3.42\pm 0.45$ & $ 0.18\pm  0.03  $\\
 && & Cool core      & $    44.39 \pm 1.36  $ & $3.83\pm 0.48$ & $ 0.25\pm  0.05  $\\
 && & Non-cool core  & $    44.06 \pm 1.27  $ & $3.92\pm 0.43$ & $ 0.21\pm  0.02  $\\
&& BCES orthogonal
    & Whole          & $    44.10 \pm 0.82  $ & $3.80\pm 0.28$ & $ 0.24\pm  0.03  $\\
 && & Undisturbed    & $    44.27 \pm 1.10  $ & $3.99\pm 0.38$ & $ 0.27\pm  0.04  $\\
 && & Disturbed      & $    44.08 \pm 1.36  $ & $3.66\pm 0.46$ & $ 0.19\pm  0.03  $\\
 && & Cool core      & $    44.44 \pm 1.37  $ & $4.18\pm 0.49$ & $ 0.26\pm  0.05  $\\
 && & Non-cool core  & $    44.10 \pm 1.53  $ & $4.50\pm 0.52$ & $ 0.23\pm  0.03  $\\
$\frac{L_{\rm 0.5-2keV,500}}{E(z)\;{\rm erg~s}^{-1}}$ & $\frac{M_{\rm gas,500}E(z)}{10^{14} M_{\odot}}$ &BCES bisector
    & Whole          & $    44.30 \pm 0.44  $ & $1.10\pm 0.03$ & $ 0.06\pm  0.01  $\\
 && & Undisturbed    & $    44.36 \pm 0.54  $ & $1.09\pm 0.04$ & $ 0.06\pm  0.01  $\\
 && & Disturbed      & $    44.30 \pm 0.42  $ & $1.15\pm 0.03$ & $ 0.01\pm  0.01  $\\
 && & Cool core      & $    44.32 \pm 0.56  $ & $1.08\pm 0.04$ & $ 0.07\pm  0.01  $\\
 && & Non-cool core  & $    44.28 \pm 0.26  $ & $1.17\pm 0.02$ & $ 0.01\pm  0.01  $\\
&& BCES orthogonal
    & Whole          & $    44.30 \pm 0.43  $ & $1.10\pm 0.03$ & $ 0.06\pm  0.01  $\\
 && & Undisturbed    & $    44.22 \pm 0.53  $ & $1.08\pm 0.04$ & $ 0.06\pm  0.01  $\\
 && & Disturbed      & $    44.30 \pm 0.41  $ & $1.15\pm 0.03$ & $ 0.01\pm  0.01  $\\
 && & Cool core      & $    44.18 \pm 0.55  $ & $1.07\pm 0.04$ & $ 0.07\pm  0.01  $\\
 && & Non-cool core  & $    44.28 \pm 0.26  $ & $1.17\pm 0.02$ & $ 0.01\pm  0.01  $\\  
\hline
  \end{tabular}
  \end{center}
  \hspace*{0.3cm}{\footnotesize}} 
\end{table*}

\section{Scaling relations using $L^{\rm co}_{\rm 0.5-2keV}$}
\label{a:lx}

\begin{figure*}
\begin{center}
\includegraphics[angle=270,width=16cm]{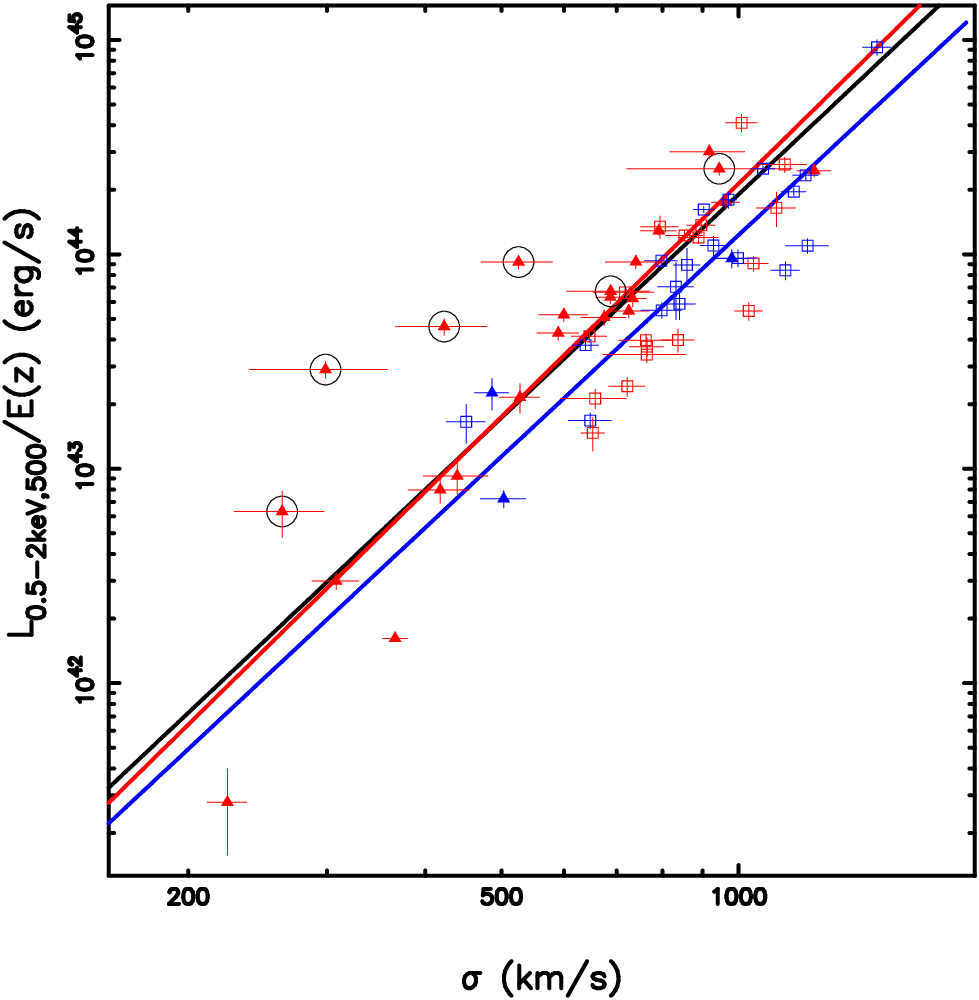}

\includegraphics[angle=270,width=4cm]{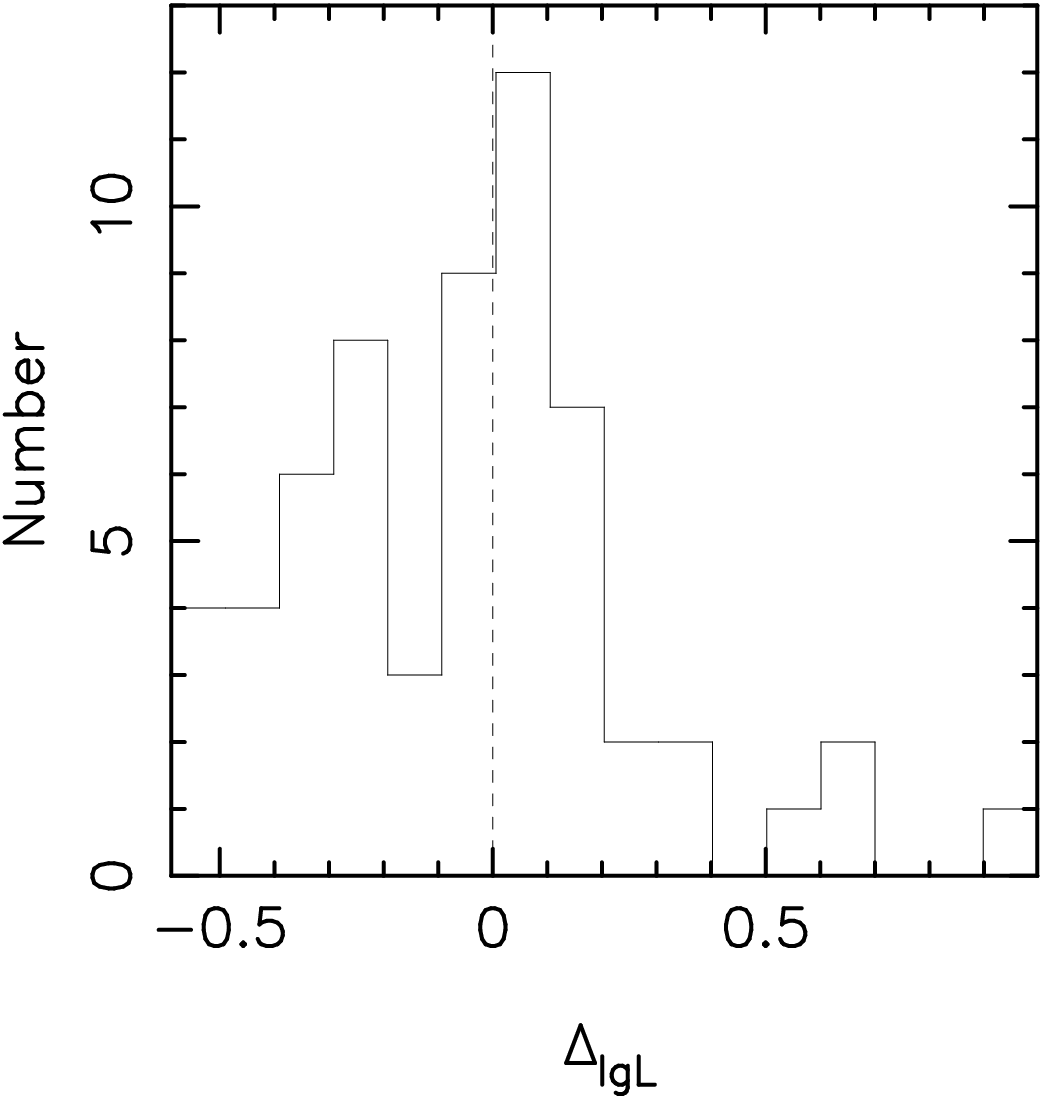}
\includegraphics[angle=270,width=4cm]{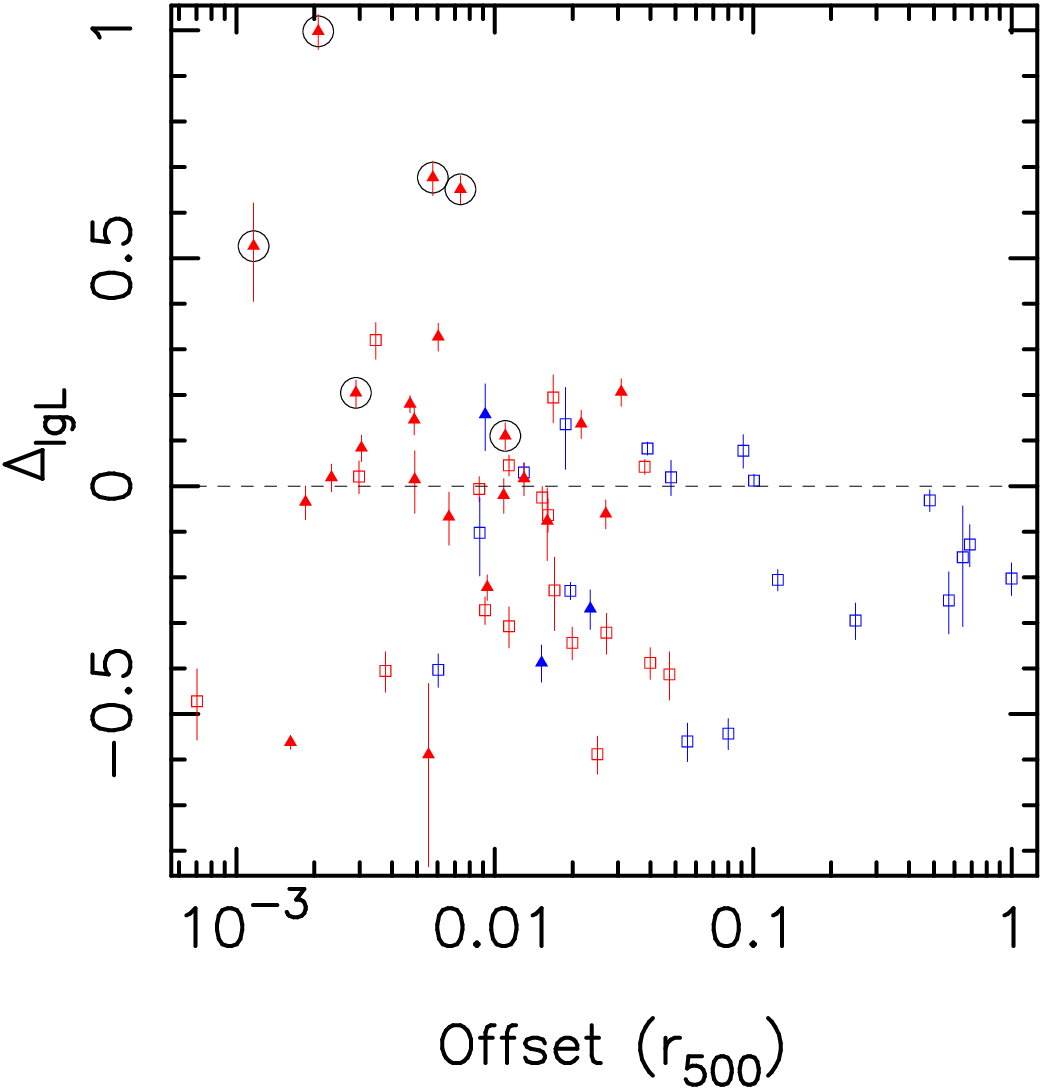}
\includegraphics[angle=270,width=4cm]{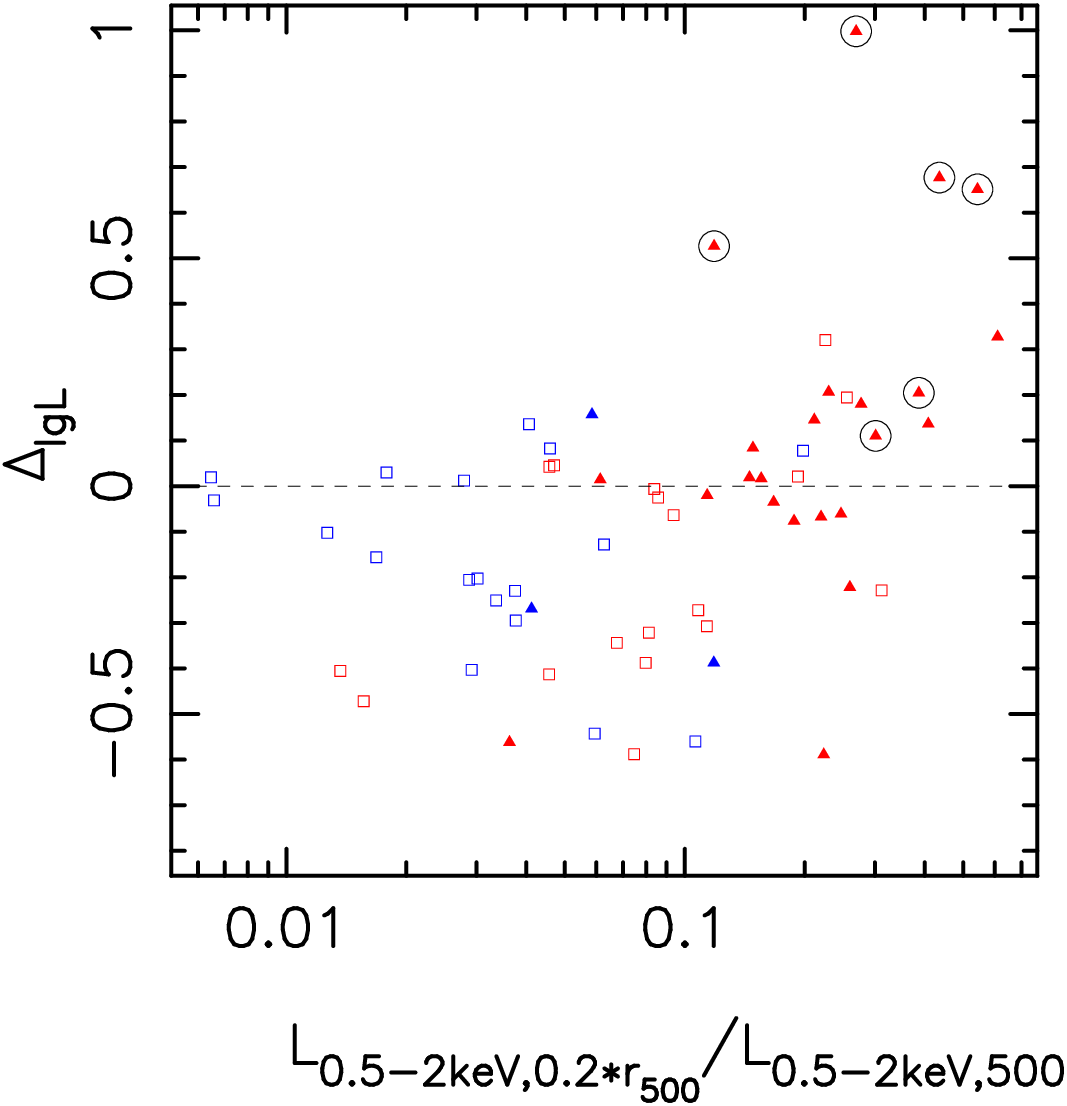}
\includegraphics[angle=270,width=4cm]{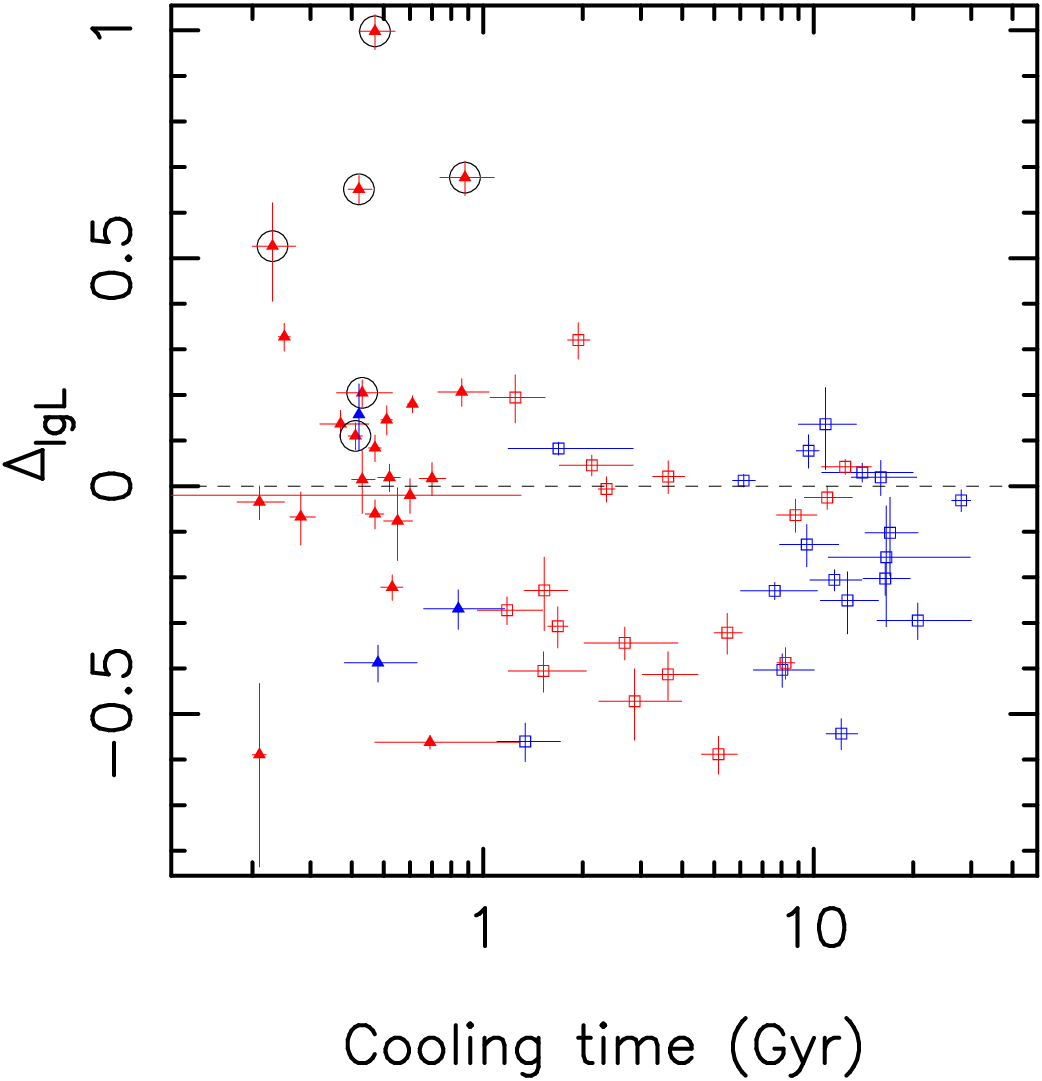}
\end{center}
\caption{{\it Upper panel:} X-ray luminosity in the 0.5--2~keV band
    vs. velocity dispersion with luminosity corrected for the cluster core
    ($L^{\rm co}_{\rm 0.5-2keV}$). {\it Lower left panel:} Histogram of
    residuals in logarithmic space from the best-fit $L^{\rm co}_{\rm
      0.5-2keV}-\sigma$ relation for the 62 clusters using the BCES bisector
    method.  {\it Lower 2nd panel:} Residual vs. offset between the X-ray
    flux-weighted center and BCG position. {\it Lower 3rd panel:} Residual
    vs. fraction of the X-ray luminosity within $0.2r_{500}$. {\it Lower right
      panel:} Residual vs. central cooling time. The colors, lines, and
    symbols have the same meaning as those in Fig.~\ref{f:lbv}.
  \label{f:dlxv}}
\end{figure*}

\begin{figure*}
\begin{center}
\includegraphics[angle=270,width=16cm]{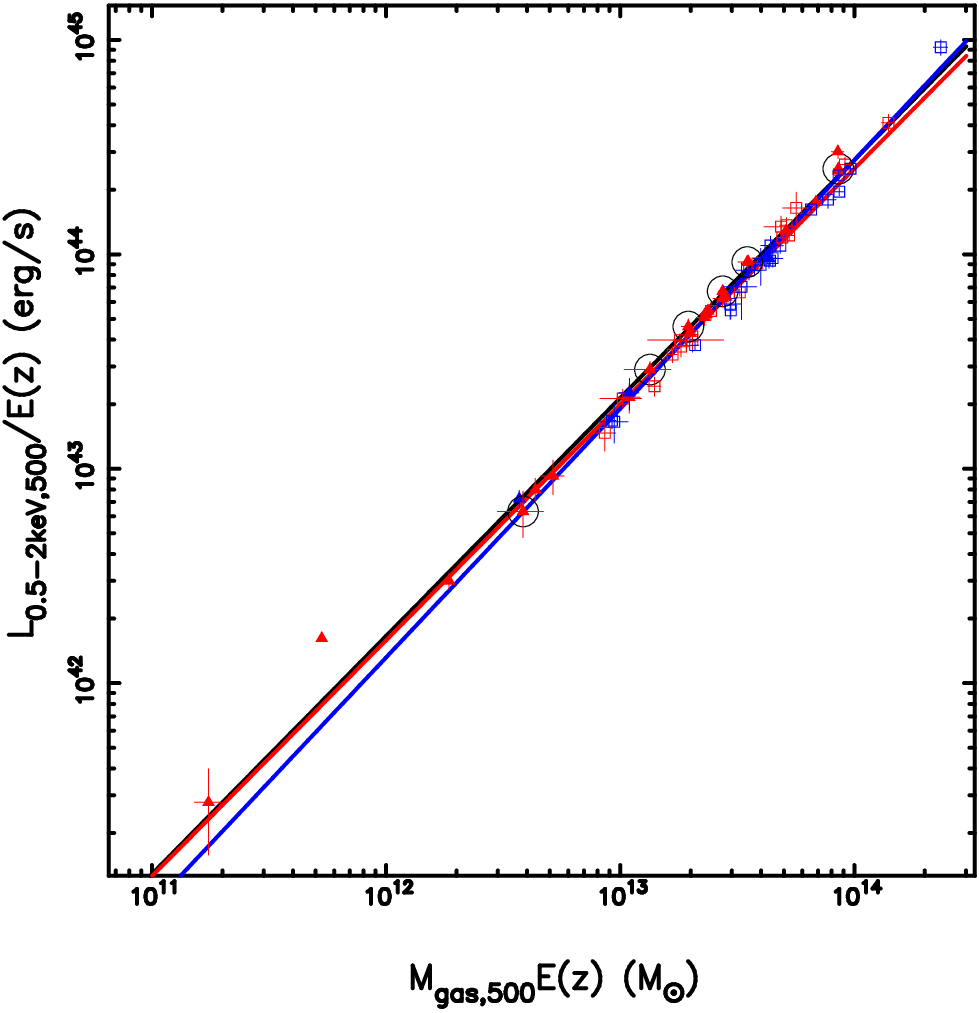}

\includegraphics[angle=270,width=4cm]{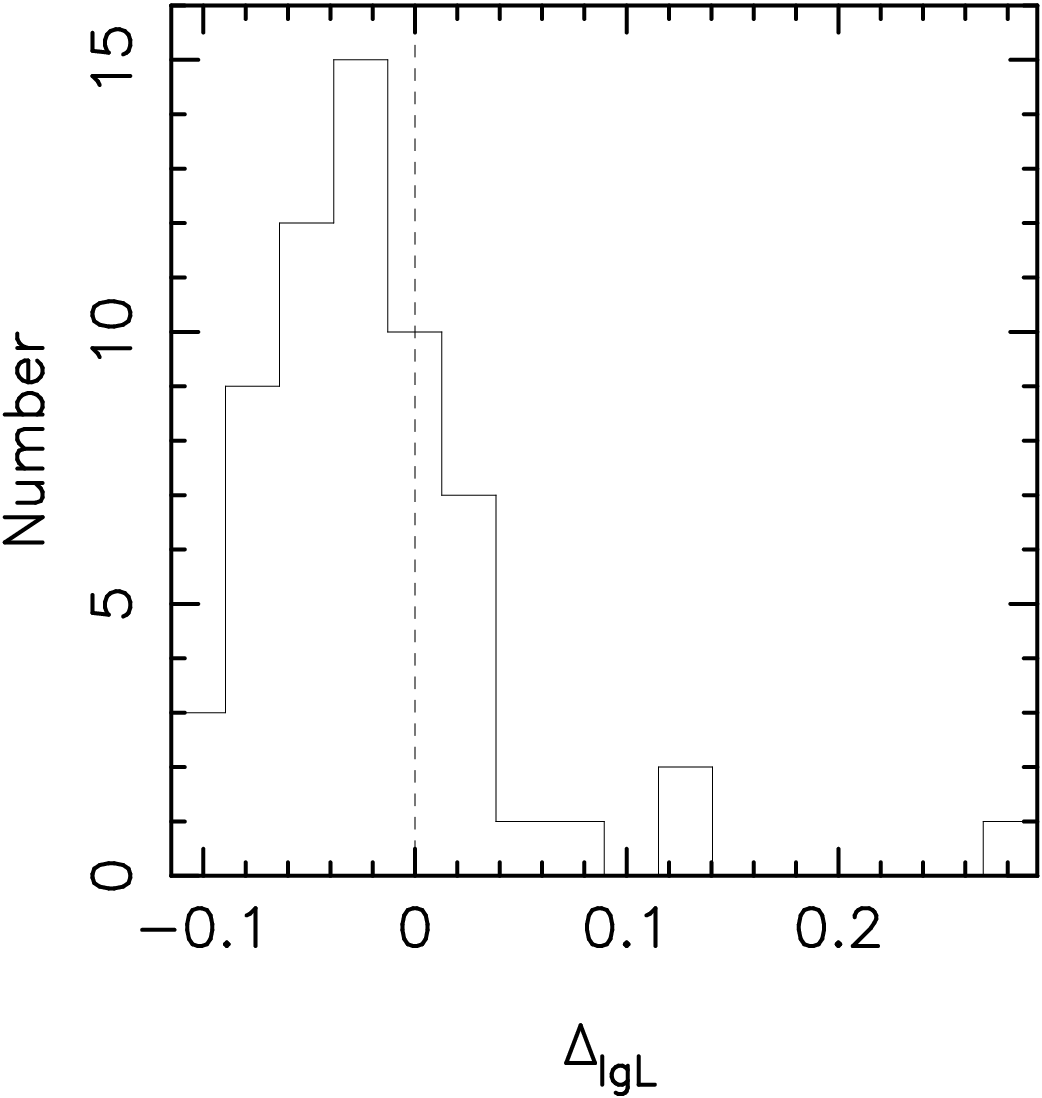}
\includegraphics[angle=270,width=4cm]{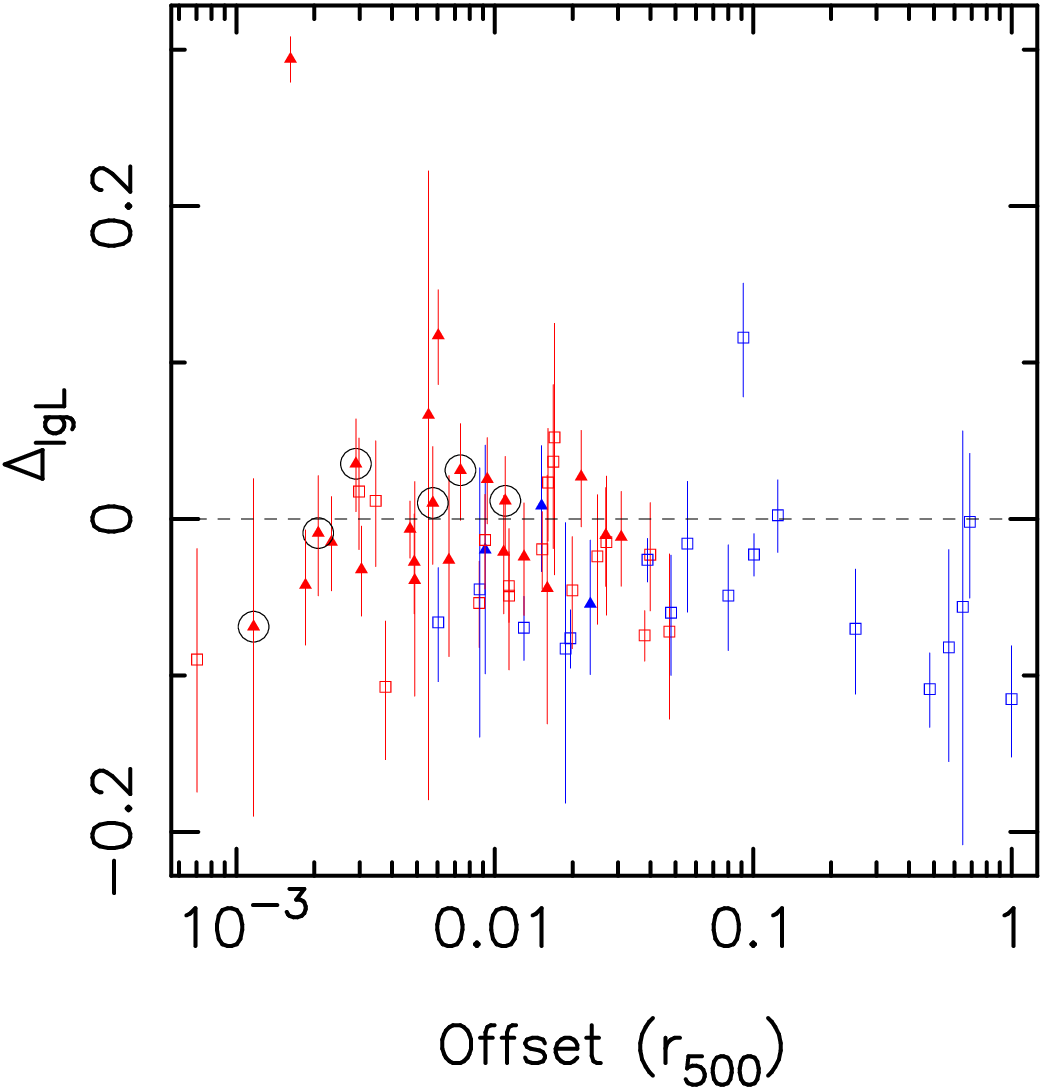}
\includegraphics[angle=270,width=4cm]{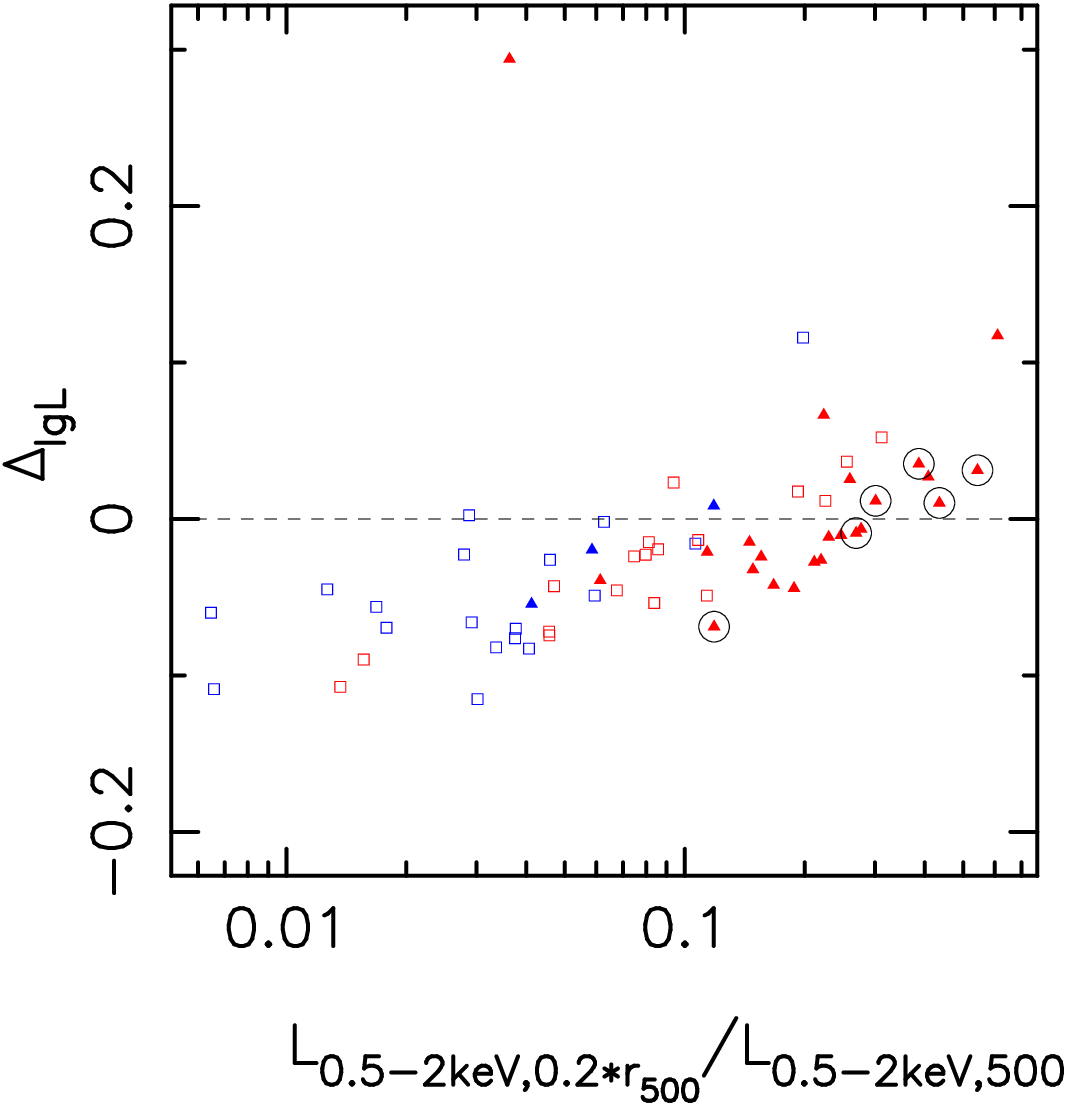}
\includegraphics[angle=270,width=4cm]{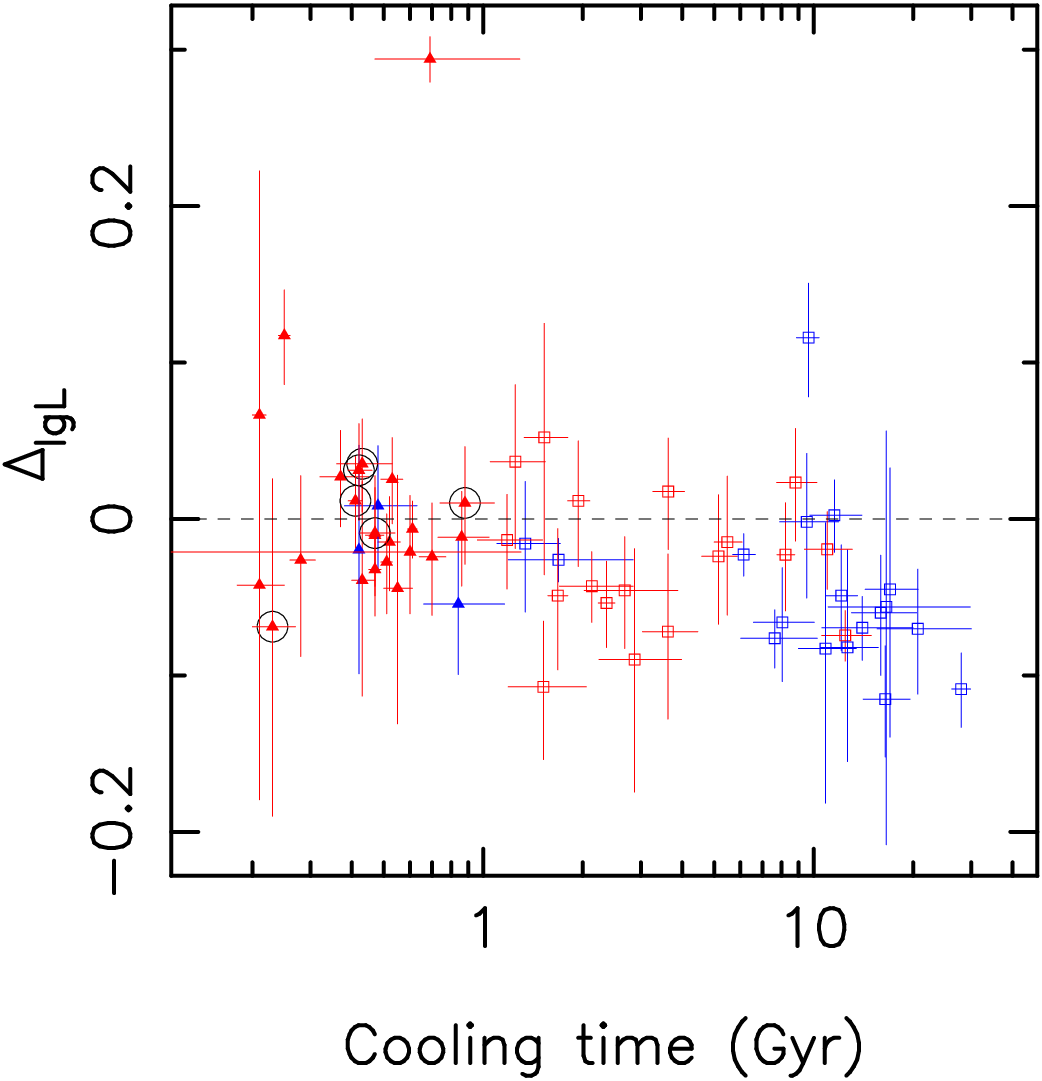}
\end{center}
\caption{{\it Upper panel:} X-ray luminosity in the 0.5--2~keV band
    vs. gas mass with luminosity corrected for the cluster core ($L^{\rm
      co}_{\rm 0.5-2keV}$). {\it Lower left panel:} Histogram of residuals in
    logarithmic space from the best-fit $L^{\rm co}_{\rm 0.5-2keV}-M_{\rm
      gas}$ relation for the 62 clusters using the BCES bisector method.  {\it
      Lower 2nd panel:} Residual vs. offset between the X-ray flux-weighted
    center and BCG position. {\it Lower 3rd panel:} Residual vs. fraction of
    the X-ray luminosity within $0.2r_{500}$. {\it Lower right panel:}
    Residual vs. central cooling time. The colors, lines, and symbols have the
    same meaning as those in Fig.~\ref{f:lbv}.
  \label{f:dlxmg}}
\end{figure*}

We present the corresponding scaling relations using the 0.5--2~keV band
luminosity corrected for the cluster central regions, $L^{\rm co}_{\rm
  0.5-2keV}$, in Figs.~\ref{f:dlxv}--\ref{f:dlxmg}. The best fits are listed
in Table~\ref{t:scaling}.

\section{\emph{XMM-Newton} images of the sample}
\label{a:image}

\begin{figure*}
\begin{center}
\includegraphics[angle=0,width=18cm]{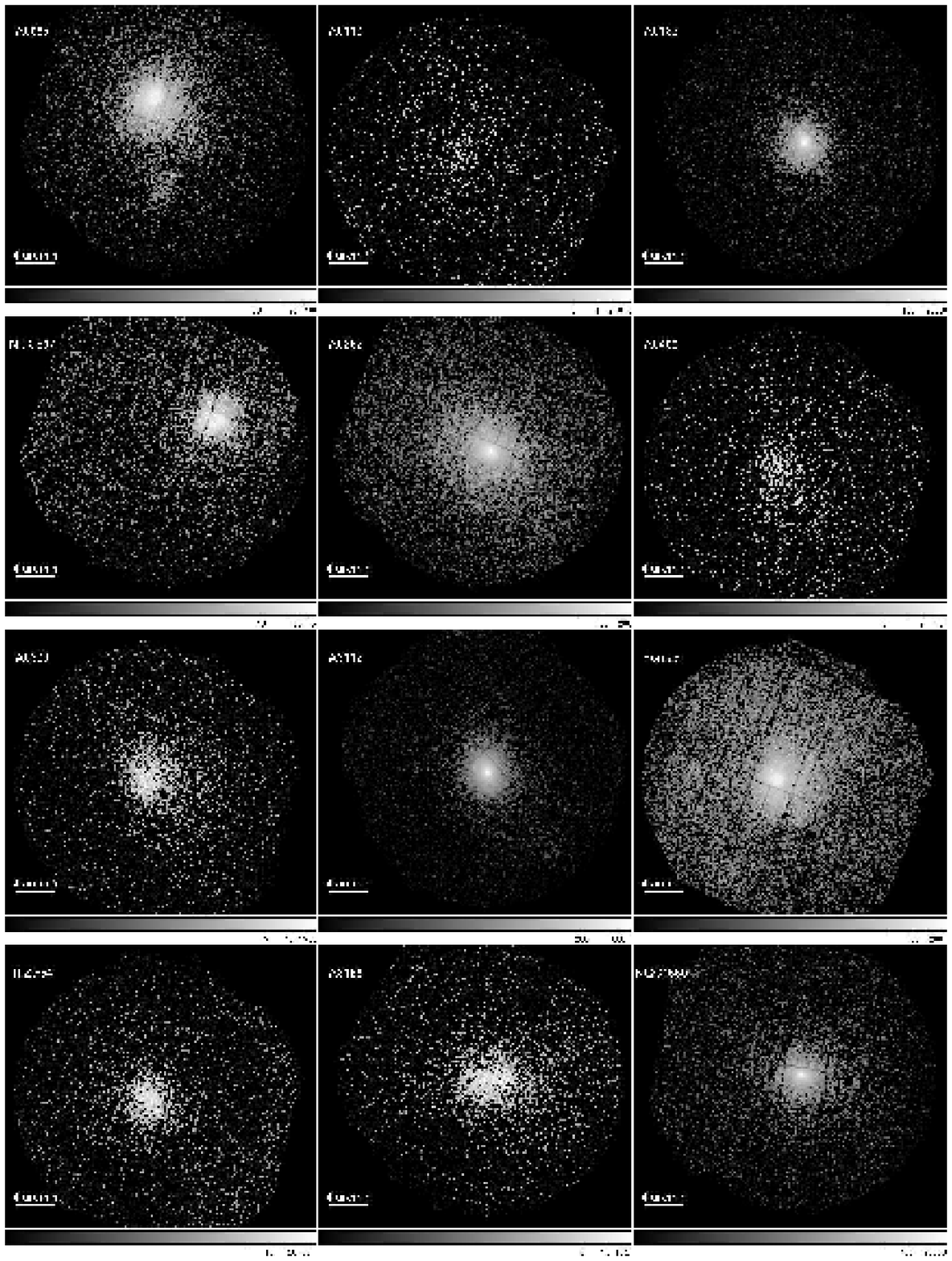}
\end{center}
\caption{Combined MOS and pn images of the clusters in the 0.7--2~keV band,
where point sources have been excised and refilled with values from neighboring
pixels.
\label{f:clid1}}
\end{figure*}

\begin{figure*}
\begin{center}
\includegraphics[angle=0,width=18cm]{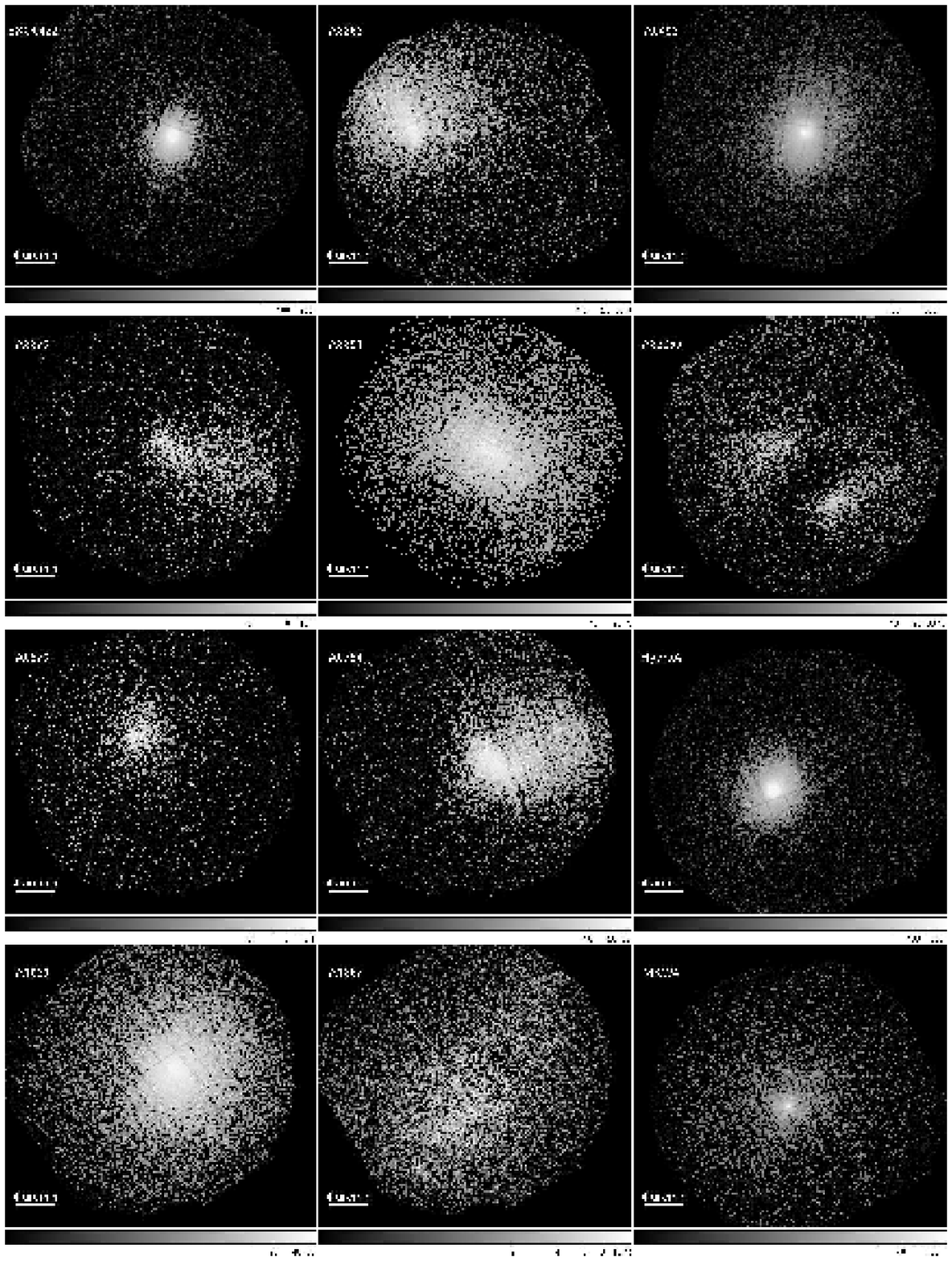}
\end{center}
\caption{Combined MOS and pn images of the clusters in the 0.7--2~keV band,
where point sources have been excised and refilled with values from neighboring
pixels.
\label{f:clid2}}
\end{figure*}

\begin{figure*}
\begin{center}
\includegraphics[angle=0,width=18cm]{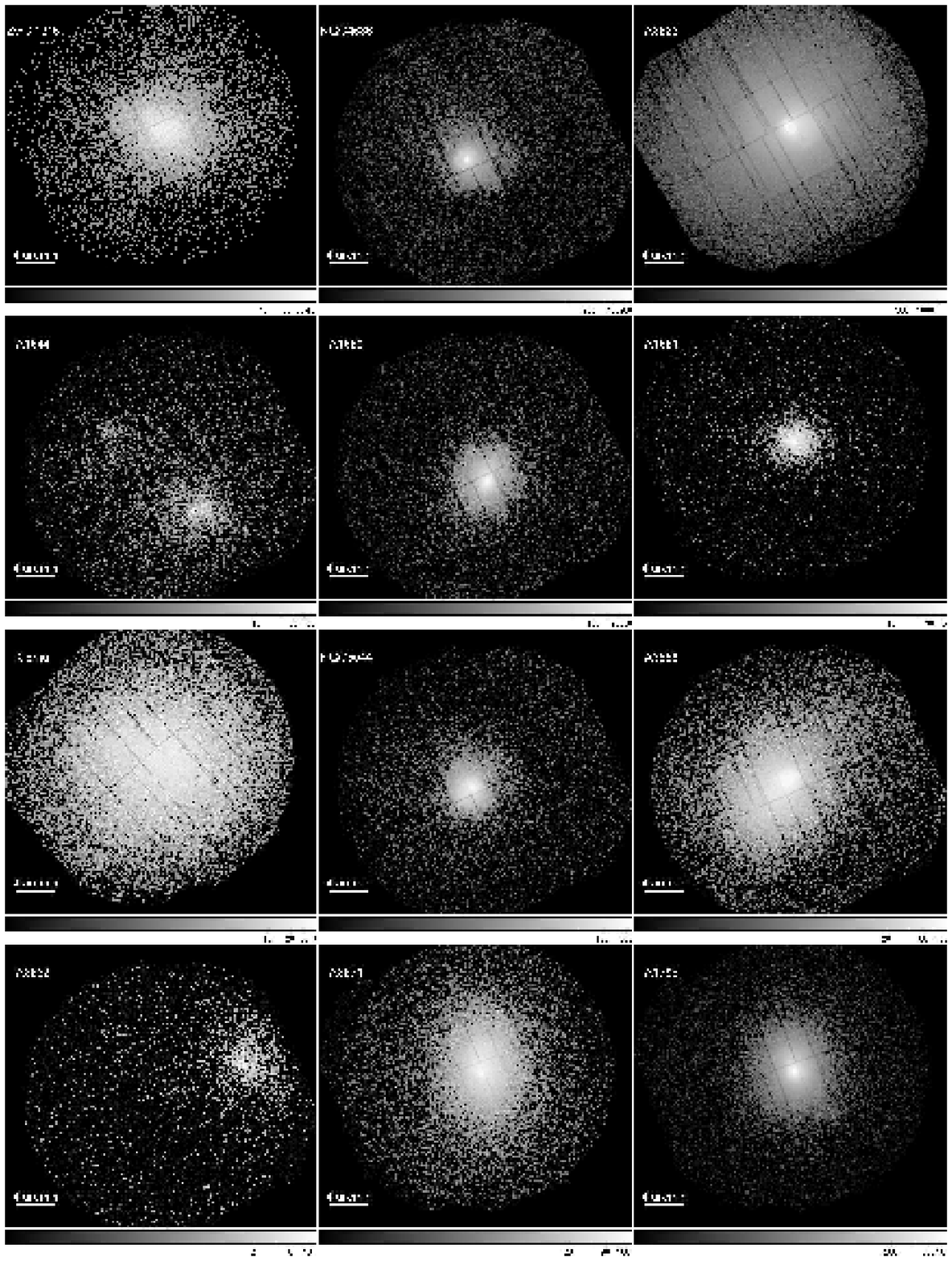}
\end{center}
\caption{Combined MOS and pn images of the clusters in the 0.7--2~keV band,
where point sources have been excised and refilled with values from neighboring
pixels.
\label{f:clid3}}
\end{figure*}

\begin{figure*}
\begin{center}
\includegraphics[angle=0,width=18cm]{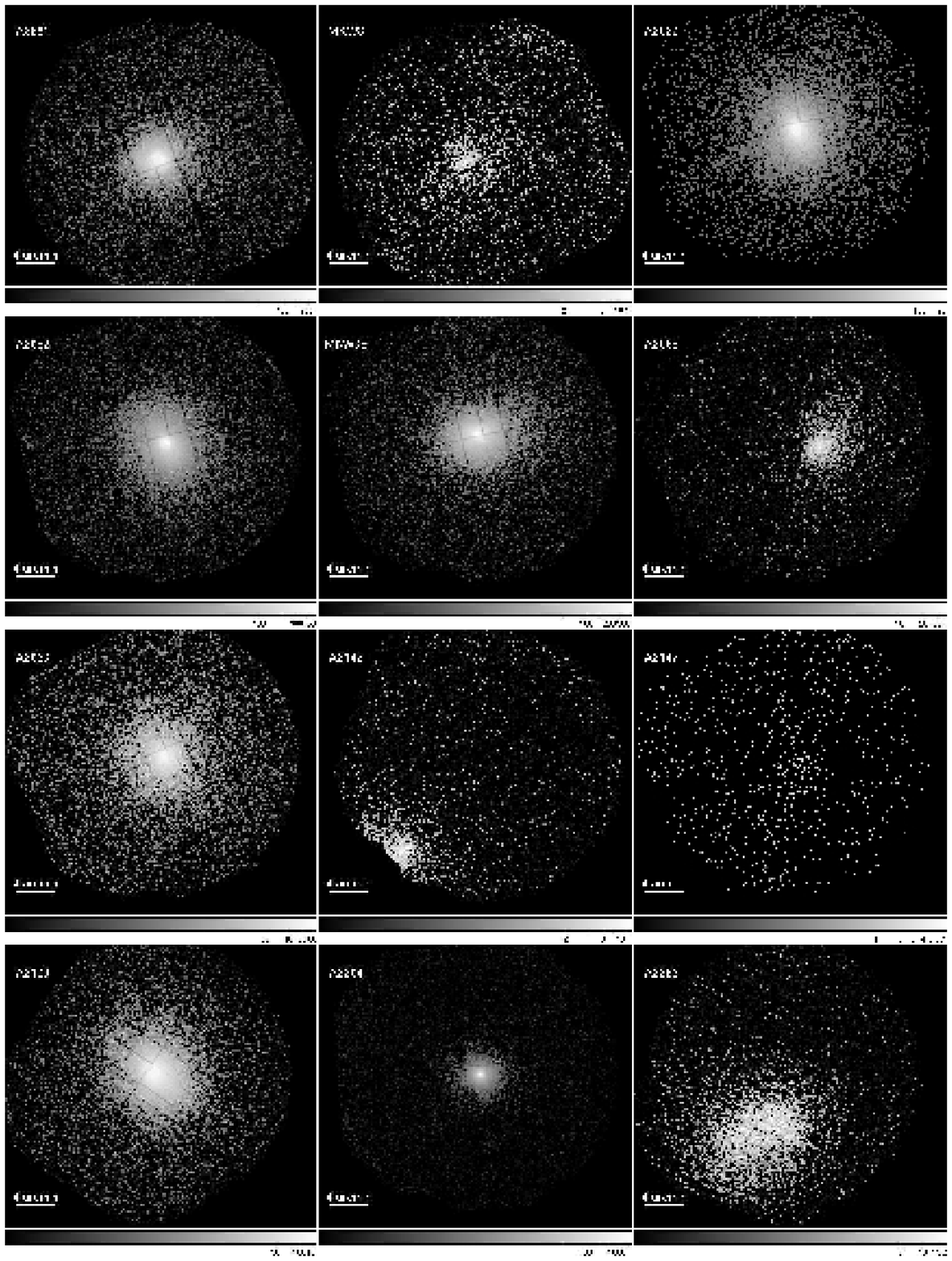}
\end{center}
\caption{Combined MOS and pn images of the clusters in the 0.7--2~keV band,
where point sources have been excised and refilled with values from neighboring
pixels.
\label{f:clid4}}
\end{figure*}

\begin{figure*}
\begin{center}
\includegraphics[angle=0,width=18cm]{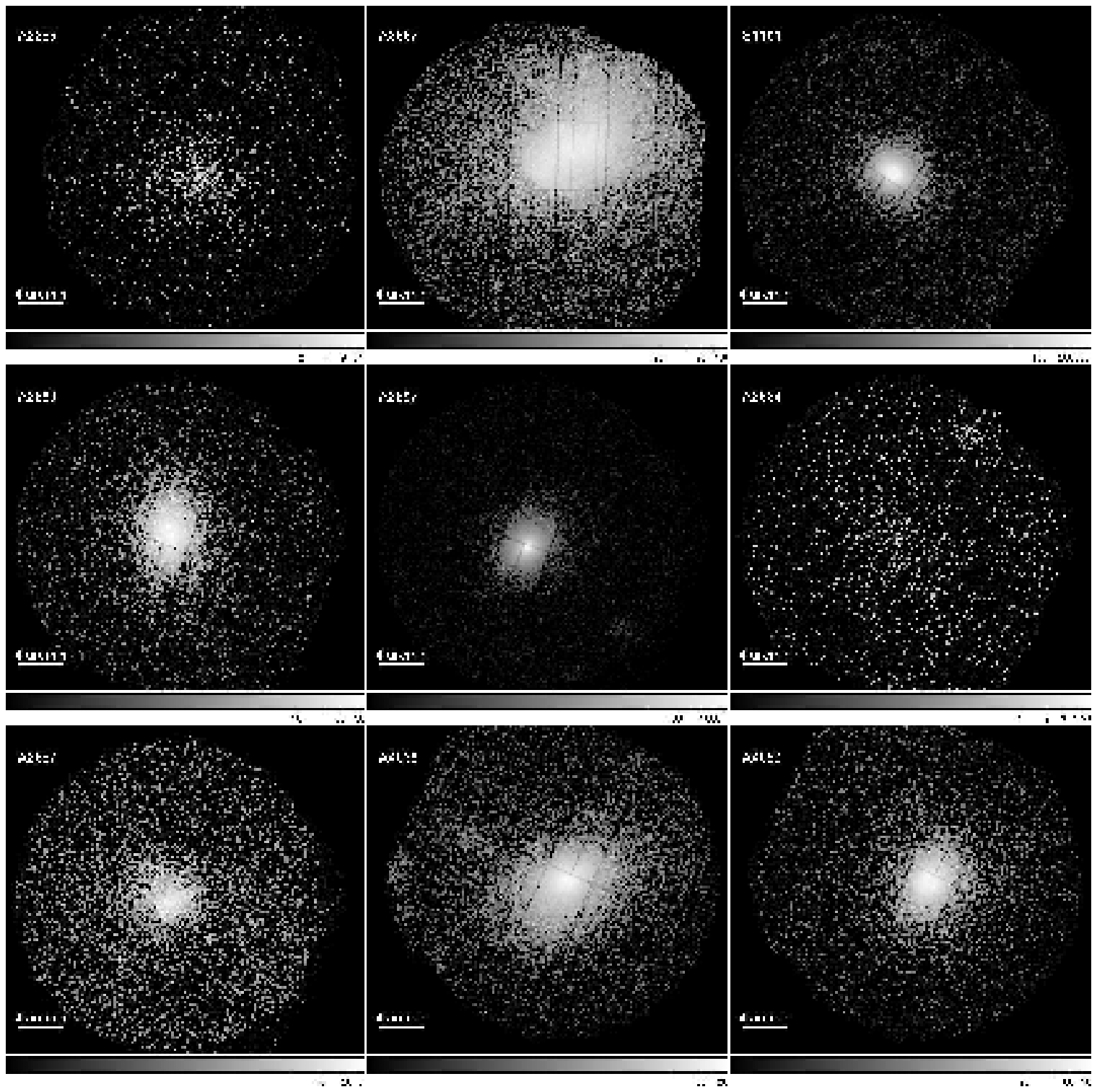}
\end{center}
\caption{Combined MOS and pn images of the clusters in the 0.7--2~keV band,
where point sources have been excised and refilled with values from neighboring
pixels.
\label{f:clid5}}
\end{figure*}

As the soft band is insensitive to the cluster temperature and has data of
high signal-to-noise ratio, we use the MOS and pn combined image in the
$0.7-2$~keV band to illustrate the X-ray morphological substructure of each
cluster (Figs.~\ref{f:clid1}--\ref{f:clid5}). X-ray point-like sources are
identified and subtracted. The holes, where the point-like sources were, are
re-filled with the \emph{Chandra} CIAO routine ``dmfilth'' using randomization
based on the surface brightness distribution around the holes. We only
  use this image to demonstrate the existence of morphological substructure in
  the cluster. Significant substructure features shown in the image are
excised before we perform the spectral and surface brightness analysis.

As addressed in Sect.~\ref{s:sub}, 13 of the 16 clusters with large offsets
between the X-ray flux-weighted centers (see Table~\ref{t:sigma}) and BCG
positions are disturbed clusters (see Table~\ref{t:xmm}). We now comment on
these 13 clusters. The BCGs in A0399 and A1736 are slightly offset from the
main X-ray emission. The ICM in A3376, A0754, A2256, and A3667 exhibits a
comet-like tail, and their BCGs are at the opposite end from the X-ray centers
probably because of their on-going dynamical activity. A3395s is the south
component of a bi-cluster, and its BCG is at an X-ray weak bright peak. The
ICM in A1367 has multi-peaks, and the BCG is at the northwest X-ray peak, which
is not the brightest one. The BCG in A2163 (A2255) is not a dominant BCG,
which sits slightly east (west) of the X-ray center. This also applies but
less significantly to some more clusters in the sample. The ICM in Coma,
A3558, and A2065 shows some weakly disturbed features, and their BCGs are only
40--60~kpc away from the X-ray centers. A3158, A3391, and A0576 are relaxed
clusters, and their BCGs are $\simgt 40$~kpc away from the X-ray centers.
 
\section{Systematic errors in estimates of $\sigma$}
\label{a:sigmaerr}

\begin{figure*}
\begin{center}
\includegraphics[angle=270,width=8cm]{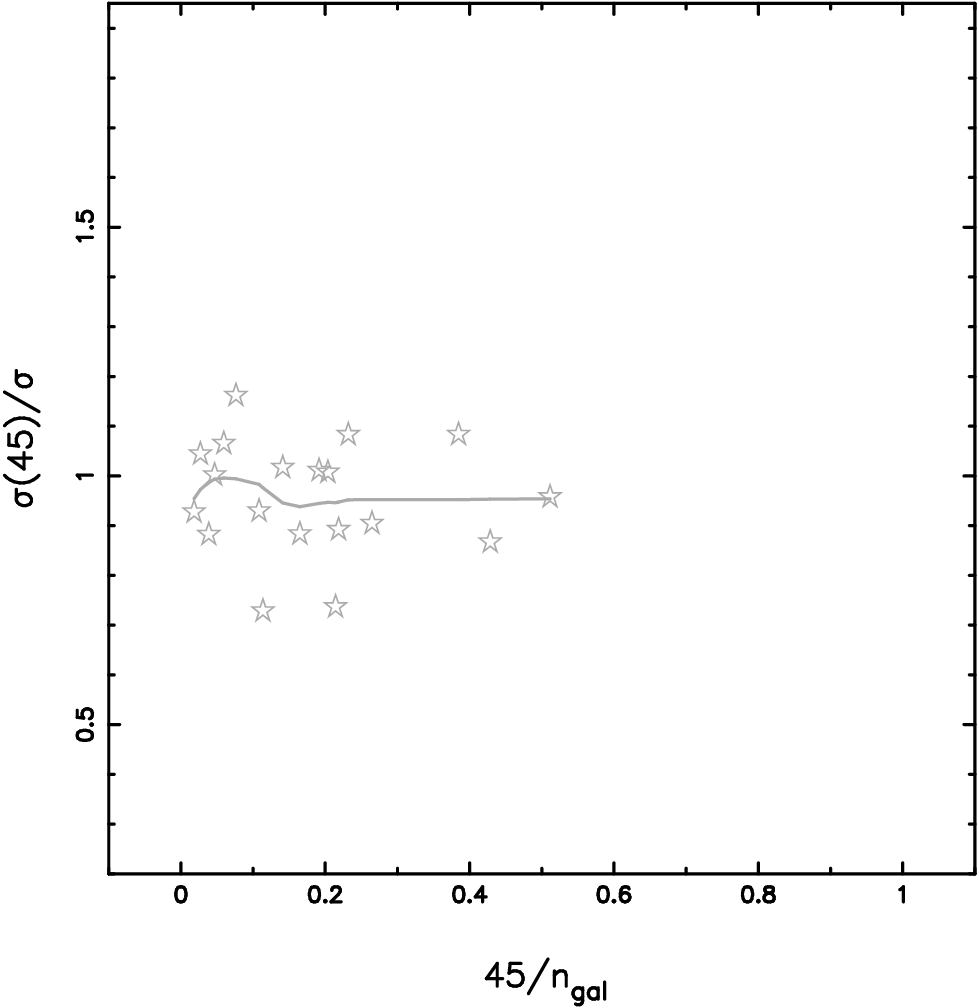}
\includegraphics[angle=270,width=8cm]{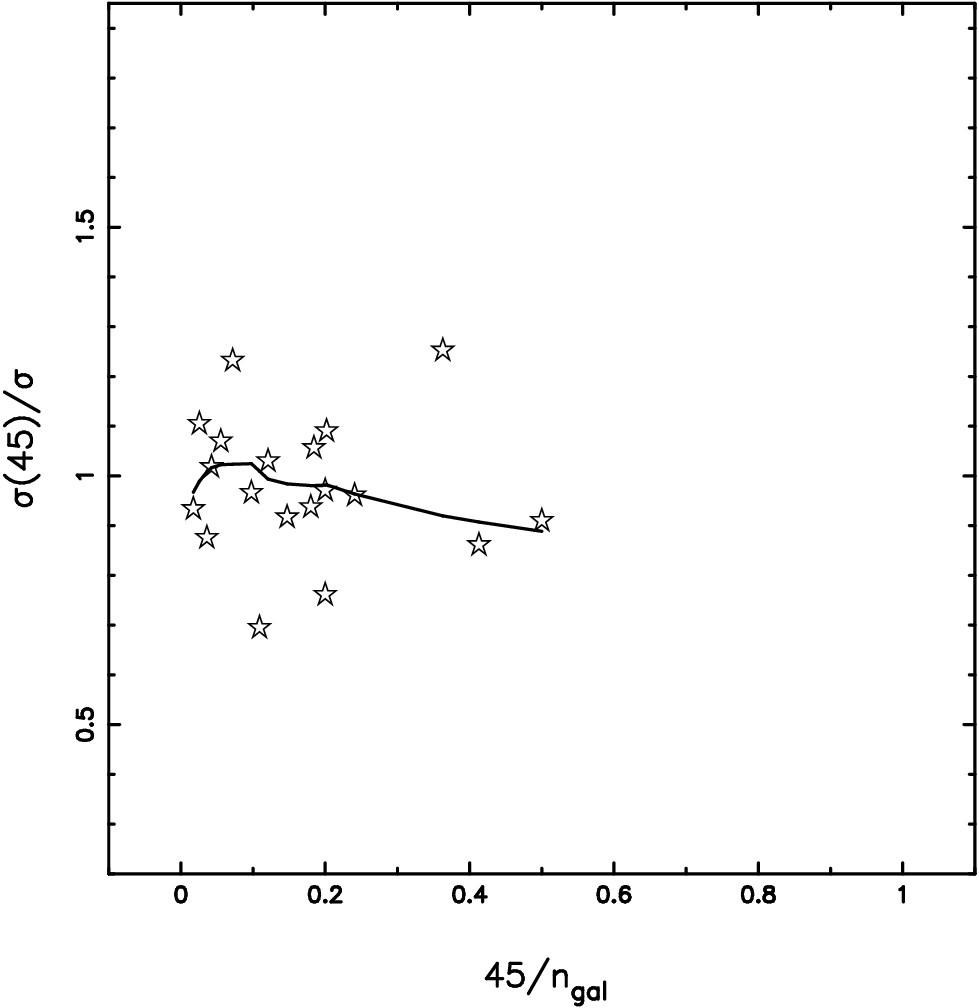}

\includegraphics[angle=270,width=8cm]{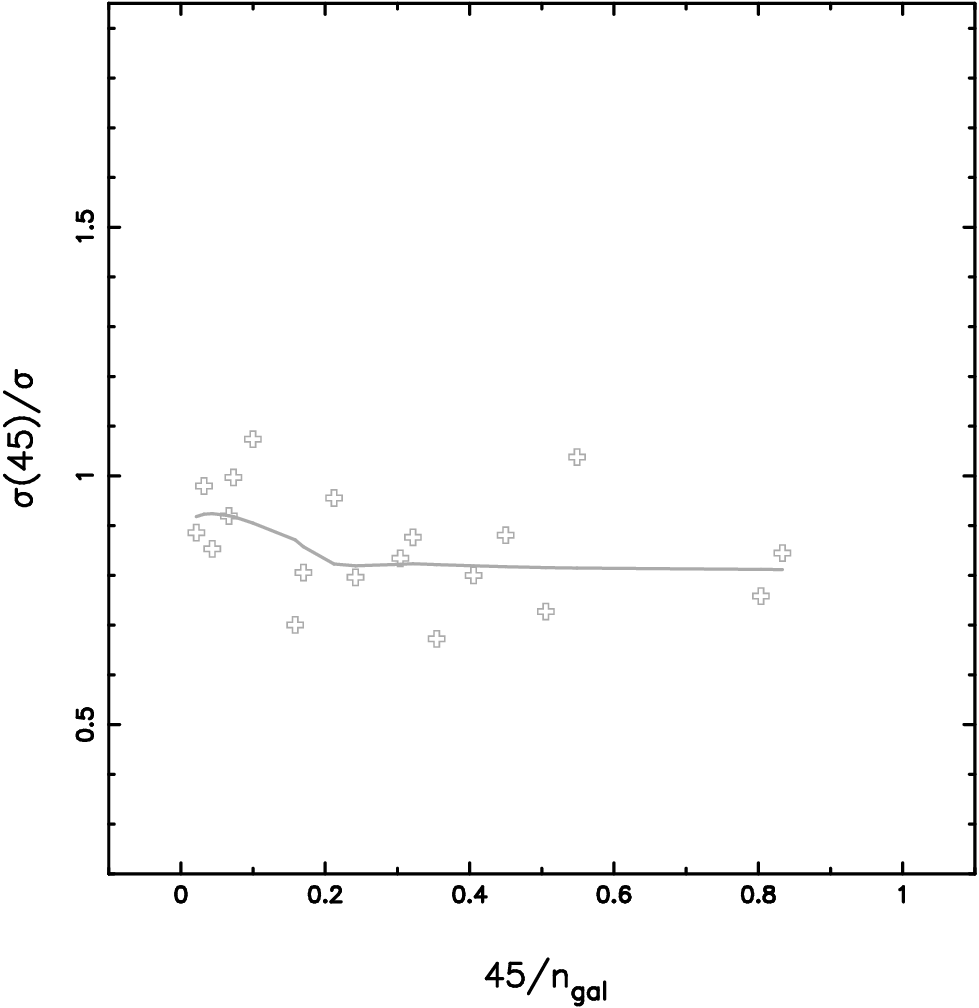}
\includegraphics[angle=270,width=8cm]{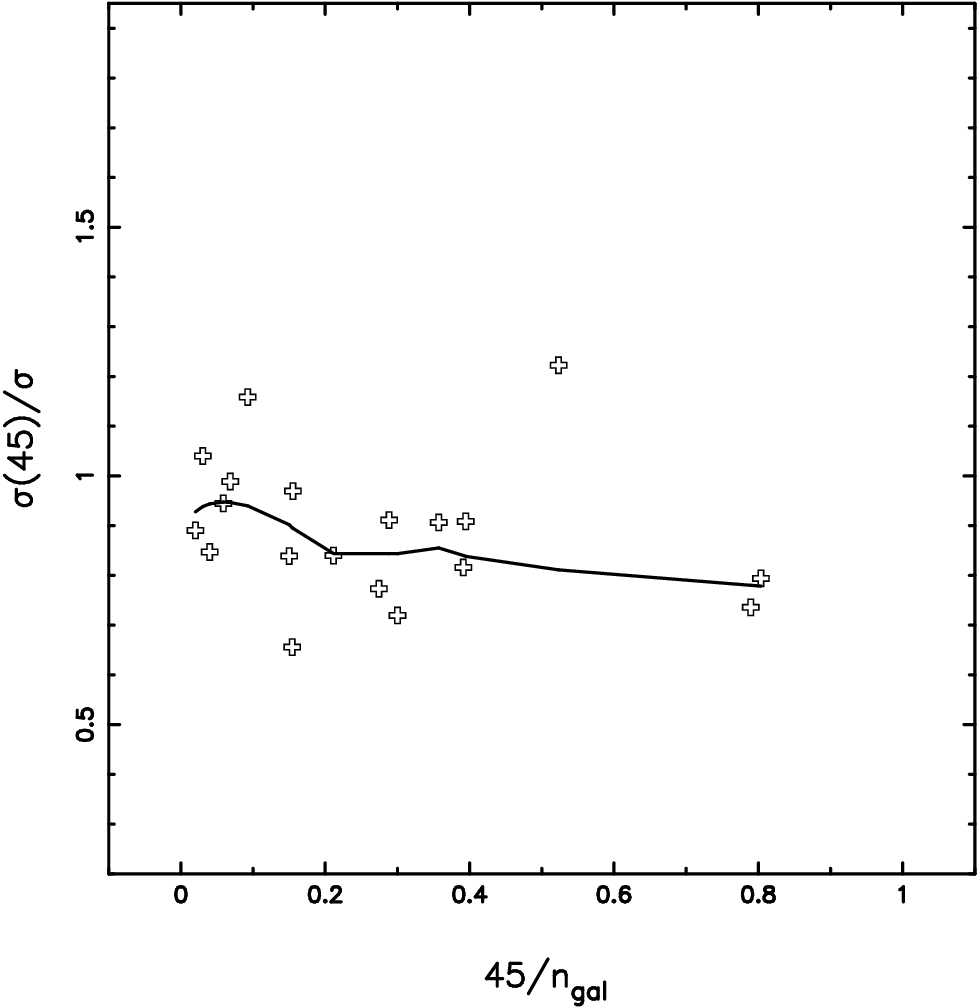}
\end{center}
\caption{Velocity dispersion measured by the 45 most massive galaxies normalized by the
  velocity dispersion within 1.2~Abell radii as a function of the
  fraction of galaxies for the simulated sample. The results are only
  based on the simulated sample of the 21 clusters. The colors and
  symbols have the same meaning as those in
  Fig.~\ref{f:sigma_dist_simu}. The curves are the local regression
  non-parametric fits.
  \label{f:sigma_45_simu}}
\end{figure*}

\begin{figure*}
\begin{center}
\includegraphics[angle=270,width=8cm]{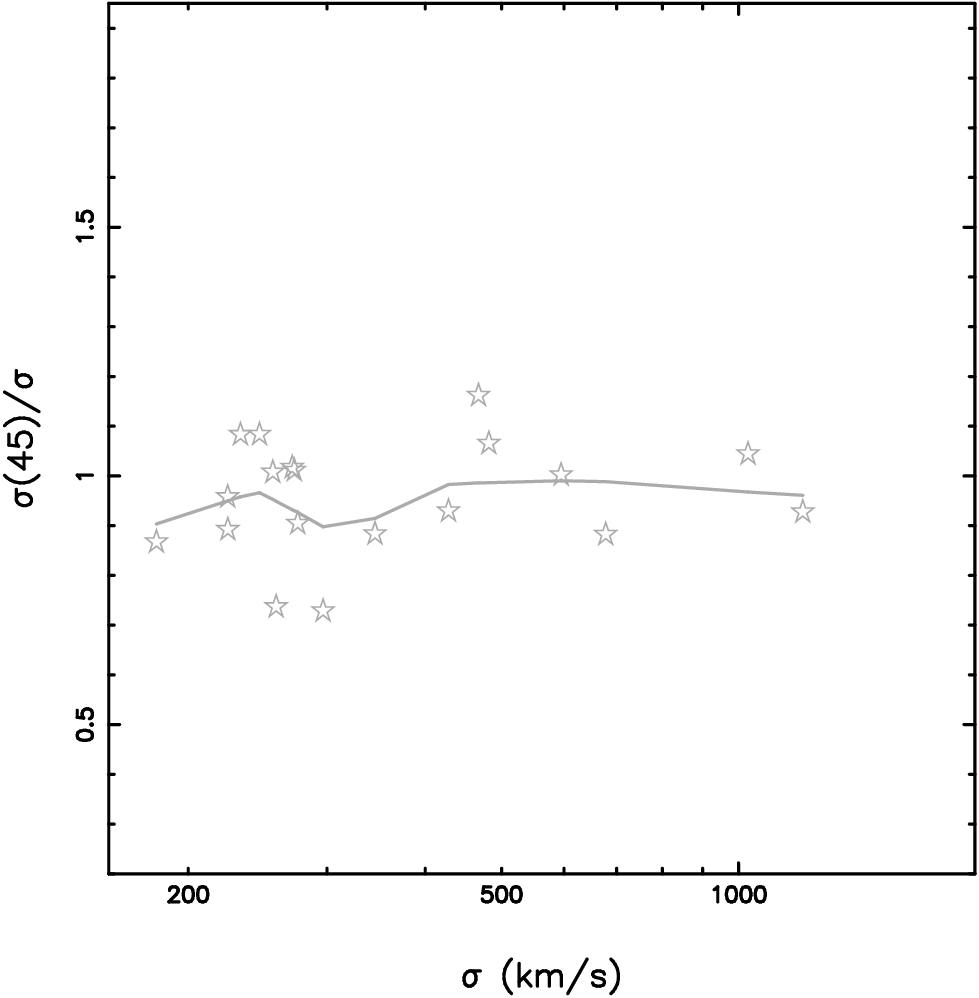}
\includegraphics[angle=270,width=8cm]{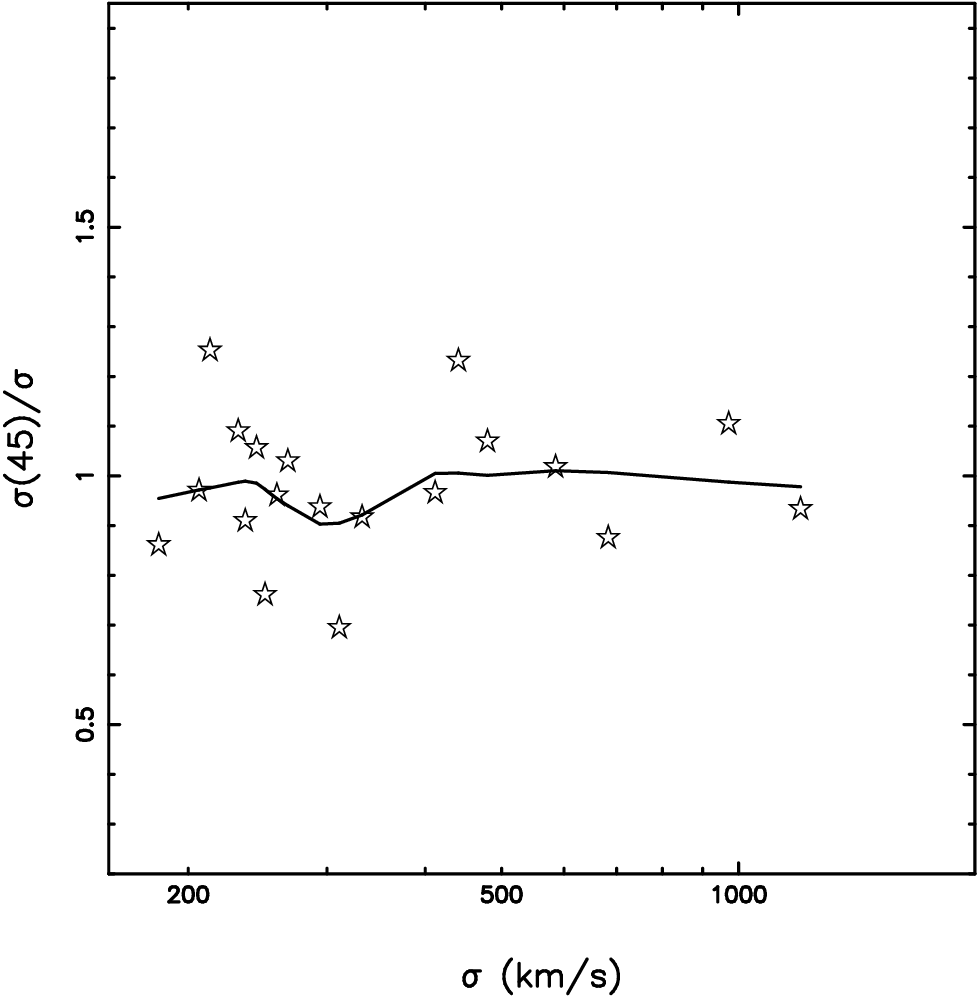}

\includegraphics[angle=270,width=8cm]{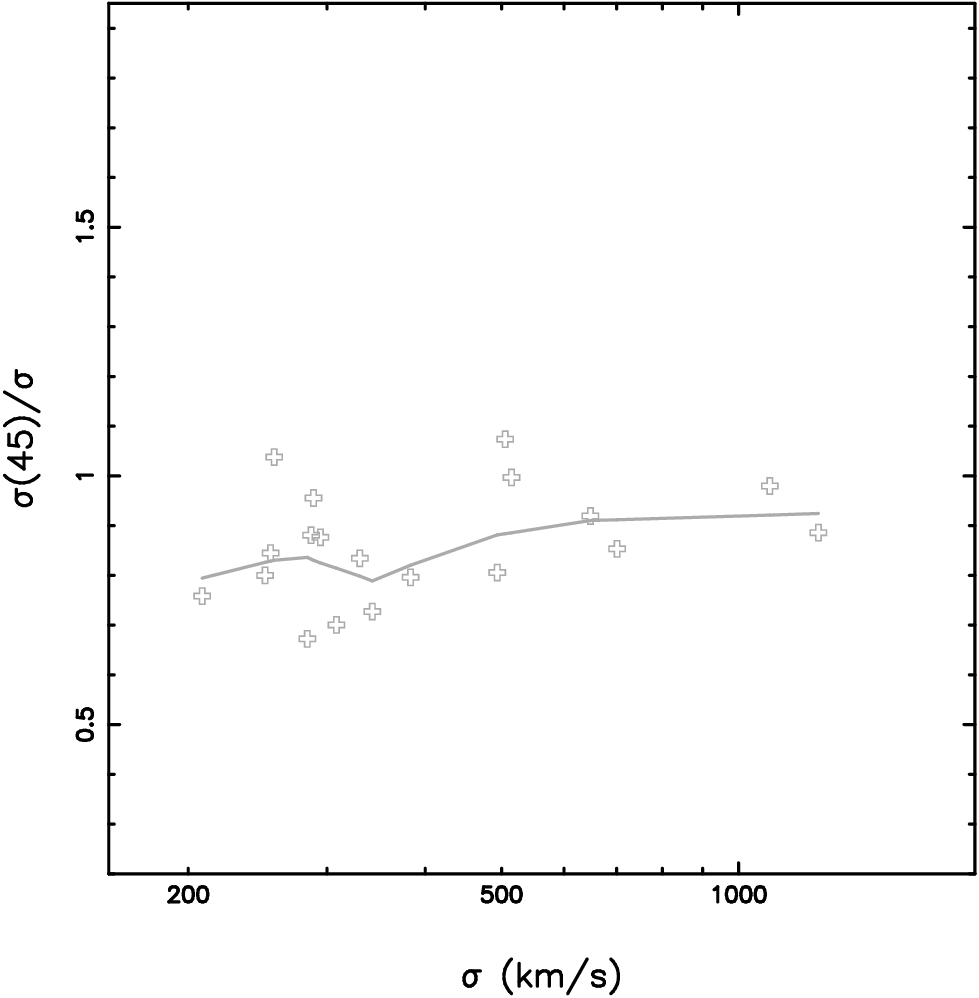}
\includegraphics[angle=270,width=8cm]{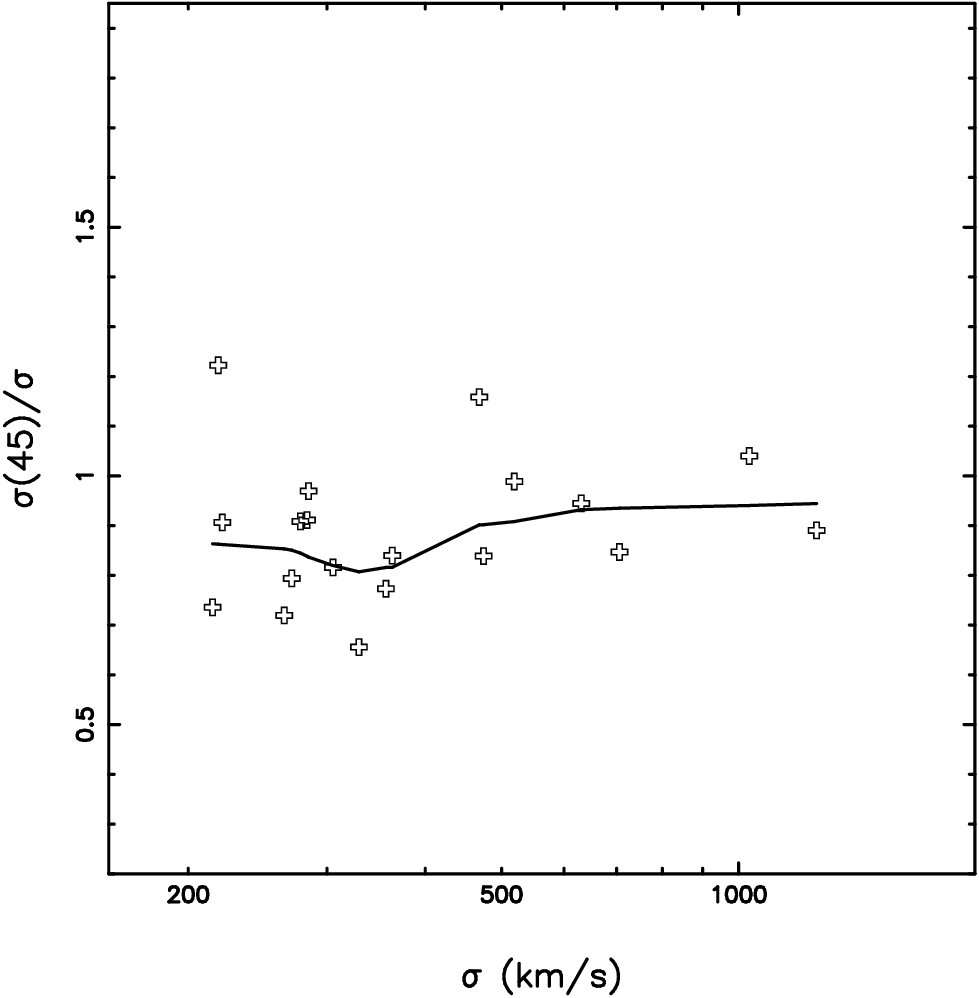}
\end{center}
\caption{Velocity dispersion measured by the 45 most massive galaxies normalized by the
  velocity dispersion within 1.2~Abell radii as a function of
  velocity dispersion for the simulated sample. The results are only
  based on the simulated sample of the 21 clusters. The colors and
  symbols have the same meaning as those in
  Fig.~\ref{f:sigma_dist_simu}. The curves are the local regression
  non-parametric fits.
\label{f:sigma_45_sigma_simu}}
\end{figure*}

\end{document}